\documentclass{aa}  

\usepackage{graphicx}
\usepackage{txfonts}

\newcommand{\ha}{H$\alpha\ $}

\begin{document}

   \title{Physics of ULIRGs with MUSE and ALMA: The PUMA project}

   \subtitle{I. Properties of the survey and first MUSE data results}

   \author{M. Perna          \inst{\ref{i1},\ref{i2}}\thanks{E-mail: mperna@cab.inta-csic.es}
                \and 
           S. Arribas
                \inst{\ref{i1}} 
                \and
           M. Pereira Santaella
                \inst{\ref{i1}} 
                \and
           L. Colina
                \inst{\ref{i1}}
                \and
           E. Bellocchi
                        \inst{\ref{i3}}
                        \and
                   C. Catal\'an-Torrecilla
                \inst{\ref{i1}} 
                \and    
           S. Cazzoli
                \inst{\ref{i4}} 
                \and    
           A. Crespo G\'omez 
                \inst{\ref{i1}} 
                \and
           R. Maiolino
                \inst{\ref{i5}, \ref{i6}, \ref{i7}}
                \and
           J. Piqueras L\'opez
                \inst{\ref{i1}} 
                \and
           B. Rodr\'iguez del Pino 
                \inst{\ref{i1}} 
                }

   \institute{Centro de Astrobiolog\'ia, (CAB, CSIC--INTA), Departamento de Astrof\'\i sica, Cra. de Ajalvir Km.~4, 28850 -- Torrej\'on de Ardoz, Madrid, Spain\label{i1}
       \and
       INAF - Osservatorio Astrofisico di Arcetri, Largo Enrico Fermi 5, I-50125 Firenze, Italy\label{i2}
       \and
       Centro de Astrobiolog\'ia (CSIC-INTA), ESAC Campus, 28692 Villanueva de la Ca\~nada, Madrid, Spain\label{i3}
       \and 
       IAA - Instituto de Astrof{\'i}sica de Andaluc{\'i}a (CSIC), Apdo. 3004, 18008, Granada, Spain\label{i4}
        \and
       University of Cambridge, Cavendish Laboratory, Cambridge CB3 0HE, UK\label{i5}
       \and
       University of Cambridge, Kavli Institute for Cosmology, Cambridge CB3 0HE, UK\label{i6}
       \and
       Department of Physics and Astronomy, University College London, Gower Street, London WC1E 6BT, UK\label{i7}
             }

   \date{Received September 15, 1996; accepted March 16, 1997}

 
  \abstract
   {Ultraluminous infrared galaxies (ULIRGs) are characterised by extreme starburst (SB) and active galactic nucleus (AGN) activity, and are therefore ideal laboratories for studying the outflow phenomena and their feedback effects. We have recently started a project called Physics of ULIRGs with MUSE and ALMA (PUMA), which is a survey of 25 nearby ($z < 0.165$) ULIRGs observed with the integral field spectrograph MUSE and the interferometer ALMA. This sample includes systems with both AGN and SB nuclear activity in the pre- and post-coalescence phases of major mergers. 
   }
   { The main goals of the project are (i) to study the prevalence of (ionised, neutral, and molecular) outflows as a function of the galaxy properties, (ii) to constrain the driving mechanisms of the outflows (e.g. distinguish between SB and AGN winds), and (iii) to identify and characterise feedback effects on the host galaxy. In this first paper, we present details on the sample selection, MUSE observations, and data reduction, and derive first high-level data products. 
   }
   {MUSE data cubes were analysed to study the dynamical status of each of the 21 ULIRGs observed so far, taking the stellar kinematics and the morphological properties inferred from MUSE narrow-band images into account.
   We also located the ULIRG nuclei, taking advantage of near-infrared (HST) and millimeter (ALMA) data, and studied their optical spectra to infer (i) the ionisation state through standard optical line ratio diagnostics, and (ii) outflows in both atomic ionised ([O {\small III}], H$\alpha$) and neutral (Na {\small ID}) gas. 
   }
   {We show that the morphological and stellar kinematic classifications are consistent: post-coalescence systems are more likely associated with ordered motions, while interacting (binary) systems are dominated by non-ordered and streaming motions. We also find broad and asymmetric [O {\small III}] and Na {\small ID} profiles in almost all nuclear spectra, with line widths  in the range [$300-2000$] km/s, possibly associated with AGN- and SB-driven winds. This result reinforces previous findings that indicated that outflows are ubiquitous during the pre- and post-coalescence phases of major mergers. }
   {}


   \maketitle
%
\section{Introduction}

Theory and observations suggest that galaxies are open systems that evolve into a quasi-stationary state, where flows of baryons determine the growth and evolution of their stellar populations and the supermassive black hole (BH) at their centre (see e.g. \citealt{Schaye2010,Bothwell2013,Sanchez2014,Hopkins2016,Naab2017,Dekel2019}). Gas accretion from the cosmic web or from merging interactions triggers star formation (SF) and activity from active galactic nuclei (AGN); in turn, the energy released by the AGN and the stars heats and pushes the surrounding interstellar medium (ISM), regulating the formation of new stars and the accretion onto the BH with feedback mechanisms (e.g. \citealt{Silk2013, Hopkins2012, Hopkins2016}). The role of these outflows is believed to be crucial because they not only regulate (and possibly quench) SF, but also redistribute dust and metals over large distances within the galaxy, or even expel them into the circumgalactic and intergalactic media (e.g. \citealt{Veilleux2005,Veilleux2020}). In this scenario, the galaxy mass function (e.g. \citealt{Somerville2015}), the mass-metallicity relation (\citealt{Maiolino2019}), and the correlation of the black hole with spheroid mass (\citealt{Kormendy2013}) are shaped by outflows.

Although this general scenario is relatively well established, many details of the processes involved remain largely unknown. An intrinsic difficulty is associated with the fact that the self-adjusted balance of gas accretion, gas outflow, and SF and AGN activity depends on the coupling between physical processes that involve multi-phase (ionised, neutral, and molecular) gas at very different physical scales, from cosmic web structures and mergers to molecular clouds (where SF occurs) and the vicinity of the BH (\citealt{Sanchez2014}).

In recent years, the study of outflows at high-$z$ has gained much interest. Because the early Universe is characterised by episodes of intense SF and  AGN activity (\citealt{Madau2014}), models predict that powerful outflows had a key role in shaping primeval galaxies and their subsequent evolution (e.g. \citealt{Ceverino2018,Pillepich2018,Dave2019}).
Although the observational constraints provided by current facilities are limited, they indeed suggest that galaxy-scale winds are highly prevalent at the peak of SF  ($z \sim  1-3$; e.g. \citealt{Genzel2014,Brusa2015a,Carniani2016,Schreiber2018,Villar2020,Kakkad2020}), and may be extremely powerful at very early epochs (e.g. \citealt{Maiolino2012, Bischetti2019}).

Local ultraluminous infrared galaxies (ULIRGs, with rest-frame [$8-1000\ \mu$m]  L$_{IR}$ in excess of $10^{12} L\odot $) are powered by strong SBs (\citealt{Genzel1998}) and/or AGN (\citealt{Nardini2010}), and therefore are ideal laboratories for studying the outflow phenomena. While scarce in the nearby universe,  luminous star-forming galaxies (SFGs) are  much more numerous at high-$z$; they are responsible for $\sim$ 50\%  of the  total star formation rate (SFR) density at $z \sim$ 2 when their number density is a factor $\sim$ 800x higher than locally (e.g.,  \citealt{Perez2005,Magnelli2013,Casey2014}). For high-$z$ luminous SFGs, however, some properties may differ from those of their local counterparts. While local ULIRGs are dominated by strong interactions and major mergers, recent morphological studies have found that distant luminous SFGs appear to be a mixture of merging interacting systems and disk galaxies (e.g. \citealt{ Arribas2012, Kartaltepe2012,Kaviraj2013,Hung2014,Alcorn2018,Wisnioski2018}), and therefore a fraction may maintain their high observed SFRs by continuous gas accretion through cold gas flows and minor mergers (\citealt{Ocvirk2008,Dekel2009}). Nevertheless, local ULIRGs offer the opportunity of investigating the properties of strong outflows and their feedback effects at similar SFR levels as observed at high-$z$, but at much higher spatial resolution and signal-to-noise ratio.

In the past, ULIRGs have been the target of numerous studies, which have demonstrated the presence of ionised (e.g. \citealt{Heckman1990, Colina1999,  Bedregal2009, Colina2012,  Westmoquette2012, Rupke2013a, Rodriguez2013, Bellocchi2013, Arribas2014, Rich2011, Rich2015}), neutral (e.g. \citealt{Martin2005,Rupke2005a,Rupke2005b,Rupke2005c, Rupke2013a, Cazzoli2016}), and molecular (e.g. \citealt{Feruglio2010,Sturm2011,Spoon2013,Veilleux2013, Rupke2013b, Cicone2014, Emonts2018,PereiraSantaella2018, Fluetsch2019}) outflows. Most of these previous works are limited to the study of a specific gas phase, however, or have resolutions that cannot trace the outflow structure in detail. 

To overcome these limitations, we have recently started a project aimed at studying the outflow phenomena in a sample of 25 representative ULIRGs, based on integral field spectroscopy (IFS) with MUSE at the Very Large Telescope (VLT) and interferometric ALMA data (PUMA: ``Physics of ULIRGs with MUSE and ALMA''). The combination of the data provided by these two facilities offers the possibility of studying the multi-phase structure of the outflow at subkiloparsec (subkpc) angular resolutions over a field of view (FOV) that covers the main body of the system. The key goals of this project include i) studying the prevalence of outflows as a function of the galaxy and BH properties, ii) establishing the relative role of the different gas phases in gas flows, iii) constraining the driving mechanisms of the outflow (e.g.  distinguish between SB- and AGN-driven winds), and iv)  identifying and characterising negative and/or positive feedback, and quantifying their effects on the host galaxy. 
The first such detailed study of the multi-phase galaxy-scale outflows has recently been presented for the closest ULIRG in our sample, Arp 220 (\citealt{Perna2020}).


\begin{figure*}[h]
\centering
\includegraphics[width=18.cm,trim= 0 0 0 25,clip]{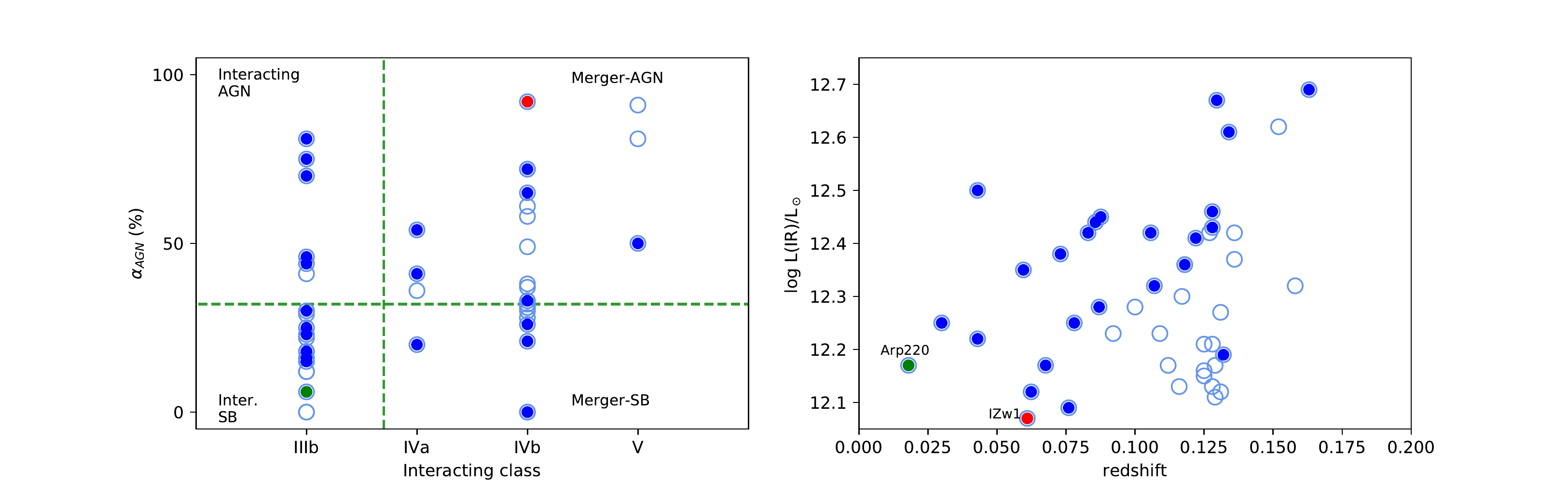}
\caption{\small  Sample properties. Parent sample formed by the 44 targets that meet the selection criteria (empty and solid blue circles) and the 25 objects selected for the MUSE-ALMA sample (filled blue circles), including Arp220 (green) and IZw1 (red). See text for selection details. 
{\it Left}: Interacting class classification, according to the \citet{Veilleux2002} scheme, vs. contribution of the AGN to the total luminosity, according to method 6 of \citet{Veilleux2009}. The vertical green line distinguishes between interacting galaxies and mergers in a more simplified classification; the horizontal line is used to separate SB and AGN targets.
{\it Right}: Redshift vs. IR luminosity. For IZw1 the bolometric luminosity is considered instead of L$_{IR}$.  
}
\label{Sample}
\end{figure*}


\begin{table*}
\tiny
\begin{minipage}[h]{1\linewidth}
\setlength{\tabcolsep}{4pt}
\centering
\caption{Sources in the PUMA survey}
\begin{tabular}{l c c c c c c c c c c c c }
IRAS ID (Other)   &   RA           & DEC         &  D & log $L_{IR}$ & $q$ & SB-AGN & $\alpha_{AGN}$ & \multicolumn{2}{c}{optical} & morph & IC & NS\\
      & ($^h:\ ^m:\ ^s$) & ($^{\circ}:\ ':\ '')$ & (Mpc) & ($L_\odot$) & & & & \multicolumn{2}{c}{type}  & &  & (kpc)\\
\scriptsize{(1)} & \scriptsize{(2)}   &\scriptsize{(3)}   & \scriptsize{(4)}   &\scriptsize{(5)}   & \scriptsize{(6)} & \scriptsize{(7)} & \scriptsize{(8)} & \scriptsize{(9)} & \scriptsize{(10)} & \scriptsize{(11)} & \scriptsize{(12)} & \scriptsize{(13)}\\
\hline

F00091-0738                             & 00:11:43.27 & $-$07:22:07.34  & $519$ & $12.27\pm 0.03$ & 2.81 & AGN    & $0.46 (0.58) $        & HII & --         & I & IIIb & 2.3\\
F00188-0856                             & 00:21:26.51 & $-$08:39:25.99  & $563$ & $12.41\pm 0.04$ & 2.32 & AGN    & $0.50 (0.35) $        & LI(N)ER & Sy 2 & M & V & $<$ 0.3\\
F00509+1225  (IZw1)             & 00:53:34.93 &+12:41:35.94             & $243$ & $11.30\pm 0.05$ $^1$ & 2.51 & AGN       & $0.92                 $         & Sy 1& Sy 1 & M & IVb & $<$ 0.2\\
F01572+0009  (Mrk 1014)                 & 01:59:50.25 &+00:23:40.88             & $717$ & $12.63\pm 0.03$ & 2.01 & AGN    & $0.65 (0.27)$         & Sy 1& Sy 1       & M & IVb & $< 0.4$ \\
F05189-2524                             & 05:21:01.40 & $-$25:21:45.30  & $189$ & $12.17\pm 0.01$ & 2.72& AGN     & $0.72 (0.30)$         & Sy 2 & Sy 2    & M & IVb & $<$0.1\\
07251-0248                                      & 07:27:37.61 & $-$02:54:54.25         & $390$ & $12.41\pm 0.04$ & 2.76 & SB           & $0.30 (0.39)$         & --  &  HII      & I     &IIIb  & 1.8 \\
09022-3615                                      & 09:04:12.71 & $-$36:27:01.93          & $268$ & $12.30\pm 0.01$ &2.21 &AGN ?    & $0.54 (0.09)$         & HII   & HII     & M     &IVa & $<0.1$\\
10190+1322                                      & 10:21:42.94 & +13:06:53.83         & $343$ & $12.05\pm 0.03$ & 2.41 &SB            & $0.17$                & HII   & HII     & I     & IIIb & 7.2\\
F11095-0238                                     & 11:12:03.38 &$-$02:54:22.94         & $477$ & $12.28\pm 0.03$ & 2.15 &AGN   & $0.44 (0.60)$         & LI(N)ER & ? &I  & IIIb & 1.1\\
F12072-0444                                     & 12:09:45.12 &$-$05:01:13:31         & $573$ & $12.40\pm 0.03$ & 2.55 &AGN   & $0.75 (0.41)$         & Sy 2    & Sy 2 &I       & IIIb & 2.3\\
F12112+0305                             & 12:13:46.06 &+02:48:41.53     & $329$ & $12.33\pm 0.02$ &2.65 &SB               & $0.17$                & LI(N)ER & -- & I        & IIIb & 4.18\\
13120-5453 (WKK 2031)                   & 13:15:06.32 &$-$55:09:22.82   & $139$ & $12.25\pm 0.03$ &2.91 &AGN      & $0.33 (0.01)$         & Sy 2 & Sy 2    & M     &IVb & $<0.1$ \\
F13451+1232                     & 13:47:33.36 &+12:17:24.24             & $542$ & $12.31\pm 0.04$ &$-0.34$ &AGN   & $0.82( 0.60)$  &       Sy 2 & Sy 2 & I   & IIIb & 4.3\\
F14348-1447                                     & 14:37:38.28 &$-$15:00:24.24   & $370$ & $12.35\pm 0.02$ &2.37 &SB               & $0.17 (0.04)$  &       LI(N)ER & ? & I         &IIIb & 5.45\\
F14378-3651                             & 14:40:59.01 &$-$37:04:31.93   & $302$ & $12.13\pm 0.02$ & 2.35&AGN      & $0.21 (0.01)$         & Sy 2 & Sy 2    & M     &IVa & $< 0.1$\\
F15327+2340 (Arp220)            & 15:34:57.22 &+23:30:11.50             & $82$   & $12.17\pm 0.01$ &2.61 &SB              & $0.06 (0.17)$         & LI(N)ER & LI(N)ER  & I  & IIIb & 0.37\\
F16090-0139                             & 16:11:40.42 &$-$01:47:06.35   & $592$ & $12.57\pm 0.02$ &2.34 &SB               & $0.41 (0.24)$         & LI(N)ER & HII & M  &IVa  & $<$0.7$^2$\\
F16156+0146                             & 16:18:09.37 & +01:39:21.76    & $585$ & $12.11\pm 0.06$ & 2.18 &AGN     & $0.70 (0.40)$         & Sy 2 &        -- & I     & IIIb & 8.26\\
17208-0014     & 17:23:21.96 & $-$00:17:00.86 & 189 & $12.50 \pm 0.02$ &2.70 &SB & 0.00 (0.11) & HII & LI(N)ER  & M & IVb & $< 0.1$\\
F19297-0406                             & 19:32:22.30 &$-$04:00:01.80   & $376$ & $12.40\pm 0.02$ &2.49 &SB               & $0.23 (0.01)$         & HII & ? & I & IIIb$^3$ & $1.1$\\
19542+1110                              & 19:56:35.78 &  +11:19:05.03   & $284$ & $12.12\pm 0.04$ & 2.55&SB               & $0.26 (0.04)$         & LI(N)ER & LI(N)ER  & M  &IVb & $<0.1$ \\
20087-0308                                      & 20:11:23.87 &$-$02:59:50.71   & $463$ & $12.44\pm 0.03$ & 3.01&SB               & $0.20 (0.03)$         & LI(N)ER &  LI(N)ER & M  &IVa & $<$ 0.9\\

20100-4156                                      & 20:13:29.56 &$-$41:47:35.21   & $571$ & $12.64\pm 0.02$ & 2.49&SB               & $0.26 (0.20)$         & HII &   LI(N)ER& I      & IIIb & 6.5 \\
F20414-1651                             & 20:44:18.16 &$-$16:40:16.82   & $381$ & $12.30\pm 0.08$ & 2.47 &SB              & $0.00$                & HII &-- & M & IVb & $<0.2$\\

F22491-1808                                     & 22:51:49.24 &$-$17:52:23.66   & $339$ & $12.19\pm 0.03$ & 3.03&SB               & $0.15$                        & HII &   HII & I & IIIb & 2.7\\

\hline
\label{Tsample}
\end{tabular}
\label{tab1}
\vspace{0.2cm}
\end{minipage}
{\small
{\it Notes.} Column (1): IRAS identification code (alternative name); 
(2) and (3): RA and DEC coordinates of the ULIRG nuclei from this work; for binary systems, we report the coordinates of the near-IR brightest nucleus; 
(4): Hubble distance (from the NED); 
(5): IR luminosity;
(6): q radio-IR parameter (\citealt{Condon1991});
(7): simplified AGN-SB  classification based on the information provided in columns (8) and (9); 
(8): fraction of AGN contribution to L$_{bol}$ derived from the 30 to 15 $\mu$m flux ratio ($f30/f15$). This ratio gives a good first-order approach to the contribution of the AGN to L$_{bol}$ because it agrees to within $10-15\%$ with the average value of the six methods considered in \citet{Veilleux2009}. In order to homogenise the various calibrations of this ratio, we took the results of \citet{Veilleux2013} as reference and fit a second-order polynomial to the log($f30/f15$) function. The  flux ratio values are from \citet{Veilleux2009,Veilleux2013} and \citet{Spoon2013}. Values in brackets are derived according to the method of \citet{Nardini2010}, which is based on the analysis of the $5 - 8\ \mu$m spectral range; 
(9-10) optical spectral type according to \citet{Veilleux2009,Veilleux2013} in column 9, and according to our classification obtained from nuclear spectra (Sect. \ref{Sism}) in column 10;  
(11): simplified morphology classes (I: interacting, M: mergers); see text for details.
(12): IC according to the scheme by \citet{Veilleux2002}, obtained from \citet{Veilleux2009} or derived from HST imaging;
(13): Nuclear separation from \citet{Veilleux2009}, \citet{Duc1997}, and \citet{Kim2013}. \\
$^1$ [$40-500$] $\mu$m IR luminosity from \citet{Petric2015}.  This target does not meet the IR luminosity criterion for being a ULIRG (although L$_{bol} > 10^{12}$ L$_{\odot}$), but it has been included because it is usually considered an archetypical type 1 AGN.\\
$^2$ No HST images are available for F16090-0139, therefore the NS upper limit has been obtained on the basis of our ALMA images (Pereira-Santaella et al., in prep).  \\
$^3$ F19297-0406, originally associated with a single nucleus (\citealt{Kim2013}), is here classified as a IIIb system because two nuclei in near-IR HST and ALMA (Pereira-Santaella et al., in prep) images are detected.
}

\end{table*}


\begin{table}
\footnotesize
\begin{minipage}[!h]{1\linewidth}
\setlength{\tabcolsep}{6pt}
\centering
\caption{MUSE observation resolutions and exposure times}
\begin{tabular}{lccc}
IRAS ID    &   resolution & seeing & t$_{exp}$\\
               &  ($''$ / kpc) & ($''$) & (h)\\
\scriptsize{(1)} & \scriptsize{(2)}   &\scriptsize{(3)}   & \scriptsize{(4)} \\
\hline

F00188-0856   &  $0.70$ / $1.60$ & 0.7 & 0.68\\

IZw1  &  $0.70$ / $0.80$ & 0.8 & 2.72\\

F01572+0009  &  $0.70$ / $2.00$& 0.8 & 2.04\\

F05189$-$2524   &  $0.60$ / $0.50$& 0.7 & 2.04\\

07251$-$0248   &  $0.55$ / $0.45$& 0.7 & 2.55\\

09022$-$3615   &  $0.60$ / $0.70$& 0.9 & 2.04\\

10190+1322   & $0.7$ / $1.00$& 0.8 & 2.04\\

F11095$-$0238   & $0.75$ / $1.50$ & 0.9 & 2.72\\

F12072$-$0444   & $0.80$ / $1.90$ & 0.8 & 0.68\\

13120$-$5453*   & $0.90$ / $0.60$ & 0.9 & 0.52\\

F13451+1232   & $0.75$ / $1.70$& 0.9 & 2.04\\

F14348$-$1447  & $0.70$ / $1.10$& 1.0 & 0.68\\

F14378$-$3651* &$0.90$ / $1.10$& 1.0& 0.47\\

Arp220  & $0.55$ / $0.20$& 1.0 & 0.65 \\

F16090$-$0139  & $0.70$ / $1.70$& 1.1 & 0.17\\

17208$-$0014* & $0.55$ / $0.45$& 0.4& 0.39\\

F19297$-$0406  & $0.65$ / $1.05$& 1.0 & 2.04\\

19542+1110     & $0.70$ / $0.80$ & 0.7 & 2.04\\

20087$-$0308     & $0.70$ / $1.40$& 1.1 & 2.04\\

20100$-$4156     &  $0.60$ / $1.40$ & 0.8 & 2.04\\

F22491$-$1808* & $0.50$ / $0.70$& 0.4 & 0.41\\

\hline
\end{tabular}
\label{Tlogfile}
\vspace{0.2cm}
\end{minipage}
{\small
{\it Notes.} Column (1): Target name. (2): Measured angular resolution at 7000$\AA$, and physical scale at the redshift of the detected target. (3): Mean DIMM
seeing during the time of the observation. (4): Total integration time.\\
$^{*}$: Seeing-limited MUSE observations from the programs 0102.B-0617(A) (13120-5453 and F14378-3651) and 0101.B-0368(B) (17208-0014 and F22491-1808).
}
\end{table}

In this paper, we present the MUSE data for the entire PUMA sample that are required to trace and characterise the ionised and neutral components of the outflows, while in Pereira-Santaella et al. (in prep.) we describe the associated ALMA data that we use to study the molecular phase. The current MUSE data represent a significant improvement in quality with respect to previous surveys of ULIRGs based on the use of first-generation IFS instruments such as INTEGRAL (\citealt{Garcia2009a,Garcia2009b}) and VIMOS (\citealt{Arribas2008}). 
The MUSE data presented here also provide a complementary view of outflow phenomena with respect to other MUSE surveys, for instance targeting local Seyfert galaxies (e.g. MAGNUM, \citealt{Mingozzi2019}; CARS, \citealt{Husemann2019}), whose outflows would likely originate from AGN, or galaxies with prominent nuclear rings and bars (TIMER, \citealt{Gadotti2019}), whose outflows would probably originate from AGN and/or SB activity triggered by secular processes. 

This paper is organised as follows. In Sect. \ref{Ssample} we describe the sample selection. Section \ref{Sreduction} presents the MUSE observations and data reduction. In Sect. \ref{Smaps} we describe the main morphological properties of our ULIRGs, as inferred from MUSE optical continuum and line feature structure, and from HST near-infrared (near-IR) images. 
In Sect. \ref{Ssanalysis} we describe the dynamical status of each ULIRG, taking into account the morphological properties and the stellar kinematics derived in this work. Section \ref{Sism} presents the nuclear gas properties inferred from the analysis of MUSE spectra in terms of ionisation conditions and multi-phase (ionised and neutral) outflows. Finally, Sect. \ref{Sconclusions} summarises our conclusions.
Throughout this paper, we adopt the cosmological parameters $H_0 =$ 70 km/s/Mpc, $\Omega_m$ = 0.3, and $\Omega_\Lambda =$ 0.7.

\section{Sample selection}\label{Ssample}

The PUMA project is designed to investigate the main properties of outflows in a representative distance-limited ($<$ 800 Mpc, $z<$ 0.165) sample of ULIRGs. The targets are selected to cover the widest possible range in host galaxy properties, such as (i) activity class (i.e. we consider systems with both AGN and SB nuclear activity), (ii) interaction stage (from advanced interacting pairs to mergers), and (iii) $8-1000\ \mu$m IR luminosity. 

The parent sample is assembled from the IRAS 1 Jy Survey (\citealt{Kim1998}), the IRAS Revised Bright Galaxy sample (\citealt{Sanders2003}), and the \citet{Duc1997} catalogue. From these surveys, we select ULIRGs with declination between $-65$ and $+20$ degrees, and within the required distance range (i.e. $z <$ 0.165).  We also request the targets to have mid-IR Spitzer spectroscopy so that the AGN contribution can be estimated reliably even if it is completely obscured at optical wavelengths (e.g. \citealt{Veilleux2009,Veilleux2013, Spoon2013}). More than $80\%$ of the general population of ULIRGs are in close pairs or in already coalescent systems (\citealt{Veilleux2002}). We therefore focus our sample on advanced interacting systems with nuclear projected separations smaller than 10 kpc (i.e. systems classified as IIIb, IV, and V in the \citealt{Veilleux2002} scheme). Forty-four systems fulfil all these criteria, from which we selected 23 objects that uniformly sample the parameter space discussed above. In particular, about half (12) of the systems are pre-mergers of type IIIb (interacting stage hereinafter), and the remaining targets are in the later type IV and V (merger) stages.

The AGN contribution to the total bolometric luminosity, $\alpha_{AGN}$, was derived considering the 30  to $15\ \mu$m flux ratio (following \citealt{Veilleux2013}): a median value of 32\% is obtained for the selected sample of 23 targets and for the parent sample of 44 objects.
We considered targets with $\alpha_{AGN}$ above this median value as significantly affected by the AGN.  
We also added  the Arp220 system (although its declination is slightly higher) and the  IZw1 galaxy (whose IR luminosity does not meet the ULIRG criterion, although L$_{bol} > 10^{12}$ L$_{\odot}$) to the group of  23 targets. These two systems are often considered archetypical of the local luminous SB (Arp220) and type 1 AGN (IZw1). 

We finally note that all but three sources obey the well-known radio$-$far-IR correlation, with a mean q-parameter of 2.37 (\citealt{Yun2001}). The only source with an excess in its radio emission is F13451+1232, with $q = -0.34$. This object hosts the powerful radio source 4C 12.50, with its twin-jet morphology (\citealt{Lister2003}). The systems 20087$-$0308 and F22491$-$1808 are instead associated with an IR excess ($q = 3.009$ and $3.025$, respectively; \citealt{Yun2001}), possibly indicating a dust-enshrouded AGN or compact SBs.

The properties of the final sample of 25 systems we presented so far are reported in Table \ref{tab1}. 
The left panel of Fig. \ref{Sample} shows the distribution of the PUMA systems in the $\alpha_{AGN}-$ interacting class (IC) space: the horizontal line separates SB from AGN systems, while the vertical line separates interacting from merger targets. Our selected sources uniformly sample this two-dimensional parameter space. In addition, the selected objects also uniformly sample the ULIRG luminosity range (right panel of Fig. \ref{Sample}).

\section{MUSE observations}\label{Sreduction}

MUSE observations were conducted as part of our programme ``Subkpc multi-phase gas structure of massive outflows in ultraluminous infrared galaxies'' (ESO projects 0103.B-0391(A) and  0104.B-0151(A), PI: Arribas). The observations presented in this paper were carried out with VLT/MUSE in its adaptive-optics-assisted  wide-field mode (AO-WFM; \citealt{Bacon2010}). 

MUSE observations cover a $60''\times 60''$ FOV with a sampling of $0.2''\times 0.2''$, resulting in a dataset of $\sim 90000$ individual spectra. We used the nominal instrument setup, with a spectral coverage from 4750 to 9350 $\AA$ and a mean resolution of 2.65 $\AA$ (FWHM). Because we use AO with a sodium laser guide system, the wavelength range $5800-5970$ $\AA$ is blocked to avoid contamination and saturation of the detector by sodium light.
For our sources the Na {\small ID} feature is redshifted outside this wavelength range.   

The requested observations were distributed in three 40-minute observing blocks (OBs) with a total integration time of two hours on source for all ULIRG systems. We split the exposures in each OB into four (dithered and rotated by 90$^{\circ}$) frames of 612 s each. Observations were performed with seeing $\sim 1''$; the root mean square  of the flux variation was $\sim 1\%$, as measured at night by the VLT DIMM station.

Seventy percent of the MUSE targets (i.e. 17 out of the 25) were observed as part of our programs. A few sources were not observed with the entire requested time: for instance, only one frame was obtained for the source F16090-0139. 
Additional frames were instead obtained for 07251-02448, F11095-0238, and IZw1 because their first observations were classified as grade C. In our analysis, we combined all these additional frames after verifying that including them does not affect  the quality of the final data cubes negatively.

MUSE archival data exist for four out of the eight remaining targets: 13120-5453, F14378-3651, 17208-0014, and F22491-1808 (already presented in \citealt{Fluetsch2020}). These sources were observed in  seeing-limited WFM and with $\sim 30$ -minute exposure times. Because the seeing conditions during their observations were good, we decided to include the archival data of these four targets in our analysis. We note that seven of our selected sources are also part of the sample presented in \citet{Fluetsch2020}: IZW1, 20100-1651 and 19542+1110, for which we present higher quality data, as well as 13120-5453, F14378-3651, 17208-0014 and F22491-1808.

Our PUMA survey so far consists of 21 ULIRGs observed with MUSE-WFM;  information about the MUSE data used in this work are collected in Table \ref{Tlogfile}. At the mean distance ($\sim$ 400 Mpc), the MUSE spaxel scale, resolution, and FoV correspond to 0.34 kpc, 1 kpc, and $\sim$ 100 $\times$ 100 kpc$^2$.

\subsection{Data reduction and astrometry registration}

MUSE observations were reduced using the MUSE EsoReflex pipeline recipes (muse - 2.6.2), which provide a fully calibrated and combined MUSE data cube. In the last step of our data reduction, we identified and subtracted the residual sky contamination in the final data cube using the Zurich Atmosphere Purge (ZAP) software package (\citealt{Soto2016}; see e.g. Sect. 3 in \citealt{Perna2020}).

We estimated the spatial resolution from the foreground stars in the MUSE FOV for all but Arp220 and F11095$-0238$, for which no stars are detected.
For each bright star in the FOV, we performed a 2D Gaussian fit, and derived an estimate for the angular resolution from the FWHM of the Gaussian fit. The  spatial resolution in the data cube of F11095$-0238$ was obtained from a 2D Gaussian fit of the broad-line region (BLR) emission of a quasi-stellar object (QSO) at z $\sim 2.2$ (CXOGSG J111204.2-025415); the spectral resolution of Arp220 was taken from \citet{Perna2020}. 
Because of the MUSE seeing-enhancer mode (and the good seeing conditions during the acquisition of seeing-limited MUSE data), we reached an overall average resolution of $\sim 0.7''$, corresponding to physical scales in the range $\sim 0.2 - 2$ kpc depending on the redshift of the system (Table \ref{Tlogfile}).

\begin{figure*}[t]
\centering
\includegraphics[width=18.cm,trim= 0 0 0 0,clip]{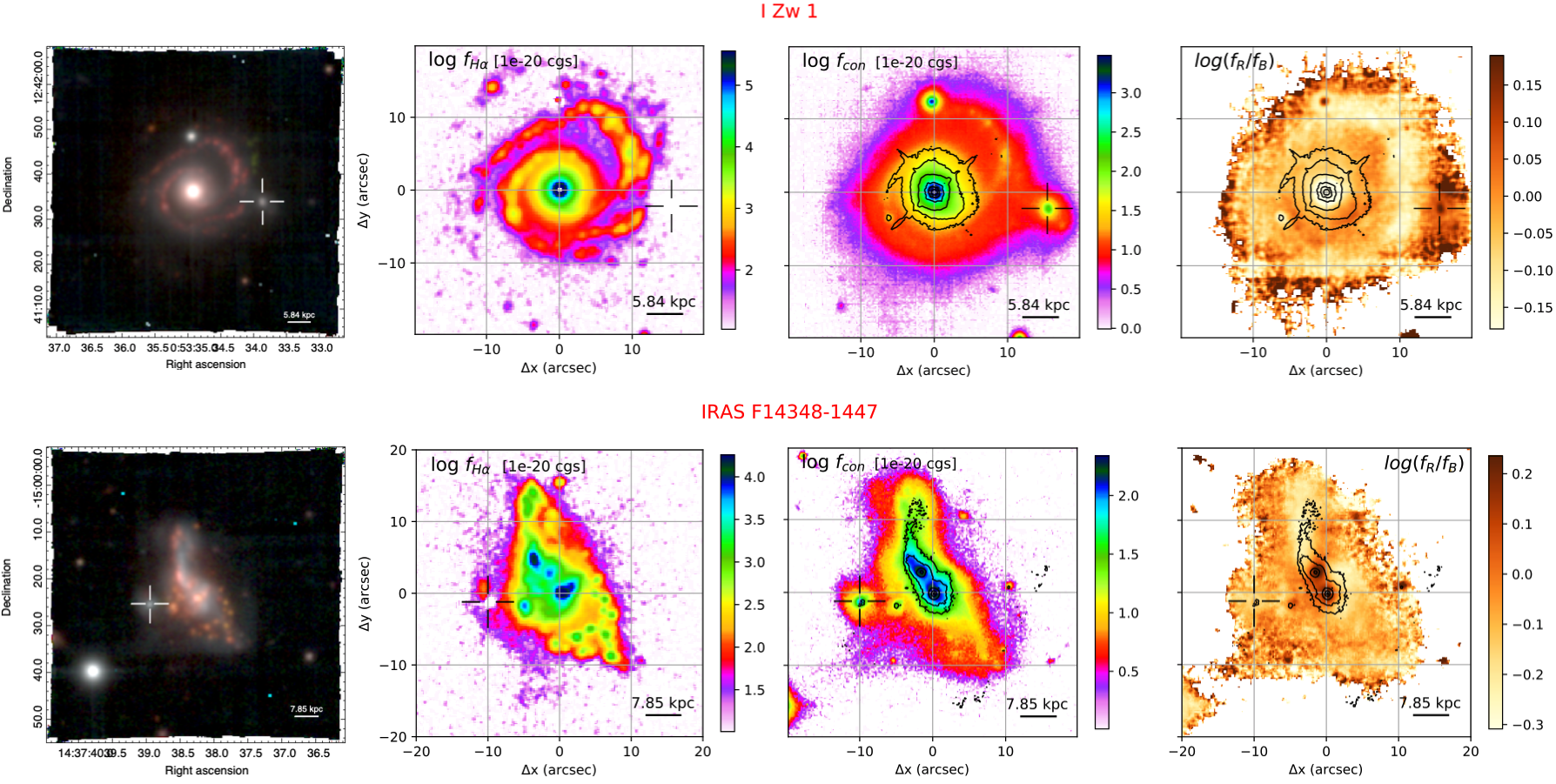}

\caption{\small IZw1 (top) and F14348-1447 (bottom) MUSE images. {\it Left column}: Colour-composite optical image, showing [O {\small III}] (green), \ha (red), and continuum emission (blue). {\it Second column}: \ha map obtained by collapsing the MUSE data cube on the emission line after subtracting the continuum using adjacent regions at shorter and longer wavelengths. 
{\it Third column}:
Red continuum ($\sim 8500\AA$) image from MUSE, with contours from HST/F160W. {\it Right column}: Continuum colour map obtained from MUSE by dividing the red continuum image ($\sim 8500\AA$) by the blue image ($\sim 4500\AA$); contours from HST/F160W. See Appendix \ref{allmaps} for further details. The crosses in the panels show the detected satellites in the MUSE FOV. }
\label{MUSE3colours}
\end{figure*}

The astrometric registration was performed using the Gaia DR2 catalogue (\citealt{Gaia2018}). 
We created continuum maps from the MUSE observations (see Sect. \ref{Smaps} and Fig. \ref{MUSE3colours}) where we measured the position of all the Gaia DR2 stars present in the MUSE FOV. 
Then, we shifted the MUSE cube astrometric solution to minimise the difference between the Gaia sky coordinates of these stars and those derived from the MUSE data. We took the proper motions of the stars as listed in the Gaia catalogue into account.
For most systems, we used between 2 and 35 Gaia stars to perform this correction. 
For 5 systems (F05189-2524, F12071-0444, F13451+1232, F14348-1447, and F11095-0238), fewer than 2 Gaia stars were available in the MUSE FOV; therefore we used HST optical images (F814W) as reference whose astrometry was tied to Gaia DR2 using the method described above. Then, we used objects detected in both the HST and MUSE images to determine the MUSE astrometric solution. HST near-IR image (F160W) astrometry was similarly tied to Gaia DR2. 
We estimate an average uncertainty in the MUSE astrometric solution of 0.15\arcsec (0.7 MUSE spaxels) based on the comparison between measured and expected coordinates of the Gaia stars.

\section{Stellar and ionised gas distributions}\label{Smaps}

In order to appreciate the MUSE data quality and its   scientific potential, in this section we present three-colour, continuum, and emission line maps built directly from the MUSE data cubes, and briefly describe the general properties that can be inferred from them. A full characterisation of the systems, similar to the one obtained for Arp220 (\citealt{Perna2020}; Catal\'an-Torrecilla et al., in prep.), will be presented in future papers. 

In particular, in this section we present the maps for two representative targets in our sample: IZw1, the type 1 AGN archetypical, hosted by an ongoing minor merger (e.g. \citealt{Scharwachter2007}), and the ULIRG F14348-1447, an SB-dominated interacting system. Colour and flux images for the remaining targets in our sample are presented in Appendix \ref{allmaps}.

\subsection{Colour-composite maps}

Colour-composites of the two representative targets are shown in Fig. \ref{MUSE3colours} (left column). These maps are built using three images derived by collapsing the MUSE data cubes along the wavelength direction. The green and red images cover the [O {\small III}]$\lambda$5007 ([O {\small III}] hereinafter) and \ha  line transitions (with underlying continuum emission), respectively. In particular, for each target, we selected the wavelength ranges covering the [O {\small III}] lines taking into account the line velocity variations observed across the MUSE FOV; the \ha maps were constructed in a similar fashion, but reducing the velocity range to minimise the contamination from the [N {\small II}] doublet. The individual wavelength ranges used to produce composite images are reported in the figure captions in Appendix \ref{allmaps}.  The blue image is instead associated with the continuum emission at bluest wavelengths. 
The blue, green, and red images have identical colour-bar scales and limits, and are combined to obtain an approximation to the actual relative contributions (and extensions) of [O {\small III}], \ha, and stellar continuum emission in the ULIRGs.

These composites provide a first-order description of the ULIRG morphologies. 
Compact and regular galactic structures are observed only in three targets: IZw1 (Fig. \ref{MUSE3colours}, top left), F00188-0856 (Fig. \ref{IRAS00188_1}), and 19542+1110 (Fig. \ref{IRAS19542_1}). The remaining ULIRGs present irregular morphologies (e.g. elongated along the merger axis; e.g. Fig. \ref{MUSE3colours}, bottom left) and  bright tidal tails extended over tens of kpc (e.g. Fig. \ref{IRAS09022_1}).

 The three-colour images also allow us to compare continuum and ionised gas distribution, providing an indication of the physical processes in the ULIRGs. 
 \ha emission generally matches the stellar continuum distributions, both in the innermost nuclear regions, tracing SB or AGN activity, and in the outer regions, tracing more diffuse ionised gas. 
 Clumpy \ha emission is also detected along tidal arms. In a few targets, the [O {\small III}] dominates continuum and \ha emission, possibly indicating AGN ionisation conditions (see e.g. IZw1 nuclear and north-west regions in Fig. \ref{MUSE3colours}, top left).

Finally, these images provide a first-order identification of foreground stars, generally associated with bright point-like sources, and galaxies at different redshifts, showing a heterogeneous mix of morphologies, colours, and sizes (see also e.g. \citealt{Perna2020}). Nine ULIRGs in our sample present between one and two nearby satellites in the MUSE FOV at projected distances from $\sim 5$ to $\sim 50$ kpc (see e.g. crosses in Fig. \ref{MUSE3colours}), and with velocities within a few 100s km/s from the ULIRG systemics (Table \ref{Tredshifts}). A more detailed description of ULIRG structures and environmental properties is provided in Sect. \ref{Smorph}, where we combine optical (MUSE) and near-IR (HST) imaging information with stellar kinematics results.

\subsection{Ionised gas and continuum emission maps}

In Fig. \ref{MUSE3colours} we also display the \ha maps (second column), the red continuum emission (third column), and the continuum colour maps (right column) for IZw1 and F14348-1447. In the last two columns, we  show the HST/F160W emission with black contours.
The \ha maps are obtained by collapsing the MUSE data cube on the emission line after subtracting the continuum  using adjacent regions at shorter and longer wavelengths; the red image is associated with the continuum emission at reddest wavelengths, but avoiding the range $9000-9300 \AA$ (observer frame), which is sometimes affected by poor background subtraction; the continuum-colour maps are instead built by dividing the red image by a blue one, derived by collapsing the MUSE data cube at wavelengths of $\sim$ 4500 $\AA$. These maps are centred on the position of the ULIRG nuclei, corresponding to the peak emission in the near-IR HST images (black contours in Fig. \ref{MUSE3colours}). For F14348-1447 (and other systems with double nuclei), the zero-position corresponds to the brightest nucleus in the HST/F160W image. H$\alpha$ and continuum emission maps for the entire sample are reported in Appendix \ref{allmaps}. 

The inspection of these images allows us a first identification of bright star-forming clumps in the outermost regions, associated with strong \ha and blue continuum emissions (e.g. along the IZw1 galactic arms in Fig. \ref{MUSE3colours}), or old stellar populations with faint (or absent) \ha and red  spectra (e.g. in the IZw1 west satellite). Tidal tails in our sample show both red and blue continuum emission (see e.g. Fig. \ref{IRAS09022_1}); blue tails are generally associated with \ha emission, suggesting a certain sign of recent SF (e.g. \citealt{Yuan2018}). 

The innermost regions of ULIRGs usually show strong \ha emission and red continuum, also associated with bright near-IR emission (black contours) in almost all our targets. This is probably due to the severe channelling of gas and dust during the merger process, which is capable of triggering intense AGN and SB activities (responsible for the \ha nuclear emission). 
An imperfect match of the MUSE optical and HST near-IR peak positions is instead observed for 09022-3615, 17208-0014, and the W nucleus of F22491-1808: these discrepancies could be explained considering an even higher nuclear extinction, associated with cold circumnuclear material (Pereira-Santaella et al., in prep.) or with dust lanes across the galaxy cores (see e.g Fig. \ref{IRAS17208_1}).  
Blue nuclear spectra are observed in IZw1 (Fig. \ref{MUSE3colours}, top right) and F01572+0009 (Fig. \ref{IRAS01572_1}), the two Seyfert 1 galaxies in our sample. 

In Appendix \ref{allmaps} we also report [O {\small III}] and [S {\small II}]$\lambda\lambda6716,31$ maps for all targets. \ha emission is generally brighter and more extended than [O {\small III}]: the former is generally detected in plumes, filaments, and extended arcs (e.g. Fig. \ref{IRAS11095_2}) as well as in compact clumps (e.g. Fig. \ref{IRAS09022_2}), possibly indicating shocks (e.g. when bright [S {\small II}] is also detected) or SF activity.

Overall, ionised gas and continuum emission maps display a great richness in spatial detail at different spatial scales. In all cases, they reveal different levels of galaxy interaction and strong nuclear activity, in line with their ULIRG nature.

\section{Dynamical status}\label{Ssanalysis}

\subsection{Stellar feature modelling}

We used the penalised pixel-fitting routines (pPXF; \citealt{Cappellari2004,Cappellari2017}) to extract the stellar kinematics. We made use of the Indo-U.S. Coud\'e Feed Spectral Library (\citealt{Valdes2004}) as stellar spectral templates to model the stellar continuum emission and absorption line systems. The models, with a spectral resolution of 1.35$\AA$, were broadened to the (wavelength-dependent) spectral resolution of the MUSE data ($\sim 2.6-2.9\AA$) before the fitting process (see e.g. \citealt{Husser2016}).  pPXF fits were performed on binned spaxels using a Voronoi tessellation (\citealt{Cappellari2003}) to achieve a minimum signal-to-noise ratio S/N $> 16$ per wavelength channel on the continuum in the vicinity of the MgIb transitions. We note, however, that in the innermost nuclear regions, a large fraction of the original spaxels remains unbinned because of the high quality of the MUSE data. 

The entire wavelength range covered with MUSE (i.e. 4700-9300$\AA$, observer frame) was used in our analysis to model the stellar component and recover the stellar kinematics after masking all optical emission lines detected in the data cubes.  Because broad emission lines alter the local continuum and the stellar absorption profiles, causing poor pPXF fits, we used wider masks in the spatial regions where broad profiles are detected;
for the two Sy 1 sources, much broader masks were used for the BLR lines in the innermost nuclear regions. 
In addition, we masked the resonant Na {\small ID} transitions because both stellar and interstellar absorption can cause these lines.
Finally, we excluded a narrow wavelength region at $\sim  7630\AA$ (observer frame) from the analysis, which is associated with strong sky-subtraction residuals, and the region 5800-5970$\AA,$ which is blocked by a filter to avoid contamination from the AO lasers (see Sect. \ref{Sreduction}). 

During the fitting procedure, we used fourth-order multiplicative Legendre polynomials to match the overall spectral shape of the data. These polynomials are generally used instead of an extinction law, which was found to produce poorer fits to the stellar continuum, and allow us to correct for small inaccuracies in the flux calibration (see e.g. \citealt{Belfiore2019}).

\subsection{Stellar velocity fields and velocity dispersion maps}

The pPXF best-fit results were used to derive the systemic redshift of the ULIRGs. In particular, for the sources with one nucleus, we derived the zero-velocity by measuring the stellar kinematics at the position of the nucleus; when two nuclei are detected instead, we set the $V_*=0$ km/s measuring the stellar kinematics at the position of the brightest nucleus in the red wavelengths. 
The target 09022-3615 does not show well-defined nuclear regions in the available HST images (covering the rest-frame UV and optical regimes); therefore, the nucleus of this target was located on the basis of ALMA millimeter continuum and CO(2-1) emission line maps (Pereira-Santaella et al., in prep.).   
Finally, for the two Sy 1 sources with strong continuum and BLR emission in the central regions (preventing the detection of stellar features), the systemic redshifts were chosen to obtain a symmetric stellar velocity gradient along their major axis, taking advantage of their well-defined rotational patterns (see below). All spectroscopic redshifts are reported in Table \ref{Tredshifts}, together with the coordinates of the nuclei. In this table, we also report the position and spectroscopic (pPXF) redshifts of the ULIRG nearby companions detected in the MUSE data cubes.

Figure \ref{ppxf} shows the IZw1 (top) and F14348-1447 (bottom) stellar kinematic maps derived from pPxF analysis. 
In Appendix \ref{allmaps} we report the stellar kinematic maps for all ULIRGs in our sample. 
An ordered disk-like rotation can be observed in IZw1, as well as in another ten ULIRGs in our sample (e.g. Figs. \ref{IRAS00188_3} and \ref{IRAS05189_3}). These galaxies show a well-defined velocity gradient axis, with line-of-sight velocity amplitudes of $\sim \pm 100$ km/s. They also present regular velocity dispersion configurations, with $\sigma_*$ of a few 100s km/s in the nuclear regions, and $\sigma_*$ close to the MUSE spectral resolution outside. However, signatures of tidal interactions are present in many of them, especially in the outermost regions (e.g. Figs. \ref{IRAS01572_3} and \ref{IRAS13120_3}). 
 
F14348-1447 (Fig. \ref{ppxf}, bottom) shows a   kinematic axis following the ULIRG elongated structure, on scales of $\sim 30$ kpc and with amplitudes up to $\sim \pm 150$ km/s. The velocity dispersion map is quite irregular, although higher $\sigma_*$ ($\sim 120$ km/s) can be found at the location of its nuclei. 20087-0308 (Fig. \ref{IRAS20087_3}) and F22491-1808 (Fig. \ref{IRAS22491_3}) show very similar characteristics. Their kinematics are therefore reasonably strongly affected by the interaction of the merging galaxies.
 
Different configurations can instead be observed in 09022-3615 (Fig. \ref{IRAS09022_3}), F12072-0444 (Fig. \ref{IRAS12072_3}),  F13451+1232 (Fig. \ref{IRAS13451_3}), and 20100-4156 (Fig. \ref{IRAS20100_3}): they show clear evidence of non-rotational motions, without preferential velocity gradient axes and with high $\sigma_*$ ($> 100$ km/s) over the entire systems. For these ULIRGs, the stellar kinematics are therefore reasonably dominated by tidal forces as well.  

Finally, F11095-0238 (Fig. \ref{IRAS11095_3}) and F16090-0139 (Fig. \ref{IRAS16090_3}) do not show clear kinematic patterns. F11095-0238 pPXF maps could suggest a nearly face-on orientation because of the small velocity amplitudes across the system and the relatively higher $\sigma_*$ in the innermost nuclear regions. The unconstrained kinematics in F16090-0139  are instead probably due to its very short exposure time (and low S/N). 

We stress here that more regular kinematics might be present on subkpc scales in the innermost nuclear regions and in the vicinity of the nuclei of binary systems (see e.g. \citealt{Medling2014}), even when large-scale motions are irregular or are dominated by streams. For instance, \citet{PereiraSantaella2018} reported circumnuclear molecular disks around the F14348-1447 SW and NE nuclei. More detailed investigation of subkpc motions in individual systems will be presented in a forthcoming paper; here we focus on the main (kpc-scale) stellar kinematic features.

\begin{figure}[t]
\centering
\includegraphics[width=9.cm,trim= 0 0 0 0,clip]{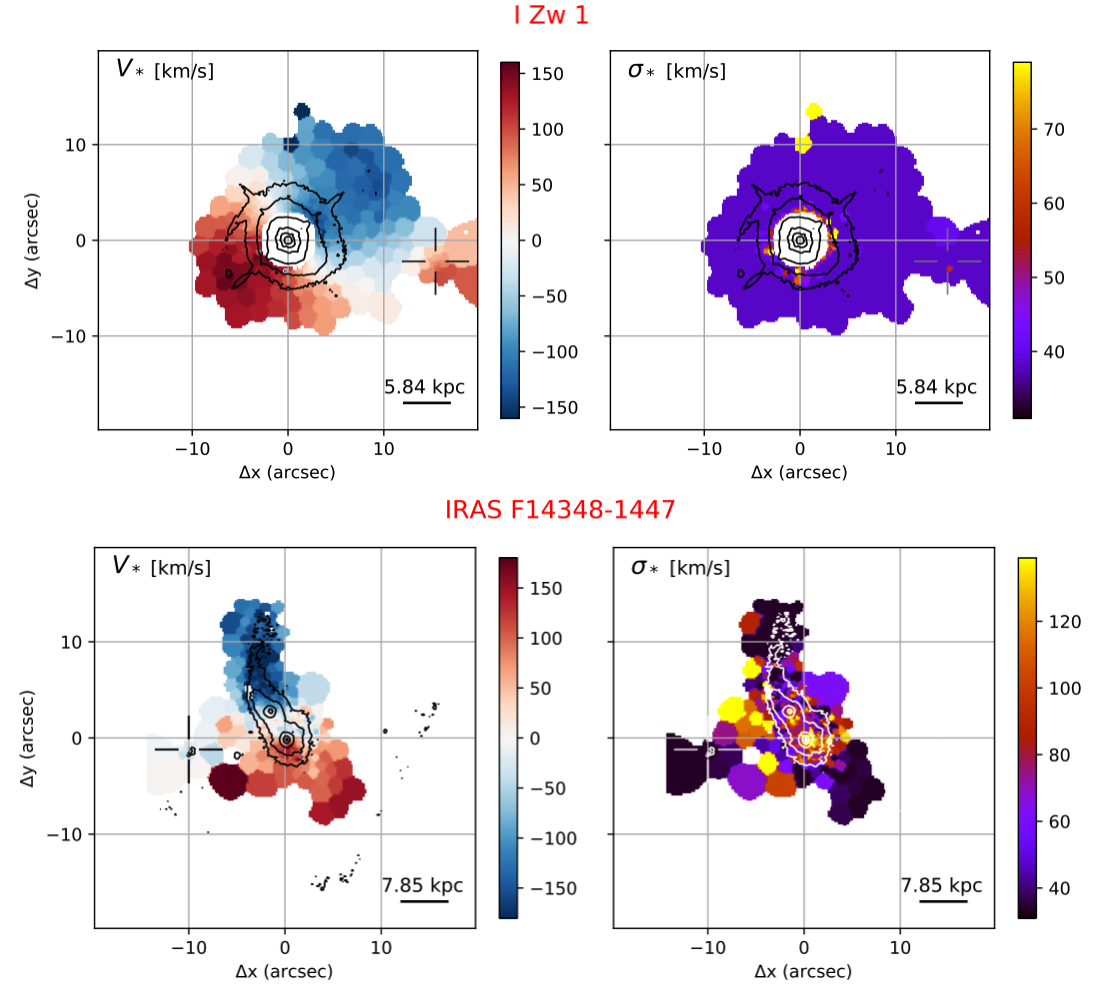}

\caption{\small pPXF stellar line-of-sight velocity (left) and velocity dispersion maps (right) for IZw1 (top) and F14348-1447 (bottom); $\sigma_*$ are not corrected for instrumental broadening. Contours from HST/F160W images, starting from 3$\sigma$ levels, equally spaced in steps of 0.5 dex. The IZw1 central pixels have been masked because of the strong  continuum and BLR emission, which prevents the detection of stellar absorption features.    
}
\label{ppxf}
\end{figure}

\begin{figure*}[h]
\centering
\includegraphics[width=16.3cm,trim= 0 0 0 0,clip]{{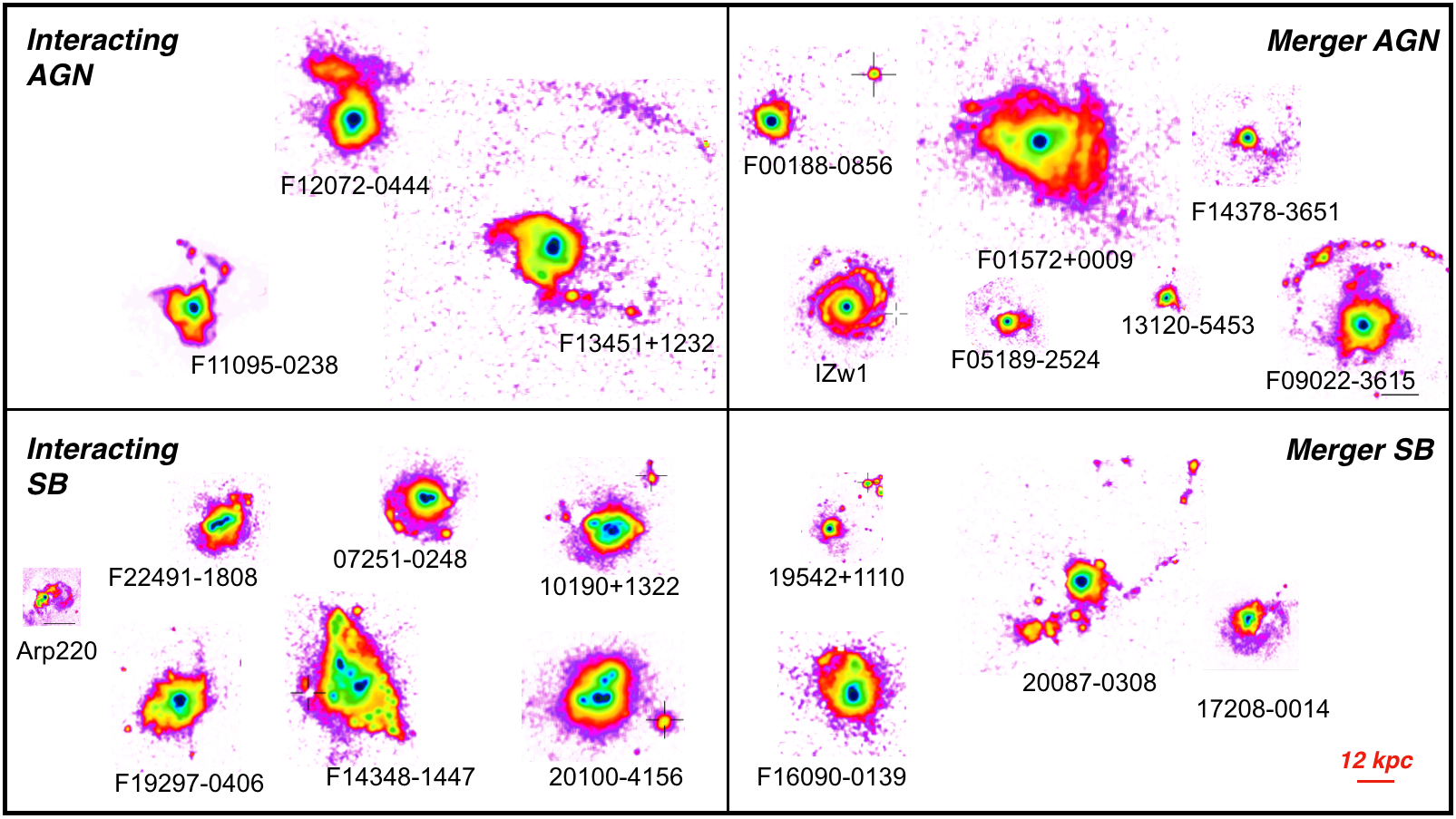}}

\caption{\small 
MUSE \ha maps for all  21 ULIRGs of the PUMA sample. All maps are taken from the images shown in Appendix \ref{allmaps}. The galaxies are displayed in a simplified version of the $\alpha_{AGN} -$ IC plane shown in Fig. \ref{Sample}, with the same physical scale; north is up.}
\label{2dha}
\end{figure*}

\begin{figure*}[h]
\centering
\includegraphics[width=16.3cm,trim= 0 0 0 0,clip]{{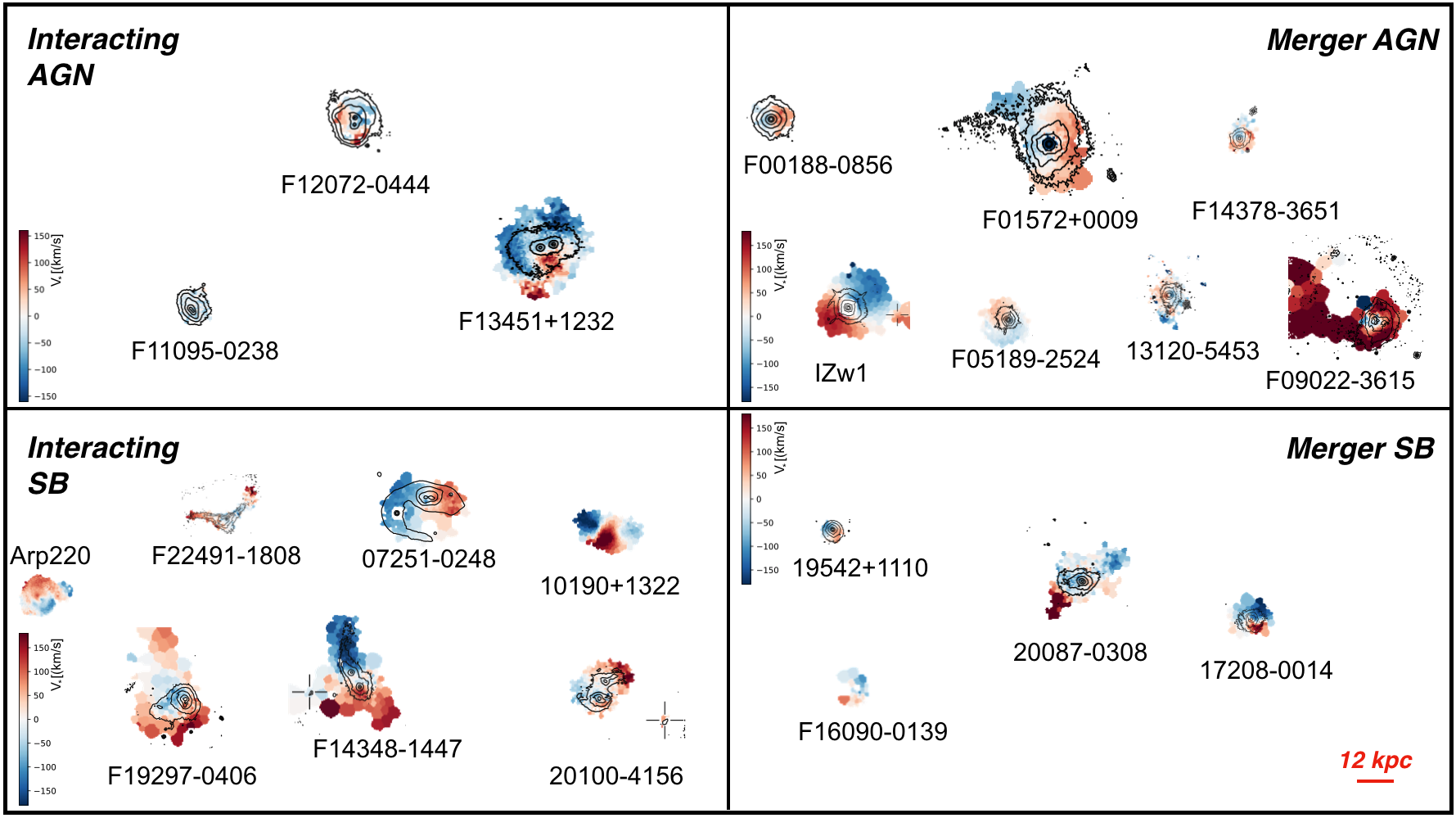}}
\caption{\small 
MUSE stellar velocity maps for all 21 ULIRGs of the PUMA sample taken from the images shown in Appendix \ref{allmaps}. The galaxies are displayed in a simplified version of the $\alpha_{AGN} -$ IC plane shown in Fig. \ref{Sample}, with the same physical scale; north is up.
}
\label{2dvs}
\end{figure*}

\begin{table*}
\footnotesize
\begin{minipage}[!h]{1\linewidth}
\setlength{\tabcolsep}{4pt}
\centering
\caption{Morphological and kinematic classification}
\begin{tabular}{lllcc} 
IRAS ID         &   overall stellar kinematics         & main morphological features                 & N$_{comp}$ & IC \\
\scriptsize{(1)}        & \scriptsize{(2)}              & \scriptsize{(3)}                         & \scriptsize{(4)}&\scriptsize{(5)}   \\
\hline

F00188-0856             & ordered motions        & 1 nucleus      & 1 & M \\ 

IZw1     & ordered motions         & 1 nucleus, spiral arms, [O {\small III}] clumps at kpc-scales    &  1 & M \\ 

F01572+0009   & ordered motions         & 1 nucleus, tails, extended \ha                        & 1 & M \\ 

F05189-2524             & ordered motions         & 1 nucleus, several tails, extended \ha            &  0 & M \\ 

07251-0248              & ordered motions         & 2 nuclei, ring-like shape, \ha clumps at kpc-scales & 0 & I\\

09022-3615              & non-ordered motions           & 1 nucleus, tail, \ha clumps at kpc-scales                &  0 & M \\

10190+1322              & ordered motions        & 2 nuclei and overlapping disks, \ha clumps at kpc-scales         & 2 &  I\\

F11095-0238             & non-ordered motions           & 2 nuclei, tails, extended \ha at kpc-scales              & 0 & I\\

F12072-0444             & non-ordered motions           & 2 nuclei, tails, diffuse \ha at kpc-scales               & 2 &  I\\

13120-5453  & ordered motions & 1 nucleus, extended structures and tails &  0 & M\\

F13451+1232             & non-ordered motions         & 2 nuclei, tails, \ha clumps at kpc-scales                &  0 &  I\\

F14348-1447             & large-scale streaming motions & 2 nuclei, \ha clumps at kpc-scales                   &  1 &  I\\

F14378-3651 & ordered motions & 1 nucleus, tails, and extended \ha shell &  1 & M\\

Arp220  & ordered motions               & 2 nuclei, streaming motions, \ha shells          & 0 &  I\\

F16090-0139             & non-ordered motions$^*$       & 1 nucleus, plume, \ha bubble                      & 0 & M\\

17208$-$0014 &  ordered motions & 1 nucleus, complex structures and kpc-scale tails & 0 & M\\

F19297-0406             & non-ordered motions           & 1 nucleus, tails              & 0 & I \\

19542+1110              & ordered motions         & 1 nucleus,  \ha clumps at kpc-scales           &   1 & M \\

20087-0308              & large-scale streaming motions & 1 nucleus, tails, \ha clumps at kpc-scales                &  0 & M\\

20100-4156              & non-ordered motions           & 2 nuclei, tail & 1 & I \\

F22491-1808 & large-scale streaming motions & 2 nuclei, tails, diffuse \ha & 0 & I \\
\hline

\end{tabular}
\label{Tmorph}
\vspace{0.2cm}
\end{minipage}
{\small
{\it Notes.}\\
Column (1): Target name.\\
Column (2): Simplified classification for the stellar motions from the pPXF analysis. \\
Column (3): Main morphological features. \\
Column (4): Number of companions in the MUSE FOV; see Table \ref{Tredshifts} for details.\\
Column (5): IC classification. 
}
\end{table*}

\subsection{Stellar velocity versus morphological classification}\label{Smorph}

Table \ref{Tmorph} summarises the morphological and stellar kinematics information we collected so far for our
ULIRGs. They are based on the current MUSE colour and flux maps, HST/F160W imaging, and pPXF stellar kinematic results.
For a more immediate visualisation, we also report in Figs. \ref{2dha} and \ref{2dvs} the MUSE \ha and stellar velocity maps that we also present in Appendix \ref{allmaps}, within a simplified version of the $\alpha_{AGN}-$IC diagram.  
These properties allow us to revise the morphological classification (with respect to the one used for the selection process in Sect. \ref{Ssample}), and confirm consistency with the derived stellar kinematics. 

According to Table \ref{Tmorph}, the observed 21 ULIRGs can be divided into four different groups, taking their IC and the main stellar kinematic features into account. The first group contains 8 sources  characterised by ordered disk-like rotations and a merger classification, with a single nucleus and small nearby companions (F00188-0856, IZw1, F01572+0009, F05189-2524, 13120-5453, F14378-3651, 17208-0014, and 19542+1110). Almost all of them show additional detached (non-rotational) velocity structures in the outermost regions, possibly due to  tidal tails. 

The second group contains three systems,  characterised by ordered disk-like motions and an interacting classification: 07251-0248, 10190+1322, and Arp220. 07251-0248 shows a disk-like rotation pattern with a kinematic centre close to the two nuclei, separated by 1.82 kpc, and a strong extended tidal tail. 10190+1322 shows two overlapping disks centred at the positions of the two nuclei, separated by $\sim 7$ kpc. Finally, Arp220 show a disturbed kpc-scale disk in the innermost nuclear regions (see also Fig. 3 in \citealt{Perna2020}). 

The third group contains three ULIRGs, mainly characterised by large-scale streaming motions along the axes connecting their nuclei (F14348-1447 and F22491-1808), or along the strong and extended tails (20087-0308). F14348-1447 and F22491-1808 are interacting, while 20087-0308 is classified as a merger system. 

Finally, the last group contains  seven systems, mainly characterised by non-ordered stellar motions: 09022-3615, F11095-0238, F12072-0444, F13451+1232, F16090-0139, F19297-0406, and 20100-4156. Two of them have a merger classification (09022-3615 and F16090-0139), and the remaining sources are interacting. 

Two sources deserve a special mention: 09022-3615 and F19297-0406. HST/F814W and HST/F225W images of 09022-3615 show two main regions of intense SF in the central part of the system, with a few bright compact ($\lesssim 0.1''$, i.e. 150 pc) clumps at the position of the two peaks in the MUSE continuum image (Fig. \ref{IRAS09022_1}). 
These findings might suggest two SB regions, separated by $\sim 1.5$ kpc and with a stellar velocity offset of $\sim 60$ km/s. In contrast, its nucleus, located $\sim 0.5''$ south of the southern SB based on ALMA data (Pereira-Santaella et al., in prep.), is highly obscured and not associated with strong optical emission. F19297-0406 was originally classified as a merger (e.g. \citealt{Kim2013}); the inspection of near-IR HST (Fig. \ref{IRAS19297_1}) and ALMA (Pereira-Santaella et al., in prep.) allowed the identification of two nuclei, separated by $\sim 1.1$ kpc. This source is therefore classified here as interacting (i.e. IIIb).  
For all remaining sources in our MUSE sample, we instead confirm the morphological classification reported in the literature.

The separation in the four sub-samples highlights that merging systems are more likely associated with ordered disk-like motions (8 out of 11, i.e. F00188-0856, IZw1, F01572+0009, F05189-2524,  13120-5453, F14378-3651, 17208-0014, and 19542+1110), while interacting systems are generally associated with non-ordered or streaming motions (7 out of 10, i.e. F11095-0238,  F12072-0444, F13451+1232, F14348-1447, F19297-0406, 20100-4156, and F22491-1808; see also e.g. \citealt{Bellocchi2016} for similar results). The morphological class (i.e. interacting or merger) therefore is consistent in general with the stellar kinematic classification (i.e. dominant non-ordered or ordered motions); we note, however, that even when rotation patterns are observed, tidal streams are also present, indicating that dynamical relaxation times depend on galactocentric distances:
Kinematic disturbances from interactions are indeed expected to fade within a few rotation cycles (e.g. \citealt{Dale2001,Kronberger2007}), and after the coalescence phase, the most external structures therefore require longer times to follow the rotation pattern and preserves irregular  tidally induced velocities.

Indications of ordered motions in merger remnants have been reported in the literature: for instance, \citet{Barrera2015} traced the stellar kinematics in CALIFA merging galaxies, and found that $\sim 90\%$ of  merger remnants exhibit disk-like motions (see their Figs. B.3 and B.4); moreover, K-band observations reported by \citet{Rothberg2004} revealed that most of the mergers in their sample show disky isophotals. Numerical simulations suggest that the number of stellar disks that survives or re-forms after an interaction is a strong function of the stellar mass ratio and gas content of the  interacting galaxies: For instance, gas-rich mergers can yield disk-dominated remnants, while for modest gas fractions, the remnants are likely to resemble spheroidal-like galaxies (e.g. \citealt{Hopkins2009a,Hopkins2009b,Naab2017}).

We finally note that the \ha flux and stellar velocity maps in the simplified $\alpha_{AGN}-$IC diagram (Fig \ref{2dha}) do not show  clear trends, but are overall consistent with the \citet{Sanders1988} evolutionary scenario. In particular, more compact and dynamically relaxed systems are more likely to be found in the merger-AGN class, while interacting galaxies more likely present  extended structures with several additional nuclear  clumps and knots of SF. 

\begin{figure*}[t]
\centering
\includegraphics[width=19.5 cm,trim= 40 0 0 20,clip]{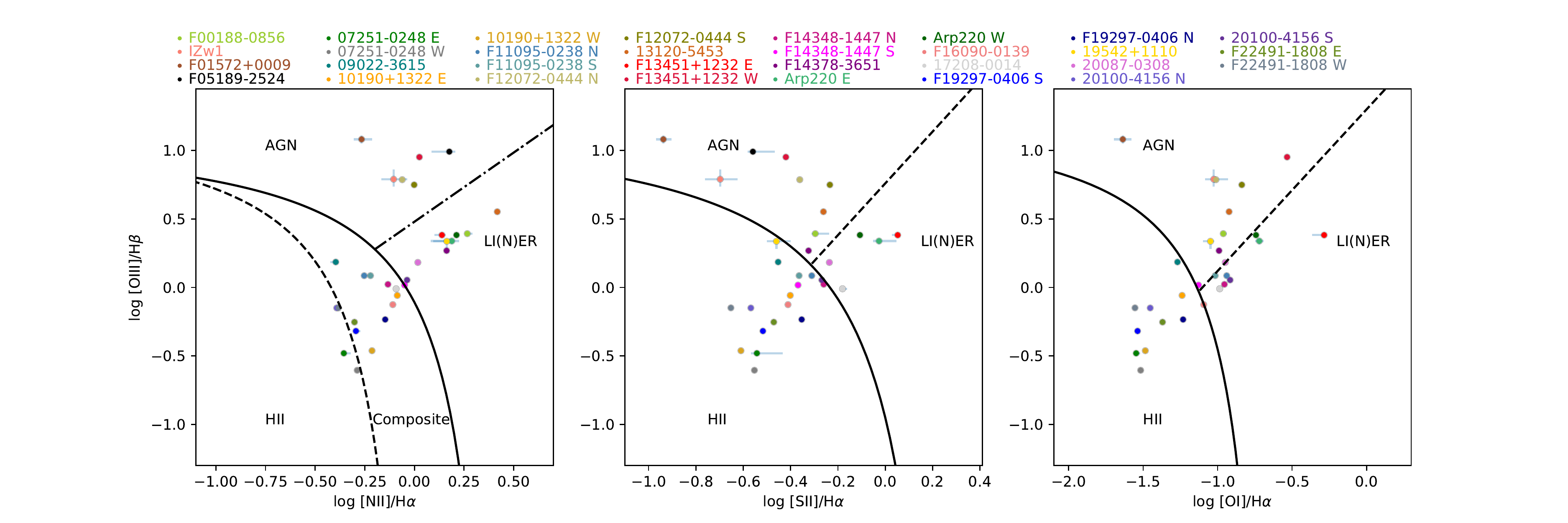}

\caption{\small Standard diagnostic BPT diagrams showing our flux ratio measurements obtained from the ULIRG nuclear spectra. Each nucleus is identified with a distinct colour, as labelled. Black curves separate HII-, composite-, AGN- and LI(N)ER-like line ratios, as labelled in the three panels (see text for details). 
}
\label{BPT}
\end{figure*}

\section{ISM nuclear properties: Outflows and ionisation conditions}\label{Sism}

In this section, we report the general ISM properties of the ULIRG nuclear regions obtained from analysing MUSE data. In particular, we revise the optical classification of the nuclear spectra and infer the possible presence of atomic outflows, taking advantage of the unprecedented quality of MUSE data. This information is required to understand whether the PUMA sample, selected to characterise multi-phase outflows,  does host gas flows that might be due to SB and AGN winds. 

Thirty-one nuclear spectra were extracted from circular apertures centred at the position of the ULIRG nuclei, with $r < 0.4''$ (i.e. considering our average spatial resolution). As a first step, we used pPXF to model the stellar contribution (Sect. \ref{Ssanalysis}), and after subtracting it, we used our own suite of python scripts to simultaneously fit all ISM prominent features. In particular, we modelled all emission lines with a combination of Gaussian profiles to account for both narrow and BLR features in type 1 AGN, or more in general, for asymmetric profiles with broad wings, that may trace perturbed kinematics and outflows. Na {\small ID} absorption is instead fit with a model parameterised in the optical depth space, as has been described in \citet{Perna2020}. In the next section, we describe the modelling prescription for the ISM features of all but IZw1 and F01572 nuclei, the two Sy 1 whose prescription is presented in Appendix \ref{AfitSy1}.

\subsection{ISM feature modelling}

We modelled the H$\beta$ and H$\alpha$ lines, the HeI$\lambda$5876,  the [O {\small III}]$\lambda\lambda$4959,5007, [N {\small II}]$\lambda\lambda$6548,83, [S {\small II}]$\lambda\lambda$6716,31, and [O {\small I}]$\lambda\lambda$6300,64 doublets with Gaussian profiles.  
We constrained the wavelength separation between emission lines in accordance with atomic physics; moreover, we fixed the FWHM to be the same for all the emission lines. Finally, the relative flux of the two [N {\small II}] and [O {\small III}] components was fixed to 2.99, the relative flux of the two [O {\small I}] lines was fixed to 3.13, and the [S {\small II}] flux ratio was required to be within the range $0.44< f$($\lambda$6716)/$f$($\lambda$6731) $<1.42$ (\citealt{Osterbrock2006}).

To account for potential asymmetric line profiles, we performed each spectral fit four times at maximum, with one to four kinematic components (i.e. Gaussian sets, each centred at a given velocity and with a given FWHM). The final number of kinematic components used to model the spectra was derived on the basis of the Bayesian information criterion (BIC, \citealt{Schwarz1978}; see e.g. \citealt{Perna2019}).

We simultaneously modelled the contribution of Na {\small ID} resonant lines. When sodium emission was detected, Gaussian profiles were used, requiring a Na {\small ID} doublet line ratio between the optically thick ($f (H)/ f (K) = 1$) and thin ($f (H)/ f (K)= 2$) limits (e.g. \citealt{Rupke2015}), where $H$ and $K$ indicate the sodium transitions at 5891 and 5896 $\AA,$ respectively. The Na {\small ID} absorption contribution was instead fitted with a model parametrised in the optical depth space, following \citet{Sato2009}, 

\begin{multline}\label{eqrupke}
I(\lambda) = I_{em}(\lambda)\times f_{ABS}(\lambda)  \\
 \ \ \ \ \ \ \ = I_{em}(\lambda)\times (1- C_f \times [1-exp(-\tau_0 e^{-(\lambda - \lambda_K)^2/(\lambda_Kb/c)^2} - \\ 
2\tau_0 e^{-(\lambda - \lambda_H)^2/(\lambda_H b/c)^2})]) ,
\end{multline}

where $H$ and $K$ indicate the two sodium transitions, $C_f$ is the covering factor, $\tau_0$ is the optical depth at the line centre $\lambda_K$, $b$ is the Doppler parameter ($b=FWHM/[2\sqrt{ln(2)}]$), and $c$ is the light velocity. The term $ I_{em}(\lambda)$  in Eq. \ref{eqrupke} represents the intrinsic (unabsorbed) intensity, defined as  $I_{*} + I_{HeI}$, where  $I_{*}$ is the best-fit model obtained from the pPXF analysis, and $I_{HeI}$ is the helium line intensity (\citealt{Baron2020}). 
When more than one kinematic component was required to fit Na {\small ID} absorption features, we assumed the case of partially overlapping atoms on the line of sight, so that the total sodium profile can be reproduced by multiple components and $I(\lambda)= I_{em}(\lambda)\times  \Pi_{i=1}^n f_{ABS}^i(\lambda)$, where $ f_{ABS}^i(\lambda)$ is the $i$th component (as given in Eq. \ref{eqrupke}) used to model the sodium features (see also Sect. 3.1 in \citealt{Rupke2002}). We stress here that the kinematics of a given Na {\small ID} component (either in absorption or emission) are tied to those of a corresponding Gaussian set used to model the ISM emission lines.

Best-fit models for individual nuclear regions in our ULIRGs sample are reported in Appendix \ref{allmaps}. These fit results are used to investigate the dominant ionisation mechanism(s) for the emitting gas, and to derive the incidence of neutral and ionised outflows.

\begin{figure*}[h]
\centering
\includegraphics[width=18.3cm,trim= 30 955 0 40,clip]{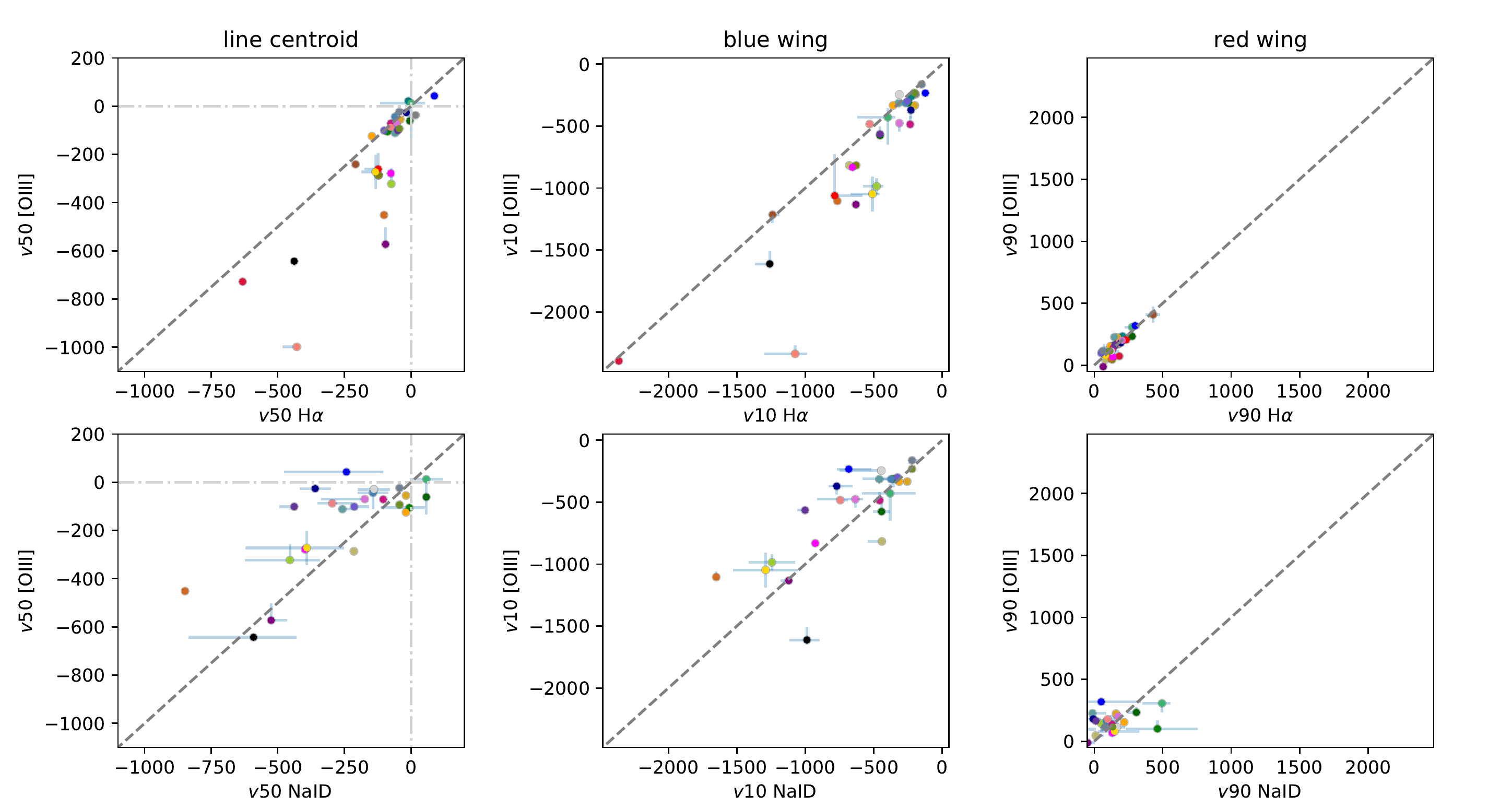}

\caption{\small Velocity-velocity diagram used to compare the non-parametric velocities of [O {\small III}], \ha, and Na {\small ID} gas for the 31 nuclei reported in Table \ref{Tism}. Each nucleus is identified with a distinct colour (see labels in Fig. \ref{BPT}). In the {\it left column}, we show the central velocities of [O {\small III}] vs. \ha (top) and of [O {\small III}] vs. Na {\small ID} (bottom); horizontal and vertical dot-dashed lines mark the zero-velocity positions, and the dashed line displays the 1:1 relation. 
In the {\it central column}, we show the $v10$ velocities, associated with the blue wings of [O {\small III}] and \ha (top), and of [O {\small III}] and Na {\small ID} (bottom); the dashed line indicates the 1:1 relation. 
In the {\it right column}, we show the $v90$ velocities, associated with the red wings of [O {\small III}] and \ha (top) and of [O {\small III}] and Na {\small ID} (bottom); the dashed line indicates the 1:1 relation. 
}
\label{kin1}
\end{figure*}

\subsection{BPT diagnostics}

Figure \ref{BPT} shows the BPT diagrams with the results from our spectroscopic analysis. All line ratios were derived by integrating the line flux over the entire fitted profiles (i.e. considering all kinematic components used to model the lines). For simplicity, hereinafter the [O {\small III}]$\lambda$5007/H$\beta$ versus [N {\small II}]$\lambda$6583/\ha (left panel), the [O {\small III}]$\lambda$5007/H$\beta$ versus [S {\small II}]$\lambda\lambda$6716,31/\ha (centre), and [O {\small III}]$\lambda$5007/H$\beta$ versus [O {\small II}]$\lambda$6300/\ha (right) are labelled [N {\small II}]-, [S {\small II}]-, and [O {\small I}]-BPT diagrams, respectively. 
In the figure, each coloured dot identifies a unique nuclear spectrum, as labelled at the top. 

The curves drawn in the [N {\small II}]-BPT diagram correspond to the theoretical boundary for extreme SBs (\citealt{Kewley2001}) and the empirical relation (\citealt{Kauffman2003}) used to separate purely SF galaxies from composite AGN-SF galaxies and AGN-/LI(N)ER-dominated systems (e.g. \citealt{Kewley2006,Belfiore2016}); the dot-dashed line is from \citet{CidFernandes2010}, and is used to separate LI(N)ERs and AGN. The curves in the [S {\small II}]- and [O {\small I}]-BPT diagrams correspond to the optical classification scheme of \citet{Kewley2006, Kewley2013}, and are also used in this case to separate SF galaxies from AGN and LI(N)ER systems, as labelled in the figures.

For the sources for which at least two BPT diagnostics indicate the same region, we constrained the dominant ionisation mechanism responsible for the line emission: 11 out of 31 nuclei are associated with SF, 9 out of 31 with AGN, and 6 out of 31 with LI(N)ER ionisation. For the remaining five spectra, the BPT diagnostics provide inconsistent results, i.e. the three diagnostics indicate three different  ionisation mechanisms.

Our flux ratio measurements are distributed over all the different regions of the BPT diagrams, and the inferred classification is overall consistent with that reported in the literature (Table \ref{Tsample}), although the classifications in the literature were obtained from long-slit (spatially integrated) spectra. 
Our optical classification is also generally consistent with the $\alpha_{AGN}$ values reported in Table \ref{Tsample}. The only exception in this case might be represented by 09022-3615, for which we observe SF-like line ratios, although the $\alpha_{AGN} = 0.54$ was derived with mid-IR diagnostics; our results are instead consistent with the small (near-IR based) AGN fraction inferred by \citet[][see Table \ref{tab1}]{Nardini2010}. These findings could suggest an extremely obscured AGN in this system (see also Sect. \ref{Smorph}).
We also note that 11 out of 17 nuclei with $\alpha_{AGN} < 0.32$ have LI(N)ER-like ratios, possibly indicating shocks induced by SB-driven outflows or gravitational interactions. In fact, while in many local galaxies, the LI(N)ER emission is associated with gas ionised by the hard radiation field of evolved (post-AGB) stars and with \ha equivalent widths $< 3$ (e.g. \citealt{Belfiore2016}), this does not  apply to our ULIRG nuclear regions.

\begin{figure*}[h]
\includegraphics[width=18.cm,trim= 45 541 23 484,clip]{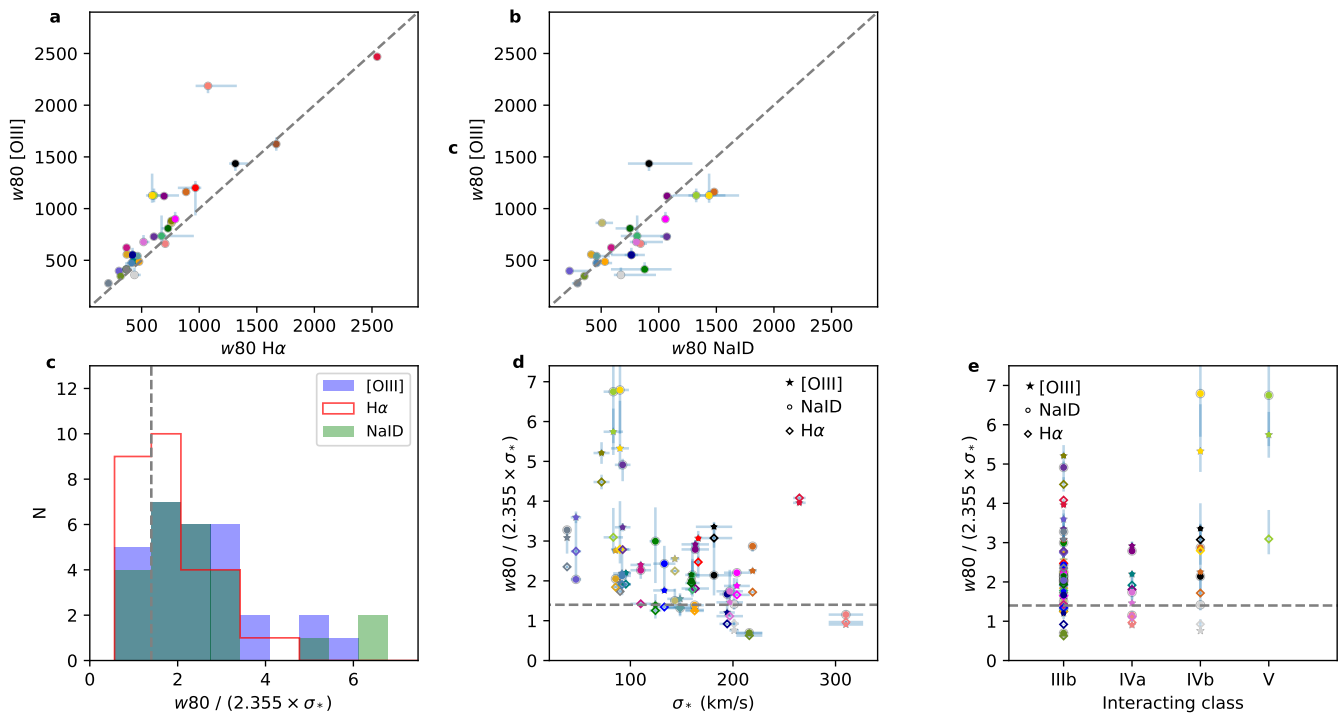}
%
\caption{\small {\it Panels a and b}: $w80$ velocity-velocity diagram for [O {\small III}] vs. \ha ({\it a}), and [O {\small III}] vs. Na {\small ID} ({\it b}). Each nucleus is identified with a distinct colour (see labels in Fig. \ref{BPT}). Dashed lines display the 1:1 relation. 
{\it Panel c}: Distributions of $\eta = w80/(2.355\times \sigma_*)$ for the three kinematic tracers, as labelled in the legend; the dashed vertical line marks the position $\eta = 1.4$ (see text). {\it Panel d}: [O {\small III}] (stars),  \ha (diamonds), and Na {\small ID} (circles) $\eta$ measurements as a function of $\sigma_*$. None of the velocity dispersions and $w80$ measurements are corrected for instrumental broadening; this correction would affect the only two $\sigma_*$ measurements close to the MUSE spectral resolution ($\sim 50$ km/s), associated with the 20100-4156 N and F22491-1808 W nuclei, further increasing their $\eta$ values. {\it Panel e}: [O {\small III}] (stars),  \ha (diamonds), and Na {\small ID} (circles) $\eta$ measurements as a function of the IC. The dashed horizontal line in panels d and e marks the position $\eta = 1.4$.
}
\label{kin2}
\end{figure*}

\subsection{Multi-phase outflow incidence}\label{Soutflow}

Almost all nuclear spectra show asymmetric and broad line profiles, possibly indicating strongly perturbed gas kinematics in both neutral (Na {\small ID}) and ionised (e.g. [O {\small III}], H$\alpha$) atomic components.
To characterise the overall kinematic properties of the ISM gas in a homogeneous way (i.e. to avoid any dependence on the number of distinct kinematic components used to model the line features in individual spectra), we used the non-parametric velocities $v10$, $v50,$ and $v90$, defined as the 10th, 50th, and 90th percentile velocities of the fitted line profiles, respectively, and the line width $w80$, that is, the difference between the 85th and 15th percentile velocities (see e.g. \citealt{Harrison2014}). All velocities are derived with respect to the systemic velocity, as inferred from the stellar velocities measured from pPXF (see Sec. \ref{Ssanalysis}). 
[O {\small III}] and \ha line features are used as tracers for the ionised gas kinematics; instead, ISM neutral atomic gas kinematics are traced by the Na {\small ID} absorption component, detected in almost all nuclear spectra and showing the broadest profiles compared with the Na {\small ID} emission component (detected in only a few targets).

Figure \ref{kin1} (first column) shows the velocity-velocity diagrams (VVD) that compare the central velocities ($v50$) of [O {\small III}] and H$\alpha$ in the top panel and the VVDs of [O {\small III}] and Na {\small ID} in the bottom panel. Almost all line centroids are blueshifted with respect to the systemic velocity, indicating a significant contribution from approaching emitting ([O {\small III}] and H$\alpha$) and absorbing (Na {\small ID}) material. This is confirmed by the fact that generally, the line profiles show very prominent blue wings, and faint and less extended red wings, as also shown in the central and right columns of Fig. \ref{kin1}, presenting $v10$ and $v90$  VVD, respectively.

In general, [O {\small III}] shows more extreme velocities than those of H$\alpha$ (see also e.g. \citealt{Bae2014}, \citealt{Venturi2018}, and \citealt{Cicone2016} for similar results; but see also \citealt{Rodriguez2019}). The Na {\small ID} velocities in turn appear slightly higher than those of the [O {\small III}], although the former are associated with higher uncertainties:  $\langle v10$ [O {\small III}] / $v10$ \ha $\rangle = 1.27$ and $\langle v10$ Na {\small ID} / $v10$ [O {\small III}] $\rangle = 1.16$. 
The high uncertainties in Na {\small ID} velocities are due to the fit degeneracy. In particular, they could originate from a blending between the HeI (in emission) and the Na {\small ID} features when strong sodium absorption is detected at very high negative velocities (e.g. F05189-2524 and F14378-3651). Alternatively, it can originate from the modelling of Na {\small ID} P Cygni profiles (F05189-2524 and F11095-0238 NE).

In Fig. \ref{kin2} (panels {\it a} and {\it b}) we show the $w80$ VVD. In this case, we also observe that in general, Na {\small ID} $w80$ are slightly higher than [O {\small III}] $w80$, which in turn are higher than \ha line widths:   $\langle w80$ [O {\small III}] / $w80$ \ha $\rangle = 1.15$ and $\langle w80$ Na {\small ID} / $w80$ [O {\small III}]$\rangle = 1.10$.
In order to distinguish between gravitational and outflow processes that might cause the velocity shifts and the extreme $v10$ and $w80$ in the [O {\small III}], \ha and Na {\small ID} lines, we compared  their line widths with the stellar velocity dispersion at the position of the nuclei. Following \citet{Woo2016}, we assumed that a non-gravitational component dominates the gravitational component when $\eta = w80/(2.355\times \sigma_*) > 1.4$, where the stellar velocity dispersion $\sigma_*$ traces the gravitational motions. This is a conservative criterion, as originally proposed for type 2 AGN and applied here to ULIRG nuclear spectra: In fact, $\sigma_*$ may be enhanced by motions associated with the merging process in ULIRGs. 

In panel c of Fig. \ref{kin2}  we report the histograms of the $\eta$ measurements, as obtained for the three kinematic tracers; in panel {\it d}, we instead show $\sigma_*$ as a function of $\eta$, for H$\alpha$, [O {\small III}], and Na {\small ID} features.  
We derived $\eta < 1.4$ from both [O {\small III}] and Na {\small ID}  for only three nuclei: 10190+1322 E, F16090-0139, and F22491-1808 E. The \ha distribution instead shows nine nuclei with $\eta < 1.4$, again indicating that the Balmer line can be less strongly affected by non-gravitational motions. For each tracer and nucleus, we indicate in Table \ref{Tism} whether gravitational or non-gravitational motions dominate according to the chosen criterion.

\begin{figure*}[t]
\centering
\includegraphics[width=13.3cm,trim= 30 335 285 898,clip]{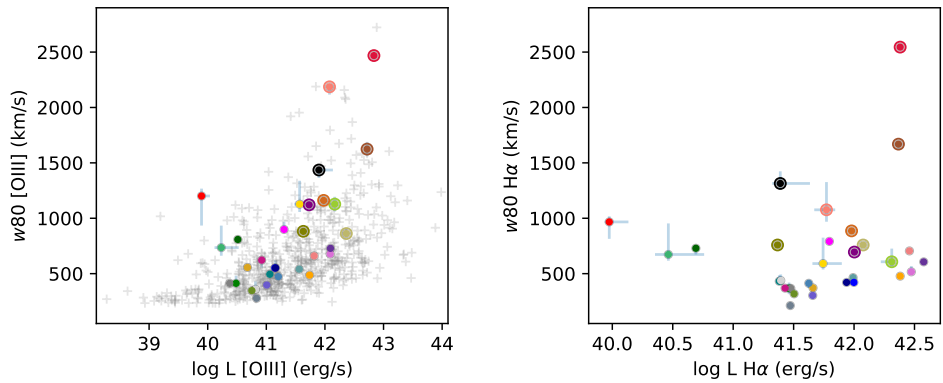}

\caption{\small   {\it Left}: [O {\small III}] non-parametric velocity $w80$ as a function of L$_{[OIII]}$ for the ULIRG nuclei presented in this work (colour-coded as in Fig. \ref{BPT}), and the X-ray/SDSS AGN (grey crosses) from \citet{Perna2017a}.  Larger symbols identify the ULIRG nuclei with AGN ionisation (see Table \ref{Tism}).  {\it Right}: \ha non-parametric velocity $w80$ as a function of L$_{H\alpha}$ for the ULIRG nuclei presented in this work.  
}
\label{kin3}
\end{figure*}

\begin{figure*}[h]
\centering
\includegraphics[width=13.3cm,trim= 30 125 285 1108,clip]{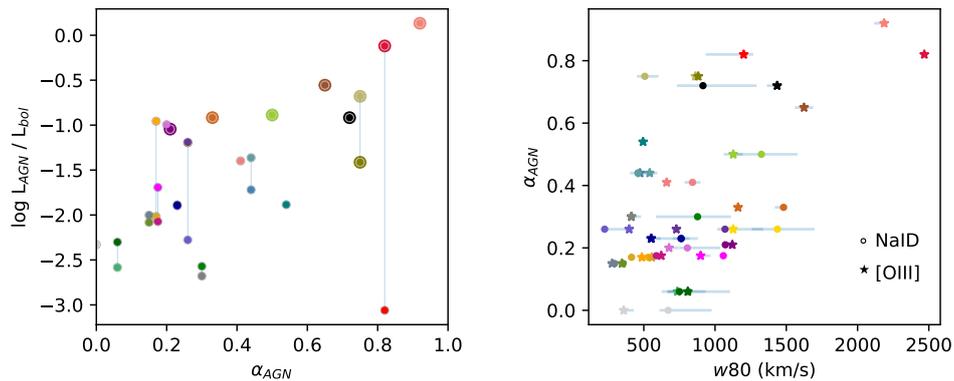}

\caption{\small  {\it Left}: (mid-IR based) AGN fraction $\alpha_{AGN}$ as a function of the ratio L$_{AGN}/$L$_{bol}$, with L$_{AGN} = 3\times 10^3$ L$_{[OIII]}$ and L$_{bol} = 1.15\times $ L$_{IR}$.  The nuclei of binary systems are connected with a vertical line; larger symbols refer to the sources with dominant AGN ionisation according to our BPT diagnostics.  {\it Right}: [O {\small III}] and Na {\small ID} non-parametric velocity $w80$ as a function of the AGN fraction $\alpha_{AGN}$, defined on the basis of mid-IR diagnostics (Sect. \ref{Sample}).  Each nucleus is identified with a distinct colour, as in Fig. \ref{BPT}. 
}
\label{kin4}
\end{figure*}

We note that the $\eta$ criterion can be applied for all nuclei for which {\it (i)} $\sigma_*$ can be derived from pPXF analysis, and  {\it (ii)}  the specific kinematic tracer for ionised and neutral gas is detected. [O {\small III}] and \ha lines are detected in all spectra, but no stellar features are present in the  
IZw1 and F01572+0009 nuclear spectra; therefore ionised gas  $\eta$ measurements were derived for all but these two Sy 1 nuclei. In contrast, Na {\small ID} absorption is not detected in 7 out of 31 nuclear spectra: the two Sy 1 targets, IZw1, and F01572+0009 show strong continuum and very broad line emission from the HeI at the position of the two Na {\small ID} transitions; moreover, although Na {\small ID} emission is detected, the strong BLR component prevents a robust analysis of a possible neutral outflow.  
The F13451+1332 W spectrum does not show sodium absorption (Fig. \ref{IRAS13451_n_w}), although a faint ($C_f = 0.11$) and broad ($b = 400$ km/s) Na {\small ID} profile in absorption has been reported in \citet{Rupke2005c}. The Na {\small ID} emission component that we used to model the observed spectrum might not be as reliable because pPXF best-fit results are not robust (no strong stellar absorbing features are detected over the entire wavelength range covered with MUSE).
Finally, for the remaining four spectra (07251-0248 W, F12072-0444 S, F13451+1232 E, and 09022-3615), the stellar Na {\small ID} contribution is responsible for the total absorption according to our pPXF analysis results.  As a consequence, we can infer the presence/absence of dominant non-gravitational motions in absorbing neutral gas  for 24 nuclei.

For the sources with $\eta <1.4$ from both [O {\small III}] and Na {\small ID} kinematics (i.e. 10190+1322 E, F16090-0139, and F22491-1808 E), we note that in at least one sources, F16090-0139, the stellar velocity dispersion might be overestimated (because of the low S/N), while the ISM features are extremely broad (e.g. Na {\small ID} $w80 \sim 840$ km/s; see Fig. \ref{IRAS16090_n}). This nucleus presents the highest $\sigma_*$ in our sample (see Fig. \ref{kin2}, panel {\it d}).
Hence, the F16090-0139 nucleus is reasonably associated with an outflow.  
We also note that the two Sy 1 nuclei display extremely broad features, likely associated with outflows.
Taking these arguments into account, we conclude that non-gravitational motions dominate the neutral (ionised) gas kinematics in 20 out of 24 (28 out of 31) nuclei.     

Finally, panel e of Fig. \ref{kin2}   shows the $\eta$ measurements as a function of the IC of each target. No clear trend is observed  for the three gas kinematic tracers (i.e. Na {\small ID}, \ha, and [O {\small III}], with Spearman coefficients of $\sim 0.1$). Despite the poor statistics, this finding suggests that nuclear winds are ubiquitous during the pre- and post-coalescence phases of major mergers.
The dominant non-gravitational component in almost all ULIRG nuclei, together with the prevalent prominent blue wings in [O {\small III}] and Na {\small ID}, with $v10$ of several 100s km/s, indicates strong multi-phase outflows driven by either AGN or SB nuclear activity.

\begin{table*}
\footnotesize
\begin{minipage}[!h]{1\linewidth}
\setlength{\tabcolsep}{4pt}
\centering
\caption{ISM nuclear properties}
\begin{tabular}{lcccccc}
IRAS ID (other)   &   optical & BPTs & \multicolumn{3}{c}{dominant motions} & max $w80$ \\
               &   class   &      & \multicolumn{3}{c}{from $\eta$ criterion} & (km/s) \\
               &          &       & \ha & [O {\small III}] & Na {\small ID} & \\
\scriptsize{(1)} & \scriptsize{(2)}   &\scriptsize{(3)}   & \scriptsize{(4)} & \scriptsize{(5)} & \scriptsize{(6)} & \scriptsize{(7)}\\
\hline

 F00188$-$0856   &  AGN & (L,A,A) & non-grav & non-grav & non-grav & $1325_{-195}^{+255}$ (Na {\small ID}) \\
\hline
IZw1 &  AGN & (A,A,A) & - & - & - & $2190_{-70}^{+10}$ ([O {\small III}])\\
\hline
F01572+0009  &  AGN & (A,A,A) & - & - & - & $1670_{-60}^{+10}$ ([O {\small III}])\\
\hline

F05189$-$2524   & AGN & (A,A,A) & non-grav & non-grav & non-grav &  $1860_{-10}^{+130}$ ([O {\small III}]) \\
\hline

07251$-$0248 E &  SB & (S,S,S) & grav & non-grav & non-grav & $880_{-200}^{+230}$ (Na {\small ID})\\

07251$-$0248 W & SB & (S,S,S) & non-grav & non-grav & - & $410\pm 5$ ([O {\small III}]) \\
\hline

09022$-$3615   & SB & (C,S,S) & non-grav & non-grav & - & $495\pm 5$ ([O {\small III}])\\
\hline

10190+1322 E  & SB & (C,S,S) & grav & grav & grav & $530_{-20}^{+5}$ (Na {\small ID})\\

10190+1322 W  & SB & (C,S,S) & non-grav & non-grav & non-grav & $555_{-5}^{+5}$ ([O {\small III}])\\
\hline 

F11095$-$0238 NE  & ? & (C,S,L) & non-grav & non-grav & non-grav & $610_{-70}^{+10}$ ([O {\small III}])\\

F11095$-$0238 SW  & ? & (C,S,L) & grav & non-grav & grav & $540\pm 5$ ([O {\small III}]) \\
\hline 

F12072$-$0444 N  & AGN & (A,A,A) & non-grav & non-grav & non-grav & $860\pm 5$ ([O {\small III}])\\

F12072$-$0444 S  & AGN & (A,A,A) & non-grav & non-grav & - & $880\pm 5$ ([O {\small III}])\\
\hline 

13120$-$5453   & AGN & (L,A,A) & non-grav & non-grav & non-grav & $1480_{-60}^{+5}$ (Na {\small ID})\\
\hline 

F13451+1232 E  & LI(N)ER & (L,L,L) & non-grav & non-grav & - & $1200_{-270}^{+70}$ ([O {\small III}])\\

F13451+1232 W (4C 12.50) & AGN & (A,A,A) & non-grav & non-grav & - & $2545\pm 5$ (H$\alpha$)\\
\hline 

F14348$-$1447 NE  & ? & (C,S,L) & non-grav & non-grav & non-grav & $620_{-70}^{+10}$ ([O {\small III}]) \\

F14348$-$1447 SW  & ? & (L,S,A) & non-grav & non-grav & non-grav & $1060\pm 5$ (Na {\small ID})\\
\hline 

F14378$-$3651 & AGN & (L,A,A) & non-grav & non-grav & non-grav & $1120\pm 5$ ([O {\small III}])\\
\hline 

Arp220 E & LI(N)ER & (L,L,L) & non-grav & non-grav & non-grav & $810_{-140}^{+300}$ (Na {\small ID})\\

Arp220 W & LI(N)ER & (L,L,L) & non-grav & non-grav & non-grav & $810\pm 5$ ([O {\small III}])\\
\hline

F16090$-$0139  & SB & (C,S,S) & grav* & grav* & grav* & $840_{-55}^{+60}$ (Na {\small ID})\\
\hline

17208$-$0014 & LI(N)ER & (C,L,L) &grav & grav & non-grav& $435_{-5}^{+55}$ (Na {\small ID})\\
\hline

F19297$-$0406 S & SB & (C,S,S) & grav & non-grav & non-grav & $760_{-180}^{+110}$ (Na {\small ID})\\

F19297$-$0406 N & SB & (C,S,S) & grav & grav & non-grav & $630_{-55}^{+10}$ (Na {\small ID})\\
\hline

19542+1110   & ? & (L,S,A) &  non-grav & non-grav & non-grav & $1440_{-180}^{+260}$ (Na {\small ID})\\
\hline

20087$-$0308     & LI(N)ER & (L,L,L) & grav & non-grav & non-grav & $800_{-50}^{+230}$ (Na {\small ID})\\
\hline 

20100$-$4156  N   & SB & (S,S,S) & non-grav & non-grav & non-grav & $400\pm 5$ ([O {\small III}])\\

20100$-$4156  S   & LI(N)ER & (L,S,L) & non-grav & non-grav & non-grav & $560_{-380}^{+400}$ (Na {\small ID})\\
\hline

F22491$-$1808 E & SB & (C,S,S) & grav & grav & grav & $355\pm 5$ (Na {\small ID})\\

F22491$-$1808 W & SB & (S,S,S) & non-grav & non-grav & non-grav & $295_{-50}^{+10}$  (Na {\small ID})\\

\hline
\hline
\end{tabular}
\label{Tism}
\vspace{0.2cm}
\end{minipage}
{\small
{\it Notes.}\\ Column (1): Target name. \\ Column (2): Optical classification obtained from the three BPT diagrams shown in Fig. \ref{BPT}. The classification is defined only when two or three diagnostic diagrams indicate the same ionisation mechanism; when the three diagnostics indicate three different mechanisms, a question mark is placed in this column.\\ Column (3): Ionisation class from the  [N {\small II}]-, [S {\small II}]-, and [O {\small I}]-BPT diagostics (S: HII, C: composite, A: AGN, and L: LI(N)ER line ratios).\\  
Columns (4-6): Dominant motions from $\eta = w80/(2.355\times \sigma_*)$, considering the H$\alpha$, [O {\small III}], and Na {\small ID} line widths; see Sect. \ref{Soutflowcorr} for details. \\
Column (7): Maximum $w80$ measured in the \ha and [O {\small III}] emission lines as well as the Na {\small ID} absorption features. \\ 
$^\text{*}$The asterisk indicates that the stellar velocity dispersion in the F16090-0139 nucleus is probably overestimated ($\sigma_* = 310$ km/s) because of the low S/N.   
}
\end{table*}

\subsection{AGN- and SB-driven outflows}\label{Soutflowcorr}

In Sect. \ref{Smorph} we reported a good consistency between the morphological and stellar kinematic classifications, with merging  systems more  likely  associated with ordered disk-like motions (in 8 out of 11 of our ULIRGs) and interacting systems with non-ordered or streaming motions (in 7 out of 10 ULIRGs).
The ubiquitous presence of atomic outflows (Sect. \ref{Soutflow}) suggests that nuclear winds are not related to a specific phase or to more/less dynamically relaxed systems, but are common in all final stages of the merger process (as has been reported in the literature; see e.g. Sect. 10.5 in \citealt{Perna2020}).   
In this section, we therefore investigate the possible connection between AGN and SB activity and the atomic outflow velocities.  

Figure \ref{kin3} (left) shows the distribution of the [O {\small III}] $w80$ measurements against the  [O {\small III}] luminosity. In the same figure (right), we also report the \ha $w80$ versus L$_{H\alpha}$. The [O {\small III}] and \ha luminosities were corrected for extinction, considering the measured Balmer decrement and a \citet{Cardelli1989} extinction law. The [O {\small III}] measurements show a positive trend with increasing luminosity (with a Spearman coefficient of 0.6), while \ha does not show any significant correlation; we also note that our ULIRGs cover similar region in the [O {\small III}] diagram as the X-ray/SDSS AGNs analysed in \citet{Perna2017a}, reported in our figure with grey pluses.  This might suggest that {\it (i)} the AGN activity, traced by L$_{ [OIII]}$, is responsible for the outflows in the ULIRG nuclei, or that {\it (ii)} AGN- and SB-driven winds are hardly distinguishable in this diagram.

To test these two scenarios, we studied the correlation between two different estimates for the AGN fraction: (i) $\alpha_{AGN}$, the AGN contribution to the L$_{bol}$ according to mid-IR fluxes (Sect. \ref{Sample}), and (ii) $L_{AGN}/L_{bol}$, the ratio between the [O {\small III}]-based AGN luminosity (inferred assuming a bolometric correction of $3\times 10^3$; \citealt{Heckman2004}), and the bolometric luminosity L$_{bol}$, given by $1.15\times  L_{IR}$ (\citealt{Veilleux2009}). A positive correlation between these quantities is expected if [O {\small III}] traces the AGN power. In Fig. \ref{kin4} (left), we report $\alpha_{AGN}$ as a function of $L_{AGN}/L_{bol}$. 
A significant correlation (with a Spearman coefficient of 0.6) is found when we exclude the most deviating measurement, namely the red dot associated with F13451+1232 E (whose AGN fraction is reasonably overestimated; see below). Therefore this finding is in principle consistent with the hypothesis that the AGN activity causes the outflows we observe in almost all ULIRG nuclei.

However, there are a number of caveats concerning the relation in the left panel of Fig. \ref{kin4}. 
Both L$_{bol}$ and $\alpha_{AGN}$ are based on spatially integrated IR emission measurements and refer to the entire ULIRG systems.  
We can reasonably assume that most of this IR emission comes from the nuclei (e.g. \citealt{Lutz1999}, Pereira-Santaella et al., in prep.), and that all quantities in the figure are related to the same spatial regions, at least for the targets with a single nucleus. For the binary systems, we used the same L$_{IR}$ and $\alpha_{AGN}$ for both nuclei (these systems are connected with vertical lines in the figure). 
This is a reasonable assumption because the [O {\small III}] luminosities associated with the two nuclei of individual systems are comparable within a factor of a few. 
The only exception is represented by F13451+1232, with L$_{[OIII]} = 6.8\times 10^{42}$ erg/s (W) and $7.8\times 10^{39}$ erg/s (E). The western nucleus, associated with AGN activity (Fig. \ref{BPT}), harbours the powerful radio source 4C 12.50 (\citealt{Lister2003}), and might cause most of the IR emission. If this is the case, the only deviating point in Fig. \ref{kin4} (left),  associated with F13451+1232 E, should have a much smaller $\alpha_{AGN}$. 
However, a detailed investigation is required to confirm all assumptions mentioned so far. Moreover, BPT diagnostics revealed that AGN dominates SB and LI(N)ER ionisation mechanisms in only nine nuclei (large symbols in Fig. \ref{kin4}, left); in the remaining nuclei, the [O {\small III}] luminosity is therefore a poor tracer for the AGN power. All these arguments weaken the significance of the trend reported in the left panel of Fig. \ref{kin4}.

We also considered the correlation between the Na {\small ID} and [O {\small III}] line widths and $\alpha_{AGN}$ because broader line profiles are generally reported in the literature for AGN-driven outflows (e.g. \citealt{Cazzoli2016,Kakkad2020}). These measurements are reported  in the right panel of Fig. \ref{kin4} and do not show a clear trend (Spearman correlation of 0.4). We therefore conclude that although a significant contribution of [O {\small III}] might be due to AGN,  a combination of SB- and AGN-driven winds might cause the observed atomic outflows. More detailed analyses of the outflow nature will be presented in future papers, however, together with detailed spatially resolved properties of neutral and ionised outflows (e.g. \citealt{Perna2020}). 

We finally note that most of the nuclei with neutral outflows have relatively low AGN fractions; instead, at $\alpha_{AGN} > 0.5$, Na {\small ID} outflows are present in only one nucleus, F05189-2524 (out of eight, see the right panel of Fig. \ref{kin4}), confirming the difficulties in observing neutral outflows in systems with strong AGN (e.g. \citealt{Perna2017a,Bae2018,Nedelchev2019}). A detailed spatially resolved analysis is required, however, to exclude neutral outflows in the more external regions of these powerful AGN (e.g. \citealt{Rupke2017}; Perna et al., in prep.).

\section{Conclusions}\label{Sconclusions}

The project called Physics of ULIRGs with MUSE and ALMA (PUMA) is a survey of 25 nearby ULIRGs observed with MUSE and ALMA. This is a representative sample that covers the entire ULIRG luminosity range, and it includes a combination of systems with AGN and SB nuclear activity in (advanced) interacting and merging stages. 
This project represents the first such study intended to characterise the multi-phase structure of the ISM in local ULIRGs at subkpc resolutions. 
This paper is the first in a series that will explore the prevalence of ionised, neutral, and molecular outflows as a function of the galaxy and BH properties, and the nature and the (feedback) effects of such outflows on the galaxy evolution. 

In this work, we presented the first data products obtained from analysing the MUSE data of the 21 ULIRGs observed so far. We described the stellar kinematics derived with the pPXF analysis, and  the properties of stellar and ionised gas emission. Our first results are summarised below. 
Colour-composites, ionised gas, and stellar continuum emission images show a great richness in spatial details at different spatial scales. These images reveal recent galaxy interactions and strong (dust-enshrouded) nuclear activity, in line with their ULIRG nature (see e.g. Fig. \ref{MUSE3colours}).

Stellar kinematics revealed that merging systems are more likely associated with ordered disk-like motions (8 out of 11),  while binary (interacting) systems are dominated by non-ordered and streaming  motions (7 out of 10; see e.g. Fig. \ref{2dvs}). 

The sources with stellar rotational patterns are more likely found in compact mergers hosting AGN (6 out of 11); on the other hand, more extended structures with non-ordered motions are found in interacting systems, showing  several additional nuclear SF clumps and strong tidal tails (Figs. \ref{2dha} and \ref{2dvs}).

All ULIRGs show distinct velocity structures that are detached from the the inner pattern through tidal tails. This also applies to the post-coalescence mergers with clear inner rotational patterns, and it confirms that the external structures require longer times to  reach a dynamically relaxed configuration.

In the second part of the paper, we analysed the 31 nuclear spectra extracted from the positions of the ULIRG nuclei, deriving the physical and kinematic properties of the nuclear ISM. Our first results are summarised below. 
We used BPT diagnostics to constrain the dominant ionisation mechanism that causes the optical emission lines: 11 out of 31 nuclei are associated with SF, 9 out of 31 with AGN, and 6 out of 31 with LI(N)ER ionisation. Overall, the inferred classification is consistent with the classifications reported in the literature (Table \ref{tab1}), although the latter were obtained with long-slit spectra. Our classification is also consistent with the archival $\alpha_{AGN}$ estimates, obtained from mid-IR diagnostics. For the remaining 5 nuclei, the three BPT diagnostics indicated three different ionisation mechanisms, possibly indicating a more complex mixture between SB, AGN, and shock-induced ionisation (see Table \ref{Tism}). 

Almost all nuclear spectra show asymmetric and broad line profiles in both neutral (Na {\small ID}) and ionised (e.g. [O {\small III}]) transitions. Nuclear ISM features display
     velocity dispersions $>0.15$ dex higher than $\sigma_*$ in $\sim 85\%$ of the nuclear spectra. Following \citet{Woo2016}, we considered these enhancements with respect to $\sigma_*$ as an indication of strong 
     non-gravitational motions in the ISM component. Together with the extreme $v10$ velocities associated with [O {\small III}] and Na {\small ID} lines (Fig. \ref{kin1}), this suggests the ubiquitous presence of powerful nuclear winds in our sample. 

Most of the nuclei with neutral outflows have relatively low AGN fractions; at $\alpha_{AGN}>0.5$, Na {\small ID} outflows are instead present in one nucleus (out of 8, see Fig. \ref{kin4}, left), confirming the difficulties of observing neutral outflows in systems with strong AGN.

While in the this paper we have presented and described the general properties of the sample and the MUSE data, more detailed studies also involving ALMA data will be presented in future papers.

\begin{acknowledgements}
 
We thank the referee for an expert review of our paper.
The authors thanks Elena Valenti for her  support when preparing the observations, and G. Vietri for useful discussion on spectral analysis of type 1 AGN. 
MP is supported by the Programa Atracci\'on de Talento de la Comunidad de Madrid via grant 2018-T2/TIC-11715. MP, SA, CTC and LC acknowledge support from the Spanish Ministerio de Econom\'ia y Competitividad through the grant ESP2017-83197-P, and PID2019-106280GB-I00. 
MPS acknowledges support from the Comunidad de Madrid through the Atracci\'on de Talento Investigador Grant 2018-T1/TIC-11035 and PID2019-105423GA-I00 (MCIU/AEI/FEDER,UE).
EB acknowledges support from Comunidad de Madrid through the Attracci\'on de Talento grant 2017-T1/TIC-5213. 
SC acknowledge financial support from the State Agency for Research of the Spanish MCIU through the \lq\lq Center of Excellence Severo Ochoa\rq\rq \ award to the Instituto de Astrof{\'i}sica de Andaluc{\'i}a (SEV-2017-0709).
ACG acknowledges support from the Spanish Ministerio de Econom\'ia y Competitividad through the grant BES-2016-078214. 
RM acknowledges ERC Advanced Grant 695671 ``QUENCH''  and support by the Science and Technology Facilities Council (STFC).
JPL acknowledges financial support by the Spanish MICINN under grant AYA2017-85170-R.

\end{acknowledgements}

\begin{appendix}

\section{Spectroscopic redshifts}

Table \ref{Tredshifts} display all spectroscopic redshifts we derived by modelling the stellar continuum and line features of nuclear spectra, as well as of ULIRG nearby companions detected in the MUSE data cubes. Coordinates, velocity offsets, and projected distances from the bright nucleus of each ULIRG are also reported. 

\begin{table*}
\footnotesize
\begin{minipage}[!h]{1\linewidth}
\setlength{\tabcolsep}{7pt}
\centering
\caption{ULIRG nuclei and companion (pPXF spectroscopic) redshifts and positions.}
\begin{tabular}{lccccc}
source name    &    RA            &       DEC            & z & $\Delta V$  & Projected distance\\
               & ($^h:\ ^m:\ ^s$) & ($^{\circ}:\ ':\ '')$&   & (km/s)      & ($''$/kpc)        \\
\scriptsize{(1)} & \scriptsize{(2)}   &\scriptsize{(3)}   & \scriptsize{(4)}   &\scriptsize{(5)}   & \scriptsize{(6)}   \\
\hline
\hline

F00188-0856   & 0:21:26.52 & $-$8:39:25.92 & $0.1284 \pm 0.0001$ & - & - \\
F00188-0856: c   & 0:21:25.43 & $-$8:39:18.46 & $0.1219 \pm 0.0001$ & $-60\pm 25$ & 18.0/41.6 \\
\hline

IZw1  & 0:53:34.93 & +12:41:35.94 & $0.0611\pm 0.0001$ &  - & -\\
IZw1: c & 0:53:33.88 & +12:41:33.63 & $0.0613\pm 0.0002$ &  $+60\pm 25$ & 15.5/18.4\\\hline

F01572+0009 & 1:59:50.25 & +0:23:40.87 & $0.1632\pm 0.0001$ &  - & -\\
F01572+0009: c & 1:59:48.76 & +0:23:43.49 & $0.1623\pm 0.0001$ &  $-244\pm 10$ & 21.7/61.3\\\hline

F05189-2524  & 5:21:01.40 & $-$25:21:45.30 & $0.0428\pm 0.0001$ &  - & -\\\hline

07251-0248 E & 7:27:37.61 & $-$2:54:54.25  & $0.0879\pm 0.0001$ &  - & -\\
07251-0248 W & 7:27:37.54 & $-$2:54:54.39  & $0.0881\pm 0.0001$ &  $55\pm 10$ & 1.1/1.8\\
\hline

09022-3615  & 9:04:12.71 & $-$36:27:01.93 & $0.0596 \pm 0.0001$ & - & - \\
\hline

10190+1322 W  & 10:21:42.49 & +13:06:53.83 & $0.0767\pm 0.0001$ &  - & -\\
10190+1322 E  & 10:21:42.75 & +13:06:55.61 & $0.0758\pm 0.0001$ &  $-264\pm 10$ & 4.9/7.2\\
10190+1322: c$_{SE}$  & 10:21:43.83 & +13:06:47.54 & $0.0760\pm 0.0001$ &  $-187\pm 11$ & 20.6/30.1\\
10190+1322: c$_{NW}$  & 10:21:41.88 & +13:07:06.36 & $0.0771\pm 0.0001$ &  $120\pm 11$ & 15.8/23.1\\
\hline

F11095-0238 NE   & 11:12:03.38 & $-$2:54:22.94 & $0.1064\pm 0.0001$  &  - & -\\
F11095-0238 SW   & 11:12:03.36 & $-$2:54:23.30 & $0.1065\pm 0.0002$  &  $+30\pm 18$ & 0.6/1.1\\
\hline

F12072-0444 N   & 12:09:45.12 & $-$5:01:13.31& $0.1288\pm 0.0001$ &  - & - \\
F12072-0444 S   & 12:09:45.13 & $-$5:01:14.23 & $0.1288\pm 0.0001$ &  $-7\pm 10$ & 1.0/2.3 \\
F12072-0444: c$_1$   &12:09:44.81 & $-$5:00:54.78 & $0.1302\pm 0.0006$ & $360\pm 30$ & 19.1/44.2 \\
F12072-0444: c$_2$   &12:09:44.82 & $-$5:00:52.33 & $0.1302\pm 0.0006$ & $360\pm 30$ & 21.6/49.94 \\
\hline

13120-5453 & 13:15:06.32 & $-$55:09:22.82 & $0.0310\pm 0.0001$ & - & -\\
\hline

F13451+1232 W  & 13:47:33.36 & +12:17:24.24 & $0.1218\pm 0.0002$ & - & -\\
F13451+1232 E  & 13:47:33.49 & +12:17:23.76 & $0.1218\pm 0.0002$ & $-5\pm 10$ & 2.0/4.3\\
\hline

F14348-1447 SW & 14:37:38.28 & $-$15:00:24.24 & $0.0824\pm 0.0002$ & - & -\\
F14348-1447 NE & 14:37:38.40 & $-$15:00:21.29  & $0.0822\pm 0.0002$ & $-49\pm 30$ & 3.4/5.3\\
F14348-1447: c & 14:37:38.95 & $-$15:00:25.47  & $0.0823\pm 0.0001$ & $-22\pm 8$ & 10.3/16.1\\
\hline

F14378-3651: n & 14:40:58.89 & $-$37:04:32.08 & $0.0682\pm 0.0001$ & - & -\\
F14378-3651: c & 14:41:01.15 & $-$37:04:43.94 & $0.0684\pm 0.0001$ & $48\pm 7$ &  29.7/38.9\\
\hline

F15327+2340 (Arp220): n$_W$ & 15:34:57.24 & +23:30:11.70 & $0.0181\pm 0.0001$ & - & - \\
F15327+2340 (Arp220): n$_E$ & 15:34:57.30 & +23:30:11.90 & $0.0182\pm 0.0001$& $30\pm 9$ & 1/0.37\\
\hline

F16090-0139: n   & 16:11:40.42 & $-$1:47:06.56 & $0.1337\pm 0.0002$ & - & - \\
\hline

17208-0014: n & 17:23:21.94 &$-$00:17:00.96& $0.0430\pm 0.0001$ & - & -\\
\hline

F19297-0406 S  & 19:32:22.30 & $-$4:00:01.80 & $0.0854\pm 0.0001$ &  - & -\\
F19297-0406 N  & 19:32:22.31 & $-$4:00:01.03 & $0.0853\pm 0.0001$ & $-25\pm 9$& 0.7/1.1\\
\hline

19542+1110 & 19:54:35.78 & 11:19:05.03 & $0.0624\pm 0.0002$ &  - & -\\
19542+1110: c & 19:54:34.91 & 11:19:20.94 & $0.0628\pm 0.0001$ &   $40\pm 15$ & 20.5/24.8\\
\hline

20087-0308  & 20:11:23.87 & $-$2:59:50.71& $0.1052\pm 0.0001$ &  - & -\\
\hline

20100-4156 SE & 20:13:29.56 & $-$41:47:35.21     & $0.1297\pm 0.0002$ &  - & -\\
20100-4156 NW     & 20:13:29.48 & $-$41:47:32.58 &  $0.1297\pm 0.0001$ &  $-6\pm 10$ & 2.8/6.5\\
20100-4156 c     & 20:13:28.63 & $-$41:47:38.39 &  $0.1300\pm 0.0001$ &  $90\pm 8$ & 10.5/24.3 \\
\hline

F22491-1808 W & 22:51:49.24 & $-$17:52:23.66 & $0.0776 \pm 0.0001$ & - & -\\
F22491-1808 E & 22:51:49.35 & $-$17:52:24.12 & $0.0777\pm 0.0002$ & $+35\pm 9$ & 1.8/2.7 \\

\hline
\hline
\end{tabular}
\label{Tredshifts}
\vspace{0.2cm}
\end{minipage}
{\small
{\it Notes.} Column (1): Target name. (2) and (3): Coordinates (RA and DEC). (4): Spectroscopic redshift from pPXF at the position of the stellar continuum peak. (5):  Velocity with respect to the ULIRG systemic. (6): Projected distance from the ULIRG (brightest) nucleus.    
}
\end{table*}

\section{Seyfert 1 fit analysis}\label{AfitSy1}

For the two Sy 1 nuclear spectra, we proceeded in two steps. First, we modelled the continuum with a power law and the iron emission components with the observational templates of \citet{Veron2004}. This initial fit was performed considering the wavelength ranges $4050-7100 \AA$ (for F01572+0009) and $4450-7100 \AA$ (for Izw1), but masking all other prominent emission lines, namely the BLR Balmer and He lines, as well as the NLR forbidden [O {\small III}], [N {\small II}], [S {\small II}], and [O {\small I}] features. We therefore obtained a template for the continuum and iron emission that we used in the second step. 

Then, we modelled all remaining emission lines with a combination of Gaussian profiles, as well as broken power-law functions (e.g. \citealt{Cresci2015}) for the BLR emission components, together with the continuum and iron components (for which we obtained good initial estimates for the fit parameters from the previous step). This analysis is very similar to the one presented in Sect. \ref{Sism}, except for the following aspects.
Additional emission lines were detected and included in the fit of Sy 1 spectra: the Balmer H$\gamma$ and H$\delta$, the Si II$\lambda\lambda$6347,71, the He II at 4685$\AA$, and the He I lines at 4387, 4713, 4921, 5875, 6678, and 7065$\AA$. To model the Balmer emission lines, we used two parameters for the \ha and H$\beta$ amplitudes, and derived the \ha/H$\beta$ flux ratio to measure the dust extinction using the \citet{Cardelli1989} extinction law and assuming case B recombination as well as $T_e = 10^4$ K. This dust correction was then used to infer the fluxes of H$\gamma$ and H$\delta$, considering case B recombination flux ratios H$\gamma /H\beta =0.468$ and H$\delta /H\beta =0.260$. Similarly, we considered the standard He I theoretical intensity ratios from \citet{Smits1996} to constrain the fluxes of all He I transitions: In particular, we assumed low-density plasma conditions (with $T_e = 10^4$ K and $N_e=10^3$ cm$^{-3}$) for the NLR component, and high-density conditions (with $T_e = 10^4$ K and $N_e=10^6$ cm$^{-3}$) for the BLR. Therefore we used a unique parameter for the HeI$\lambda$4471 flux amplitude, and the theoretical intensity ratios to infer the fluxes of the remaining transitions, correcting for the dust extinction (from the Balmer decrement).  
For the BLR components, we assumed that all Balmer and helium lines have the same broken power-law indices (e.g. \citealt{Nagao2006}) and kinematics. 
BLR emission components were used for all permitted transitions (see e.g. \citealt{Veron2004}).

\section{Individual ULIRGs}\label{allmaps}

\begin{figure*}[t]
\centering
\includegraphics[width=22.cm,trim= 110 380 0 25,clip]{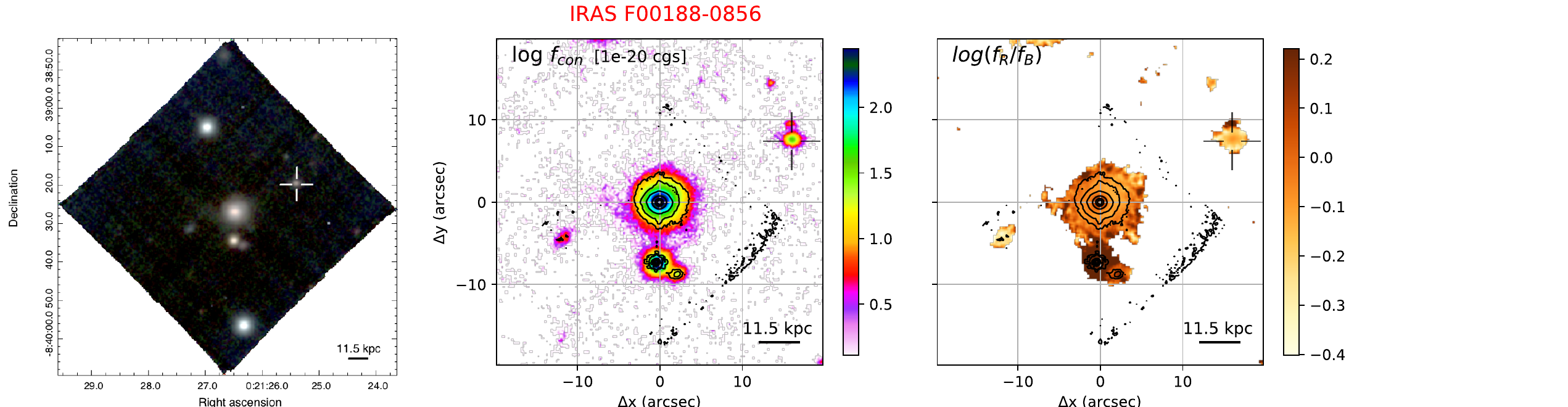}

\caption{\small IRAS F00188-0856 images from MUSE observations with TOT $= 0.68$ hr.  {\it Left}: Colour-composite optical image, showing [O {\small III}] (green, from the wavelength range $4990-5015\AA$ rest-frame), \ha (red, $6555-6574\AA$), and stellar continuum (blue, $4400-4500\AA$). {\it Centre}: Red ($7390-7530\AA$)  continuum image, with contours from HST/F160W. {\it Right}: Continuum colour map obtained from MUSE by dividing the red continuum image (in the central panel) by a blue image obtained by collapsing the stellar emission in the range $4400-4500\AA$; contours from HST/F160W. In all panels, we display the IRAS F00188-0856 companion with a cross.}
\label{IRAS00188_1}
\end{figure*}

Figures \ref{IRAS00188_1}-\ref{IRAS22491_n_w} display the three-colour composites, stellar and emission line maps, as well as the nuclear spectra for each ULIRG presented in this paper.  

\begin{figure*}[h]
\centering
\includegraphics[width=22.cm,trim= 105 200 0 218,clip]{{IRAS00188_Appendix}.pdf}

\caption{\small IRAS F00188-0856 emission line images from MUSE observations. [O {\small III}] (left, from the wavelength range $4990-5015\AA$ rest-frame), \ha (centre, $6555-6574\AA$), and [S {\small II}] (right, $6702-6742\AA$) images have been obtained by subtracting continuum emission using the adjacent regions at shorter and longer wavelengths with respect to the emission line systemics. In all panels, we display the position of the companion galaxy  with a cross.}
\label{IRAS00188_2}
\end{figure*}

\begin{figure*}[t]
\vspace{0.5cm}
\centering

\includegraphics[width=13.cm,trim= 295 20 90 398,clip]{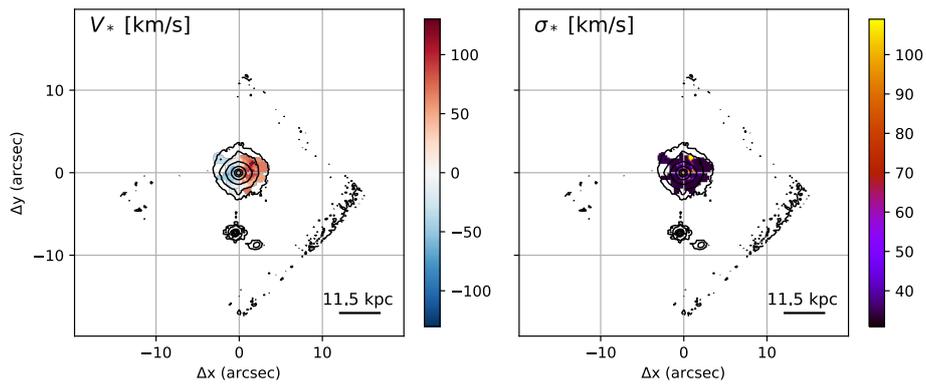}
\centering

\caption{\small IRAS F00188-0856 stellar kinematic maps from the pPXF analysis with contours from HST/F160W. The left panel shows the stellar velocity $V_*$, and the right panel displays the velocity dispersion $\sigma_*$.}
\label{IRAS00188_3}
\end{figure*}

\begin{figure*}[t]
\vspace{0.5cm}

\centering
\includegraphics[width=18.cm,trim= 0 0 0 0,clip]{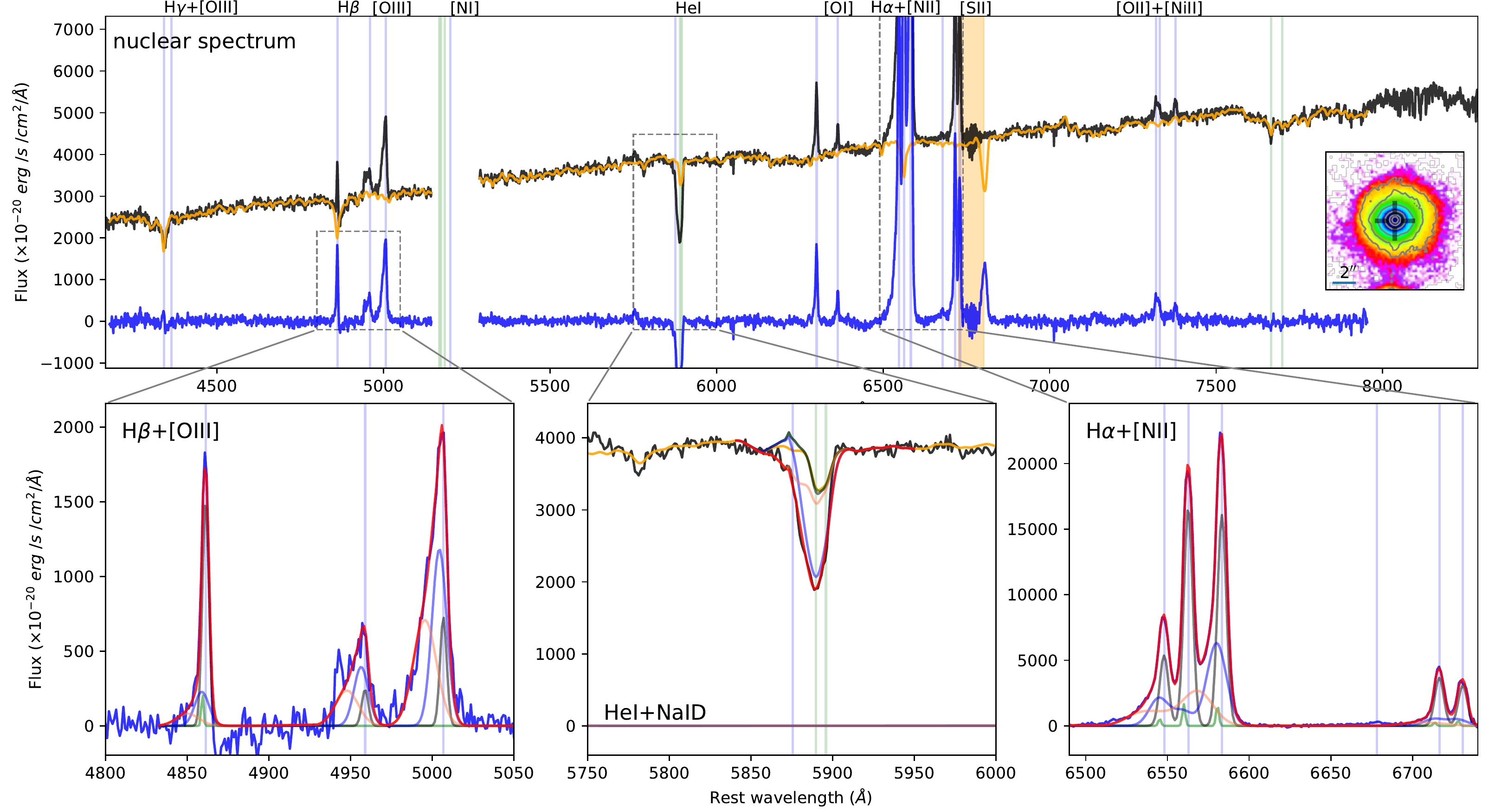}

\caption{\small {\it Top panel}: IRAS F00188-0856 nuclear spectrum (black curve), extracted from a circular aperture with $r< 0.4''$. The corresponding pPXF best-fit model profile is shown in orange. The continuum-subtracted spectrum (blue curve) is obtained by subtracting the best-fit pPXF model  from the original spectrum. The vertical blue lines mark the wavelengths of the emission lines detected in the spectrum; the vertical green lines mark the position of stellar absorption systems (i.e. from left to right: MgI triplet, Na {\small ID} and KI doublets). The regions excluded from the pPXF fit and corresponding to the most intense sky line residuals are highlighted as orange shaded areas; the portions of the spectra around 5300 $\AA$ are missing because a filter blocked the laser contamination. The inset in the top panel shows the red stellar continuum emission map in the vicinity of the nuclear regions, with  HST/F160W contours as in Fig. \ref{IRAS00188_1}; the nuclear position is shown with a black marker. {\it Bottom insets}: Multi-component best-fit analysis results for the main emission (H$\beta$ and [O {\small III}], left; H$\alpha$, [N {\small II}] and [S {\small II}], right) and absorption (Na {\small ID}, centre) line features. Na {\small ID} multi-component best-fit curves are superimposed on the observed spectrum (black curve); emission components are not required to reproduce the total profile of Na {\small ID}.}
\label{IRAS00188_n}
\end{figure*}

\clearpage 

\begin{figure*}[t]
\centering
\includegraphics[width=22.cm,trim= 100 380 0 25,clip]{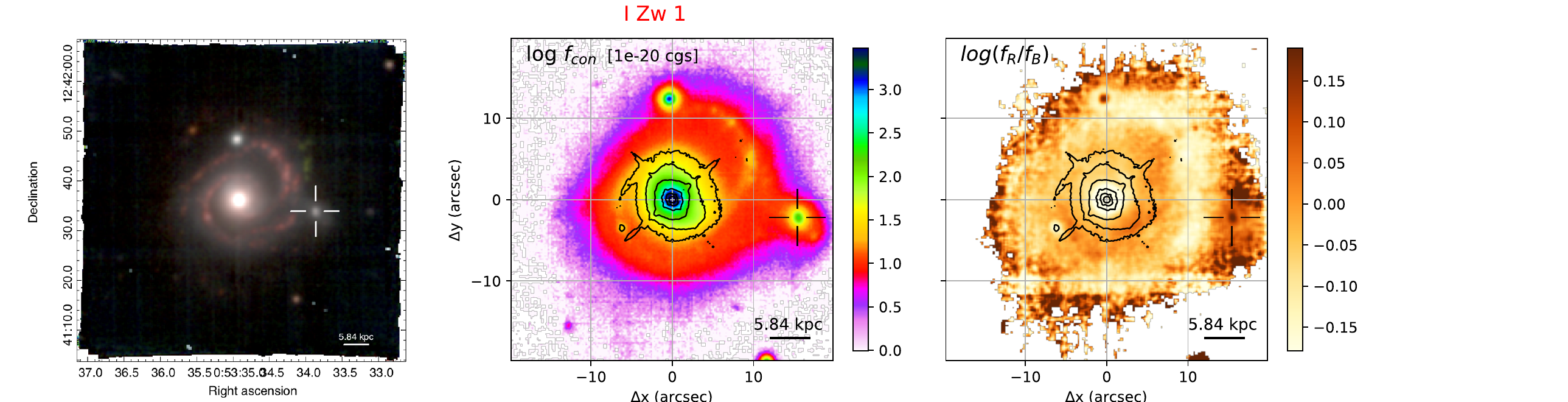}

\caption{\small IZw1 images from MUSE observations with TOT $= 2.72$ hr. {\it Left}: Colour-composite optical image, showing [O {\small III}] (green, from the wavelength range $4959-5018\AA$ rest-frame), \ha (red, $6552-6570\AA$), and stellar continuum (blue, $4570-4670\AA$). {\it Centre}: Red ($8010-8110\AA$) continuum image from MUSE , with contours from HST/F160W. {\it Right}: Continuum colour map obtained from MUSE by dividing the red continuum image (central panel) by a blue image obtained by collapsing the stellar emission in the range $4570-4670\AA$; contours from HST/F160W. In all panels, we display the  IZw1 companion galaxy with a cross.
}
\label{IZw1_1}
\end{figure*}

\begin{figure*}[t]
\vspace{0.5cm}
\centering

\includegraphics[width=22.cm,trim= 90 198 0 218,clip]{{Mrk1502_Appendix}.pdf}

\caption{\small
IZw1 emission line images from MUSE observations. [O {\small III}] (left, from the wavelength range $4985-5018\AA$ rest-frame), \ha (centre, $6552-6570\AA$) and [S {\small II}] (right, $6710-6741\AA$) images have been obtained by subtracting continuum emission using the adjacent regions at shorter and longer wavelengths with respect to the emission line systemics. In all panels, we display the position of the  non-emission line companion galaxy of IZw1 with crosses.}
\label{IZw1_2}
\end{figure*}

\begin{figure*}[t]
\vspace{0.5cm}
\centering

\includegraphics[width=13.cm,trim= 295 20 90 398,clip]{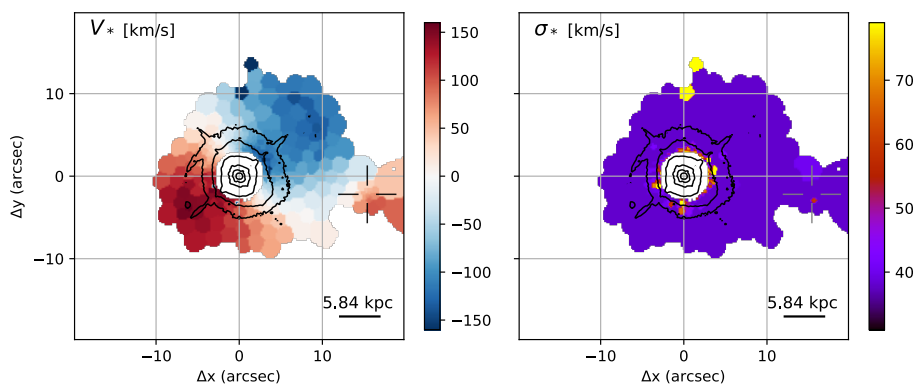}

\caption{\small {\it Left}: IZw1 stellar kinematic maps from the pPXF analysis with contours from HST/F160W. The left panel shows the stellar velocity $V_*$, and the right panel represents the velocity dispersion $\sigma_*$. The central pixels have been masked because of strong AGN continuum and BLR emission, which prevents the detection of stellar continuum and absorption features.}
\label{IZw1_3}
\end{figure*}

\begin{figure*}[t]
\vspace{0.5cm}

\centering
\includegraphics[width=18.cm,trim= 0 0 0 0,clip]{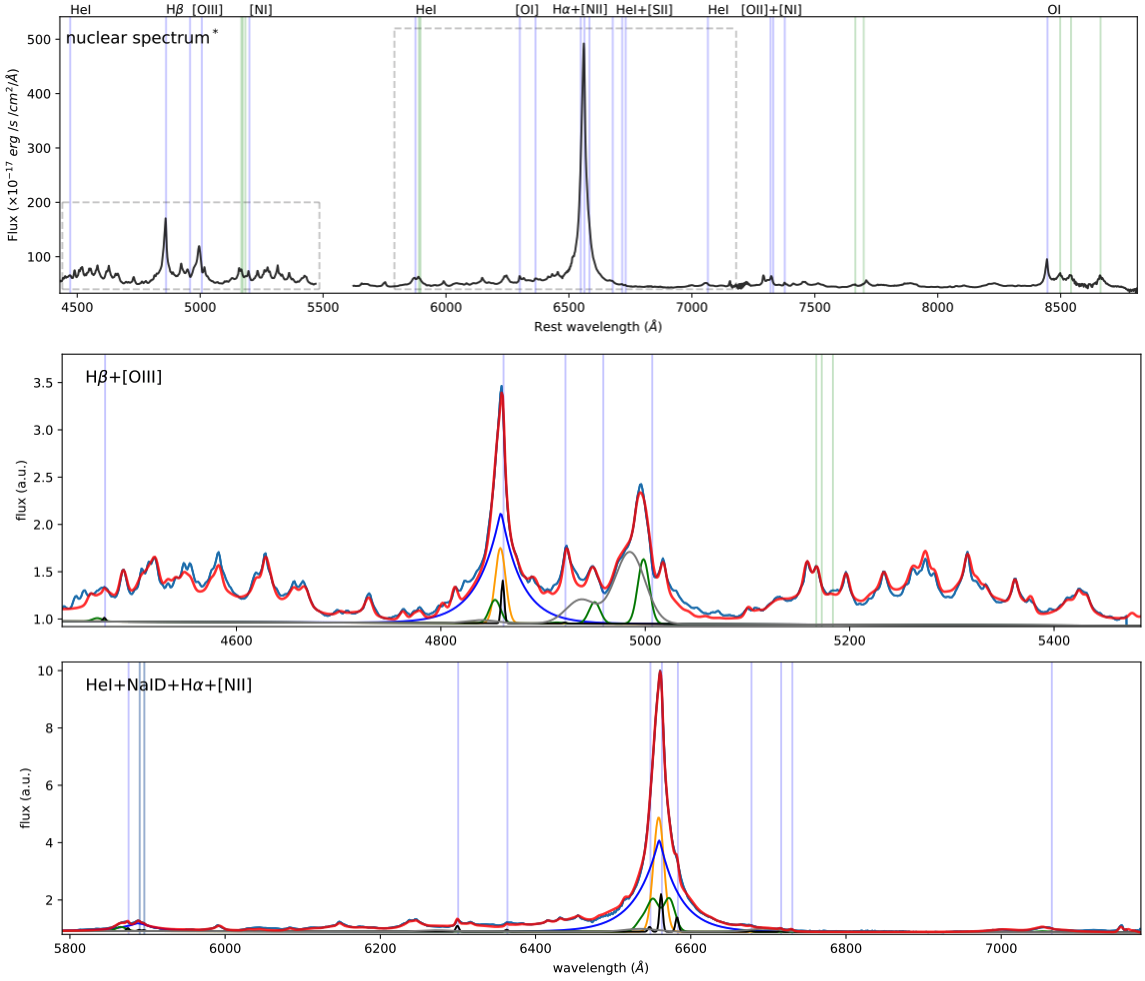}

\caption{\small {\it Top}: IZw1 nuclear spectrum (black curve) extracted from a circular aperture $r< 0.4''$ but excluding the central pixel (where \ha is saturated). The blue vertical lines mark the wavelengths of the emission lines detected in the spectrum; the green lines mark the position of stellar absorption systems (i.e. from left to right: MgI triplet, Na {\small ID} and KI doublets, CaII triplet). The latter transitions in absorption are not detected because of the strong AGN continuum and emission lines.  The portions of the spectra around 5500 $\AA$ are missing because a filter blocked the laser contamination; the portions in the two boxes are reported in the central and bottom panels with the best-fit curves.  {\it Centre}: Normalised nuclear spectrum in the vicinity of the H$\beta$+[O {\small III}] complex (blue curve), with total best-fit model (red). Individual kinematic components are shown with different colours: black, green, and grey Gaussians represent systemic and outflow components; blue and orange curves are used to model the BLR Balmer emission; and iron components are not reported. {\it Bottom}:  Normalised nuclear spectrum in the vicinity of the HeI+Na {\small ID} and H$\alpha$+NII complexes (blue curve), with superimposed best-fit model (red curve). Different kinematic components required to reproduce the spectrum are shown as in the central panel. }
\label{IZw1_n}
\end{figure*}

\clearpage 

\begin{figure*}[t]
\centering
\includegraphics[width=21.cm,trim= 100 380 0 15,clip]{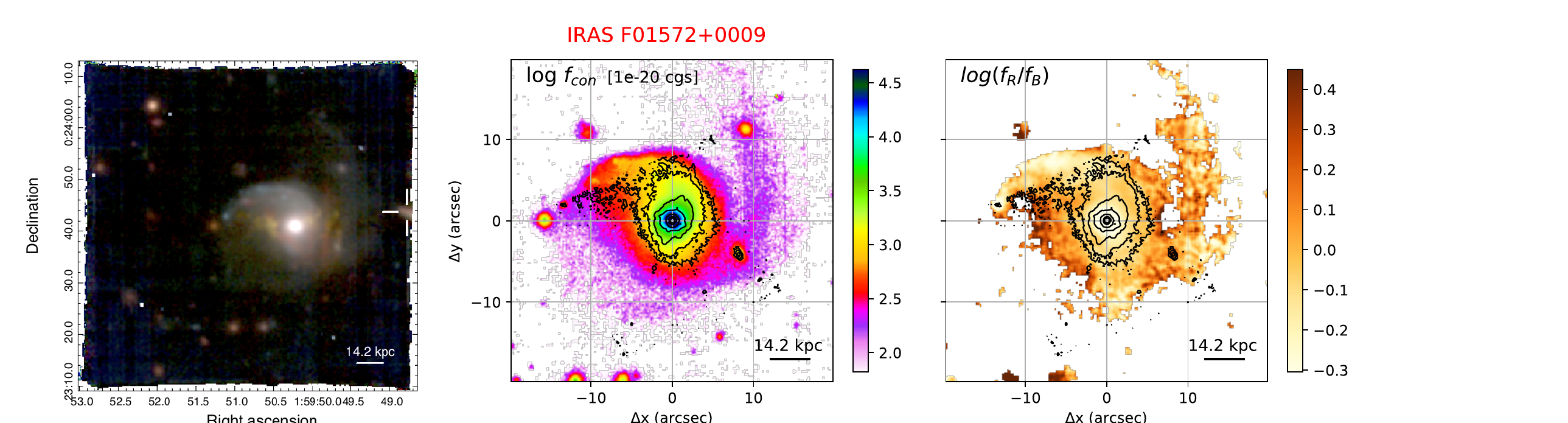}

\caption{\small 
IRAS F01572+0009 images from MUSE observations with TOT $= 2.04$ hr. {\it Left}: Colour-composite optical image showing [O {\small III}] (green, from the wavelength range $4925-4970 \AA$ rest-frame), \ha (red, $6550-6575 \AA$), and continuum (blue, $4150-4220 \AA$). {\it Centre}: Red ($7330-7400 \AA$) continuum image from MUSE with contours from HST/F160W. {\it Right}:  Continuum colour map obtained from MUSE by dividing the red continuum image (central panel) by a blue image obtained by collapsing the stellar emission in the range $4150-4220 \AA$; contours from HST/F160W. In the first panel, we display the  F01572+0009 companion galaxy with a cross.}
\label{IRAS01572_1}
\end{figure*}

\begin{figure*}[t]
\vspace{0.5cm}
\centering

\includegraphics[width=21.cm,trim= 90 197 0 218,clip]{{Mrk1014_Appendix}.pdf}

\caption{\small
IRAS F01572+0009 emission line images from MUSE observations. [O {\small III}] (left, from the wavelength range $4925-4970 \AA$ rest-frame), \ha (centre, $6550-6575 \AA$), and [S {\small II}] (right, $6700-6748 \AA$) images have been obtained by subtracting continuum emission using the adjacent regions at shorter and longer wavelengths with respect to the emission line systemics.}
\label{IRAS01572_2}
\end{figure*}

\begin{figure*}[t]
\vspace{0.5cm}
\centering

\includegraphics[width=13.cm,trim= 295 20 90 398,clip]{{Mrk1014_Appendix}.pdf}

\caption{\small {\it Left}: IRAS F01572+0009 stellar kinematic maps from the pPXF analysis with contours from HST/F160W. The left panel shows the stellar velocity $V_*$, and the right panel represents the velocity dispersion $\sigma_*$.}
\label{IRAS01572_3}
\end{figure*}

\begin{figure*}[t]
\vspace{0.5cm}

\centering
\includegraphics[width=16.cm,trim= 0 0 0 0,clip]{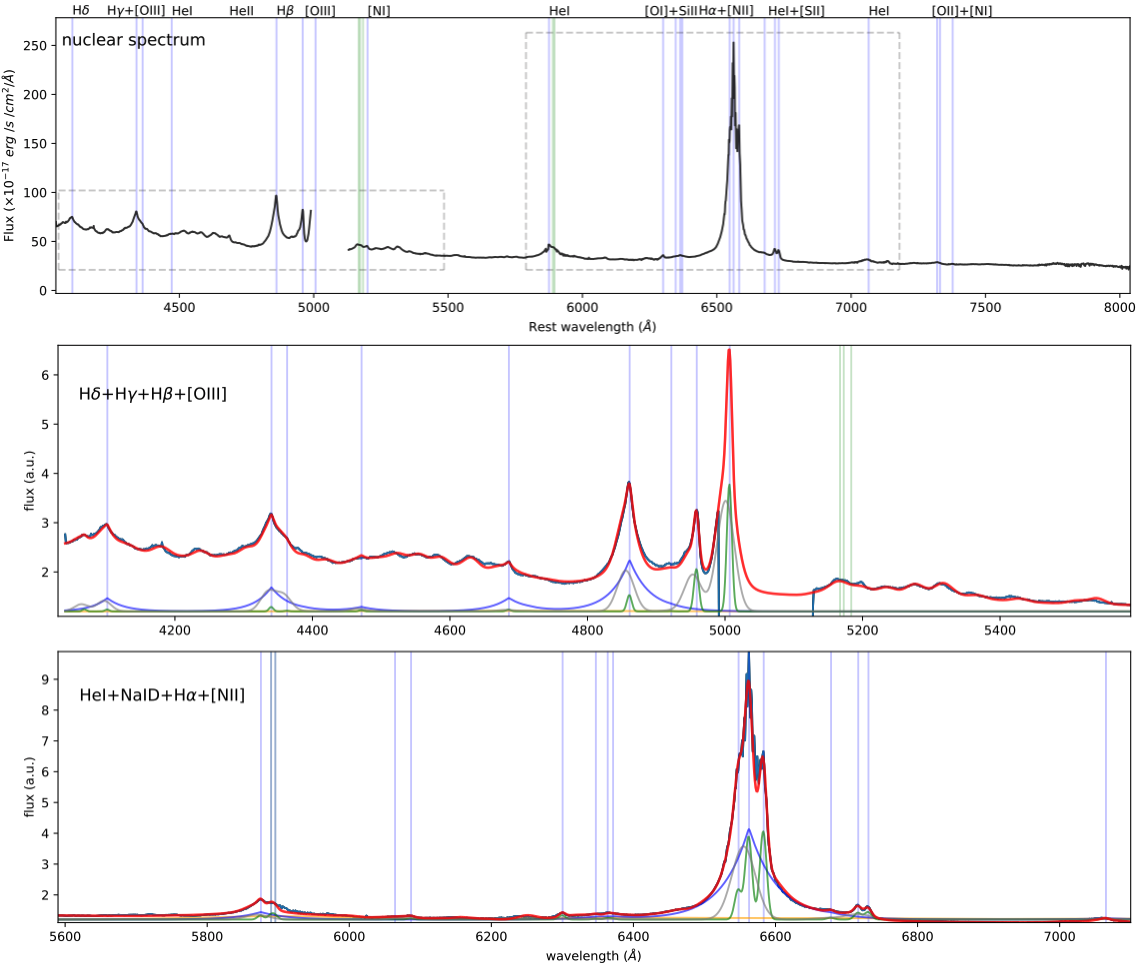}

\caption{\small {\it Top}: IRAS F01572+0009 nuclear spectrum (black curve) extracted from a circular aperture $r< 0.4''$. The blue vertical lines mark the wavelengths of the emission lines detected in the spectrum; the green lines mark the position of stellar absorption systems (i.e. from left to right: MgI triplet, and Na {\small ID} doublets). The latter transitions in absorption are not detected because of the strong AGN continuum and emission lines.  The portions of the spectra around 5050 $\AA$ are missing because a filter blocked the laser contamination; the portions in the two boxes are reported in the central and bottom panels with the best-fit curves.  {\it Centre}: Normalised nuclear spectrum in the vicinity of the H$\beta$+[O {\small III}] complex (blue curve), with total best-fit model (red). Individual kinematic components are shown with different colours: green and grey Gaussians represent systemic and outflow components, respectively; blue and orange curves are used to model the BLR Balmer emission; and iron components are not reported. The [O {\small III}]$\lambda$5007 line is mostly within the missing portion of the spectrum; this profile has been reconstructed using the [O {\small III}]$\lambda$4959 transition. {\it Bottom}:  Normalised nuclear spectrum in the vicinity of the HeI+Na {\small ID} and H$\alpha$+NII complexes (blue curve), the best-fit model is superimposed (red curve). The different kinematic components required to reproduce the spectrum are shown as in the central panel. }
\label{IRAS01527_n}
\end{figure*}

\clearpage 

\begin{figure*}[t]
\centering
\includegraphics[width=21.cm,trim= 100 380 0 15,clip]{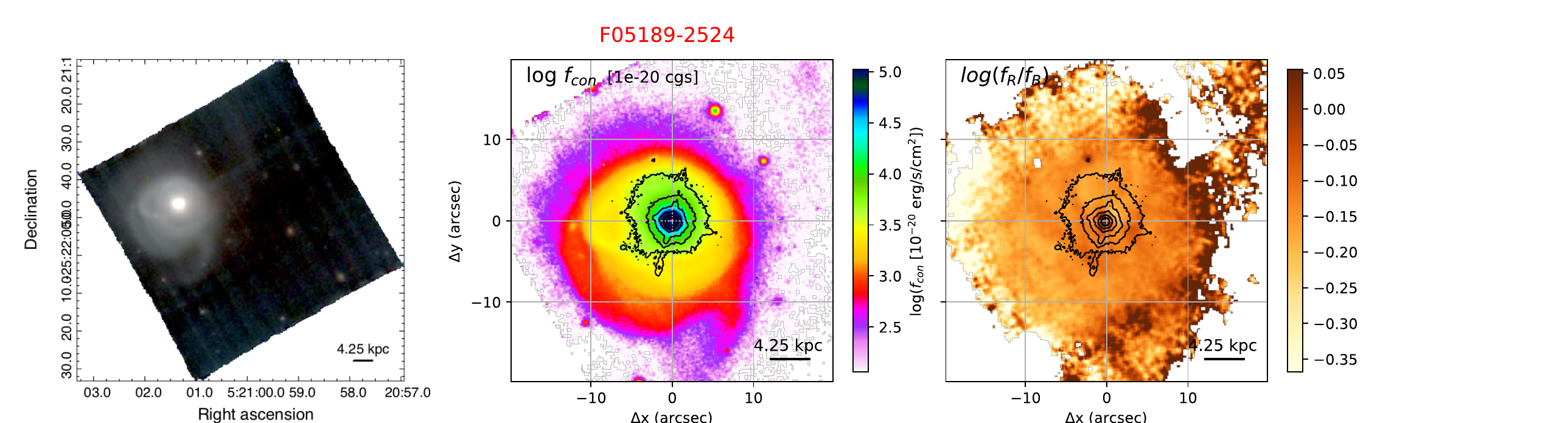}

\caption{\small IRAS F05189-2524 images from MUSE observations with TOT $= 2.04$ hr. {\it Left}: Colour-composite optical image, showing [O {\small III}] (green, from the wavelength range $4962-5022\AA$ rest-frame), \ha (red, $6554-6578\AA$), and continuum (blue, $4530-4660\AA$) emission. {\it Centre}: Red ($8025-8150\AA$) continuum image from MUSE with contours from HST/F160W. {\it Right}: Continuum colour map obtained from MUSE by dividing the red continuum image (central panel) by a blue image obtained by collapsing the emission in the range $4530-4660\AA$; contours from HST/F160W.
}
\label{IRAS05189_1}
\end{figure*}

\begin{figure*}[t]
\vspace{0.5cm}
\centering

\includegraphics[width=21.cm,trim= 90 197 0 218,clip]{{IRAS05189_Appendix}.pdf}

\caption{\small IRAS F05189-2524 emission line images from MUSE observations. [O {\small III}] (left, from the wavelength range $4962-5022\AA$ rest-frame), \ha (centre, $6554-6578\AA$), and [S {\small II}] (right, $6694-6742\AA$) images have been obtained by subtracting continuum emission using the adjacent regions at shorter and longer wavelengths with respect to the emission line systemics.}
\label{IRAS05189_2}
\end{figure*}

\begin{figure*}[t]
\vspace{0.5cm}
\centering

\includegraphics[width=13.cm,trim= 295 20 90 398,clip]{{IRAS05189_Appendix}.pdf}

\caption{\small {\it Left}: IRAS F05189-2524 stellar kinematic maps from the pPXF analysis with contours from HST/F160W. The left panel shows the stellar velocity $V_*$, and the right panel represents the velocity dispersion $\sigma_*$.}
\label{IRAS05189_3}
\end{figure*}

\begin{figure*}[t]
\vspace{0.5cm}

\centering
\includegraphics[width=18.cm,trim= 0 0 0 0,clip]{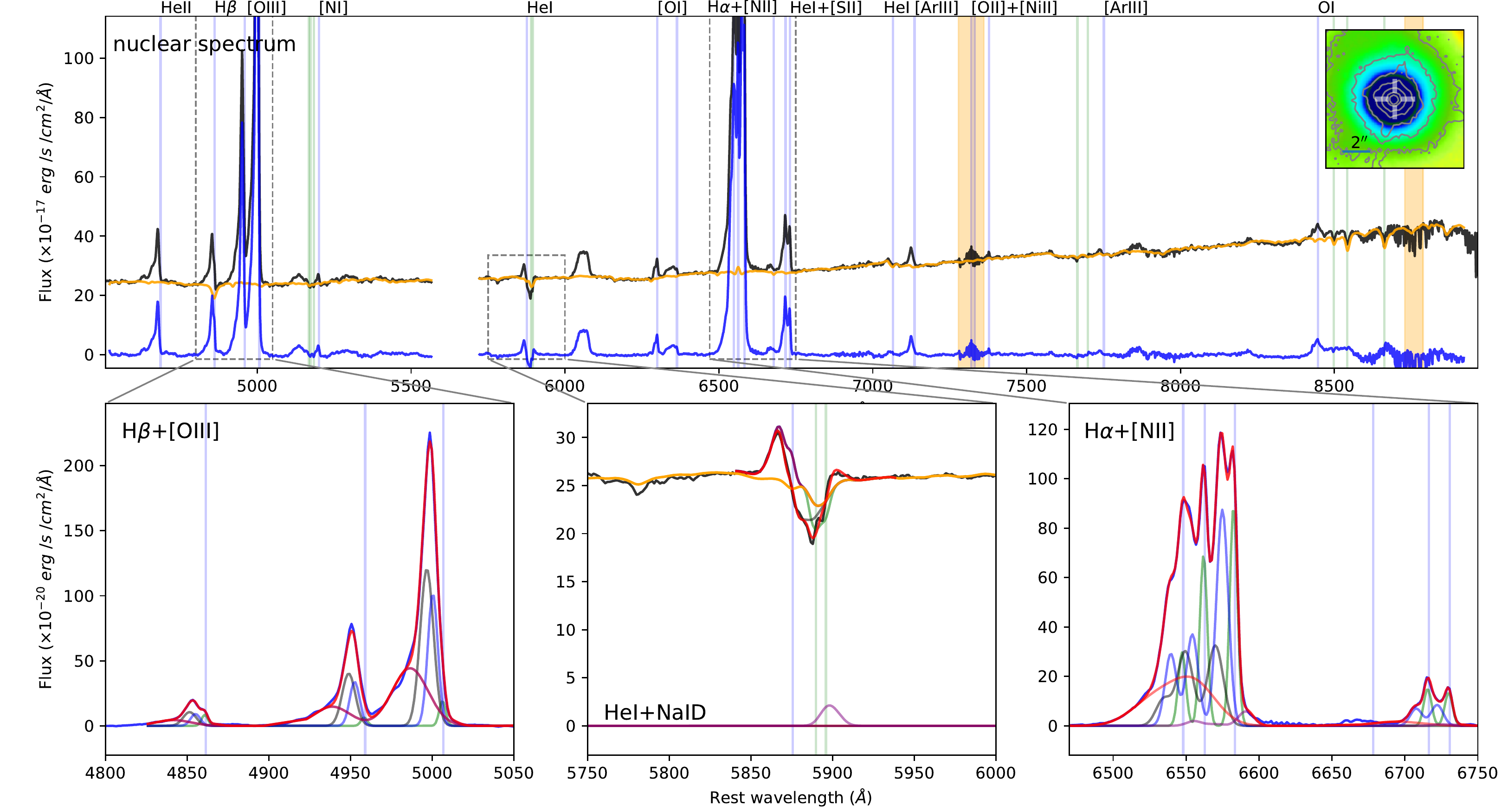}

\caption{\small {\it Top panel}: IRAS F05189-2524 nuclear spectrum (black curve) extracted from a circular aperture with $r< 0.4''$. The corresponding pPXF best-fit model profile is shown in orange. The continuum-subtracted spectrum (blue curve) is obtained by subtracting the best-fit pPXF model  from the original spectrum. The vertical blue lines mark the wavelengths of the emission lines detected in the spectrum, and the vertical green lines mark the position of stellar absorption systems (i.e. from left to right: MgI triplet, Na {\small ID} and KI doublets, CaII triplet). The regions excluded from the pPXF fit and corresponding to the most intense sky line residuals are highlighted as orange shaded areas; the portions of the spectra around 5700$\AA$ are missing because a filter blocked the laser contamination. The inset in the top panel shows the red stellar continuum emission map in the vicinity of the nuclear regions with  HST/F160W contours as in Fig. \ref{IRAS05189_1}; the nuclear position is shown with a white marker. {\it Bottom insets}: Multi-component best-fit analysis results for the main emission (H$\beta$ and [O {\small III}], left; H$\alpha$, [N {\small II}] and [S {\small II}], right) and absorption (Na {\small ID}, centre) line features. Na {\small ID} multi-component best-fit curves required to reproduce the cool, neutral absorption are superimposed on the observed spectrum (black curve); emission components are shifted downward for the sake of clarity. }
\label{IRAS05189_n}
\end{figure*}

\clearpage 

\begin{figure*}[t]
\centering
\includegraphics[width=21.cm,trim= 100 375 0 26,clip]{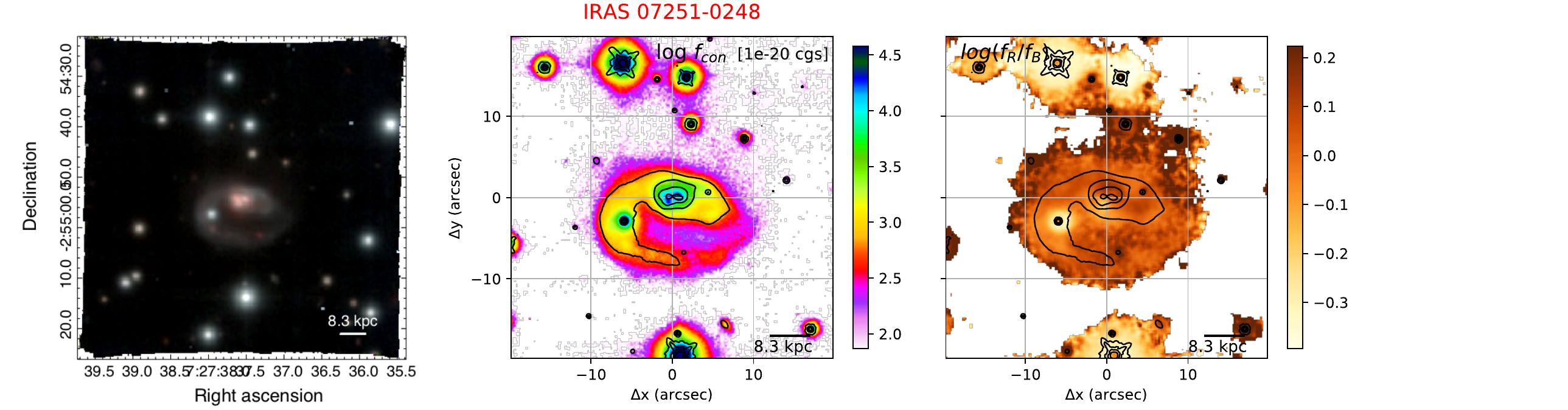}

\caption{\small IRAS 07251-0248 images from MUSE observations with TOT $= 2.55$ hr. {\it Left}: Colour-composite optical image showing [O {\small III}] (green, from the wavelength range $4995-5020\AA$ rest-frame), \ha (red, $6554-6575\AA$), and stellar continuum (blue, $4430-4540\AA$). {\it Centre}: Red  ($7440-7550\AA$) continuum image from MUSE with contours from HST/F160W. {\it Right}: Continuum colour map obtained from MUSE by dividing the red continuum image (central panel) from a blue image obtained by collapsing the stellar emission in the range $4430-4540\AA$; contours from HST/F160W.
}
\label{IRAS07251_1}
\end{figure*}

\begin{figure*}[t]
\vspace{0.5cm}
\centering

\includegraphics[width=21.cm,trim= 90 197 0 218,clip]{{IRAS07251_Appendix}.pdf}

\caption{\small IRAS 07251-0248 emission line images from MUSE observations. [O {\small III}] (left, from the wavelength range $4995-5020\AA$ rest-frame), \ha (centre, $6554-6575\AA$), and [S {\small II}] (right, $6704-6746\AA$) images have been obtained by subtracting continuum emission using the adjacent regions at shorter and longer wavelengths with respect to the emission line systemics.}
\label{IRAS07251_2}
\end{figure*}

\begin{figure*}[t]
\vspace{0.5cm}
\centering

\includegraphics[width=13.cm,trim= 295 20 90 398,clip]{{IRAS07251_Appendix}.pdf}

\caption{\small IRAS 07251-0248 stellar kinematic maps from the pPXF analysis with contours from HST/F160W. The left panel shows the stellar velocity $V_*$, and the right panel represents the velocity dispersion $\sigma_*$.}
\label{IRAS07251_3}
\end{figure*}

\begin{figure*}[t]
\vspace{0.5cm}

\centering
\includegraphics[width=18.cm,trim= 0 0 0 0,clip]{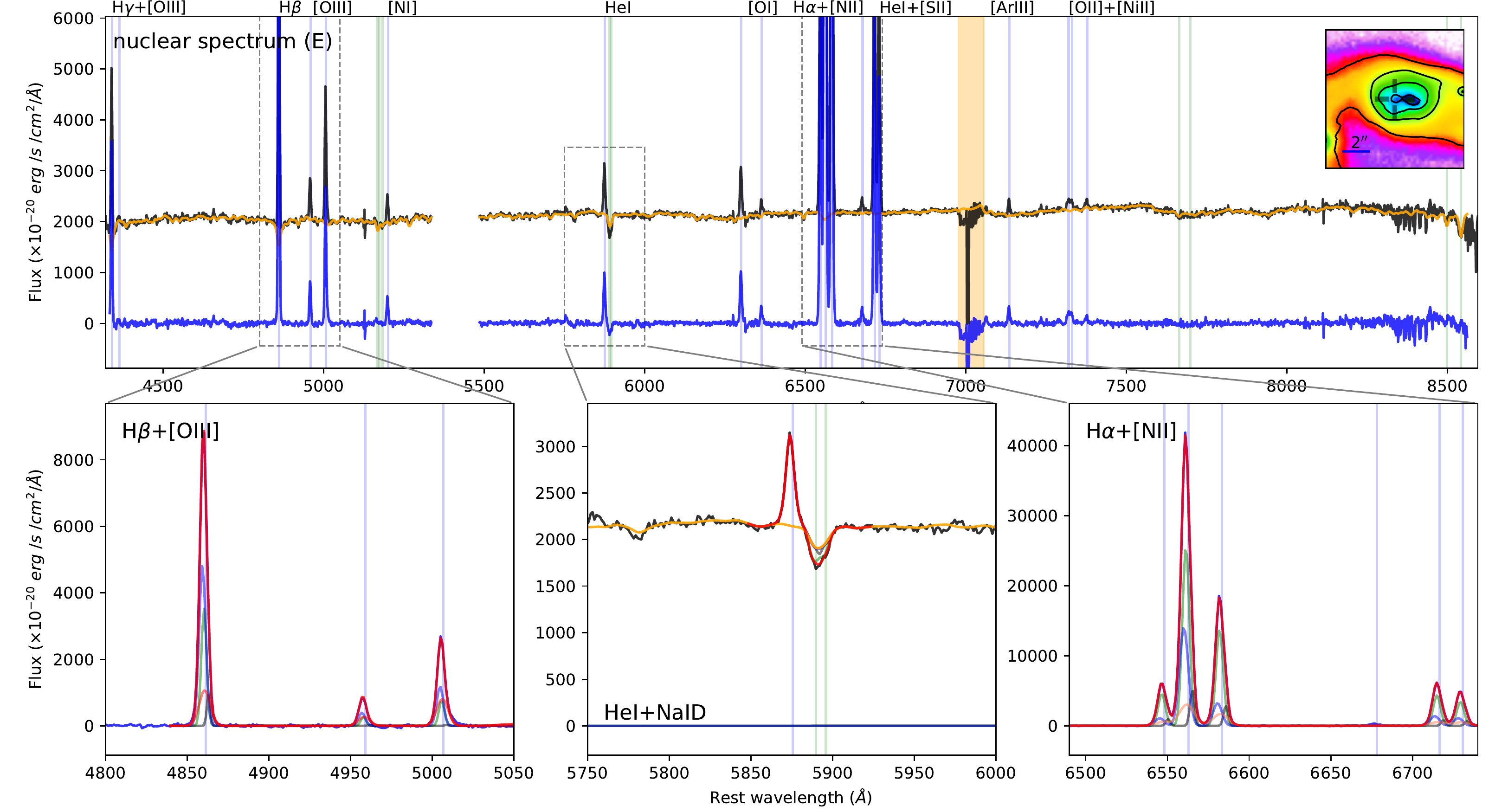}

\caption{\small IRAS 07251-0248 E nuclear spectrum extracted from a circular aperture with $r < 0.4''$, with the corresponding pPXF (top panel) and multi-component (bottom insets) best-fit models. See Fig. \ref{IRAS00188_n} for details. }
\label{IRAS07251_n_e}
\end{figure*}

\begin{figure*}[t]
\centering
\includegraphics[width=18.cm,trim= 0 0 0 0,clip]{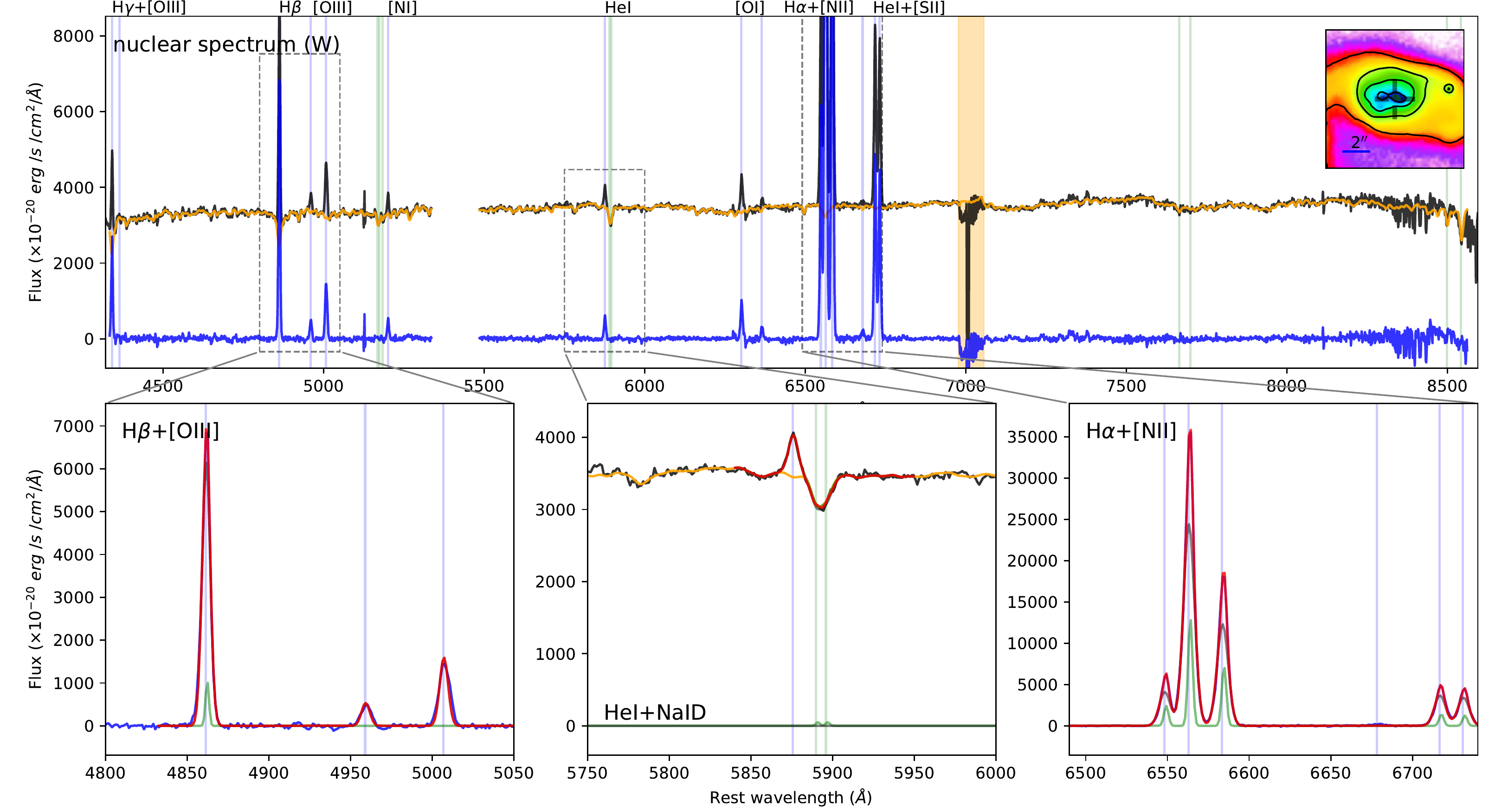}

\caption{\small IRAS 07251-0248 W nuclear spectrum extracted from a circular aperture with $r < 0.4''$, with the corresponding pPXF (top panel) and multi-component (bottom insets) best-fit models. See Fig. \ref{IRAS00188_n} for details.}
\label{IRAS07251_n_w}
\end{figure*}

\clearpage 

\begin{figure*}[t]
\centering
\includegraphics[width=21.cm,trim= 100 378 0 24,clip]{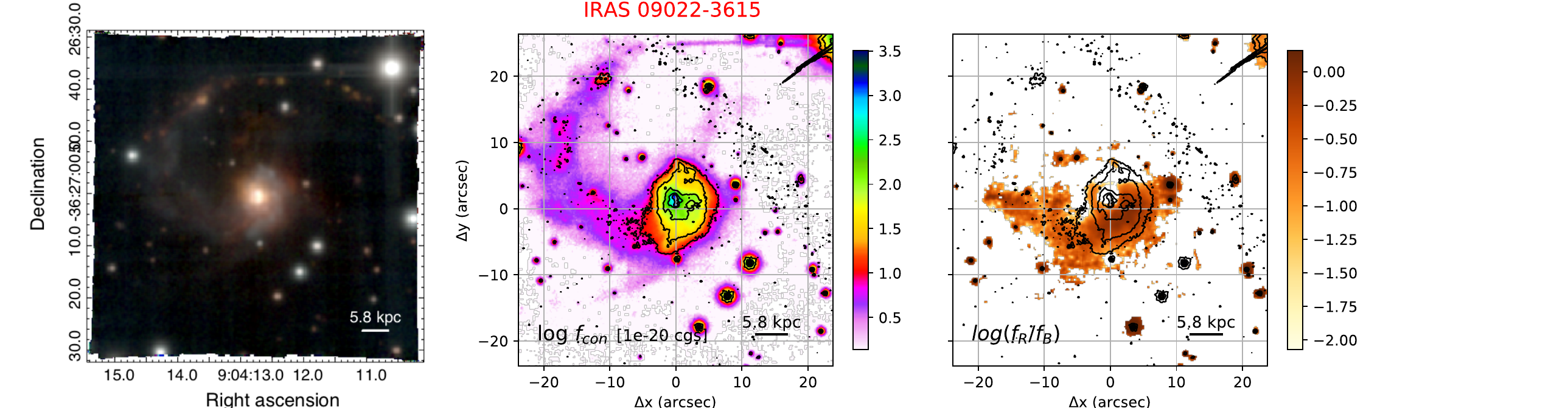}

\caption{\small IRAS 09022-3615 images from MUSE observations with TOT $= 2.04$ hr. {\it Left}: Colour-composite optical image showing [O {\small III}] (green, from the wavelength range $4995-5020 \AA$ rest-frame), \ha (red, $6552-6574 \AA$), and stellar continuum (blue, $4550-4710 \AA$). {\it Centre}: Red ($7600-7760 \AA$) stellar continuum image from MUSE with contours from HST/F814W. {\it Right}: Stellar continuum colour map obtained from MUSE by dividing the red continuum image (central panel) by a blue image obtained by collapsing the stellar emission in the range $4550-4710 \AA$; contours from HST/F814W.}
\label{IRAS09022_1}
\end{figure*}

\begin{figure*}[t]
\vspace{0.5cm}
\centering

\includegraphics[width=21.cm,trim= 90 199 0 218,clip]{{IRAS09022_Appendix}.pdf}

\caption{\small  IRAS 09022-3615 emission line images from MUSE observations. [O {\small III}] (left, from the wavelength range $4995-5020 \AA$ rest-frame), \ha (centre, $6552-6574 \AA$), and [S {\small II}] (right, $6700-6748 \AA$) images have been obtained subtracting continuum emission using the adjacent regions at shorter and longer wavelengths with respect to the emission line systemics.}
\label{IRAS09022_2}
\end{figure*}

\begin{figure*}[t]
\vspace{0.5cm}
\centering

\includegraphics[width=13.cm,trim= 297 10 90 396,clip]{{IRAS09022_Appendix}.pdf}

\caption{\small IRAS 09022-3615 stellar kinematic maps from the pPXF analysis with contours from HST/F814W. The left panel shows the stellar velocity $V_*$, and the right panel represents the velocity dispersion $\sigma_*$.}
\label{IRAS09022_3}
\end{figure*}

\begin{figure*}[t]
\vspace{0.5cm}

\centering
\includegraphics[width=18.cm,trim= 0 0 0 0,clip]{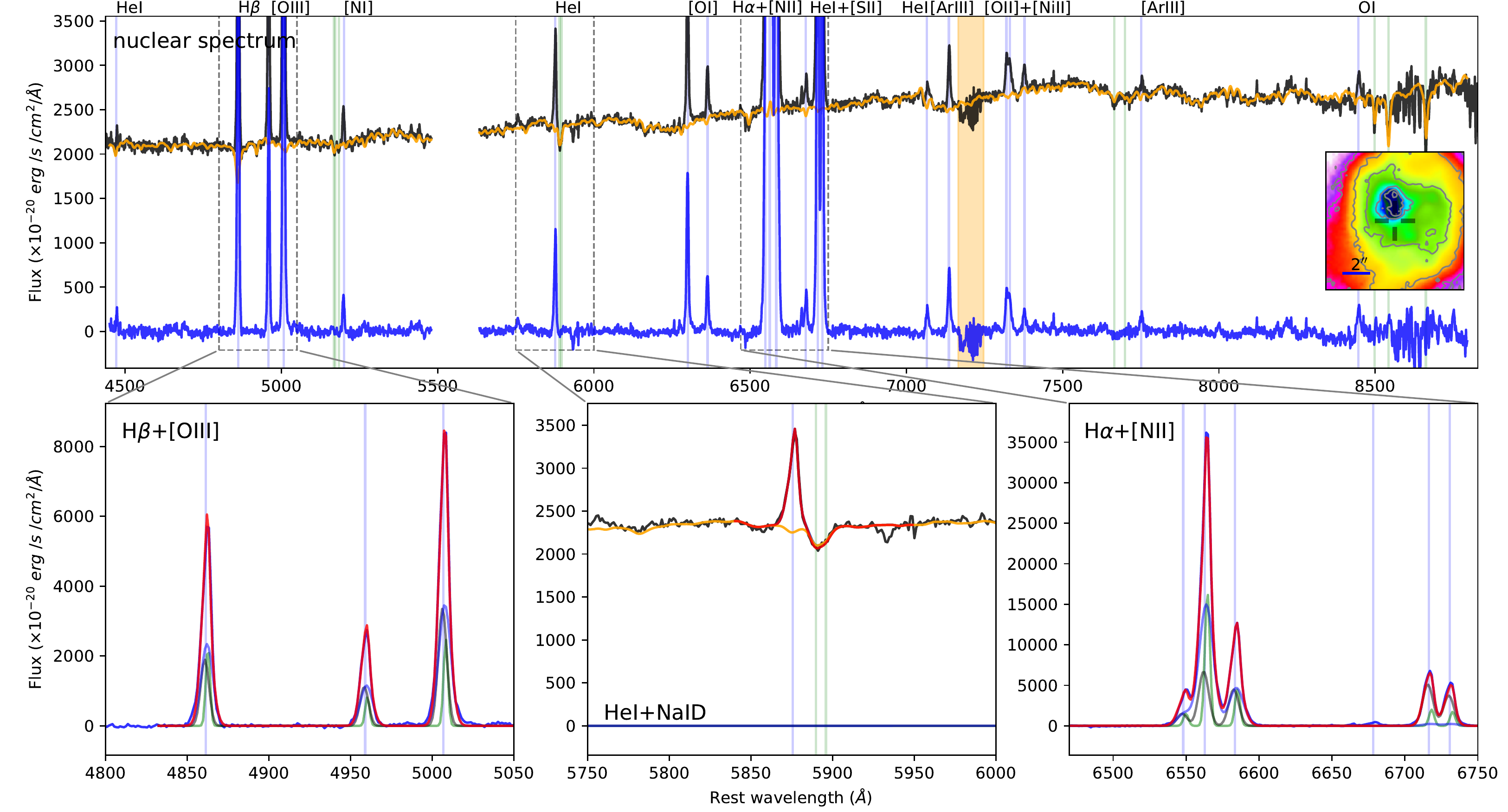}

\caption{\small IRAS 09022 nuclear spectrum extracted from a circular aperture with $r < 0.4''$, with the corresponding pPXF (top panel) and multi-component (bottom insets) best-fit models. See Fig. \ref{IRAS00188_n} for details. }
\label{IRAS09022_n_n}
\end{figure*}

\clearpage 

\begin{figure*}[t]
\centering
\includegraphics[width=21.cm,trim= 100 377 0 15,clip]{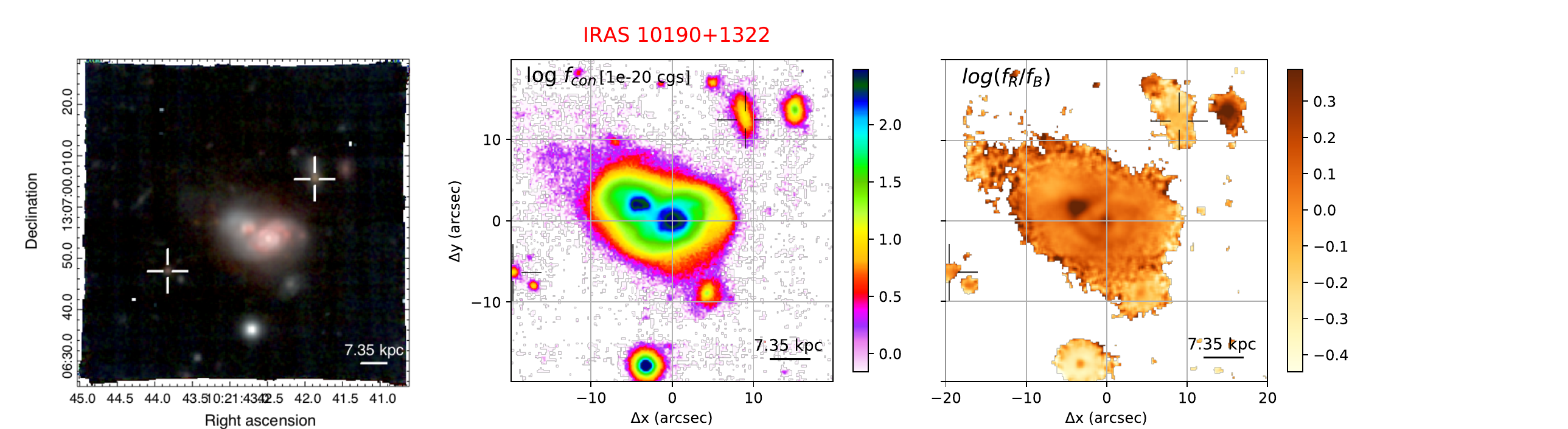}

\caption{\small IRAS 10190+1322 images from MUSE observations with TOT $= 2.04$ hr. {\it Left}: Colour-composite optical image showing [O {\small III}] (green, from the wavelength range $4988-5022 \AA$ rest-frame), \ha (red, $6556-6572 \AA$), and stellar continuum (blue, $4460-4570 \AA$). {\it Centre}: Red ($7440-7560 \AA$) stellar continuum image from MUSE with contours from HST/F160W. {\it Right}: Stellar continuum colour map obtained from MUSE by dividing the red continuum image (central panel) by a blue image obtained by collapsing the stellar emission in the range $4460-4570 \AA$. In all panels, we display the two companion galaxies of IRAS 10190+1322 with cross markers.}
\label{IRAS10190_1}
\end{figure*}

\begin{figure*}[t]
\vspace{0.5cm}
\centering

\includegraphics[width=21.cm,trim= 90 198 0 218,clip]{{IRAS10190_Appendix}.pdf}

\caption{\small IRAS 10190+1322  emission line images from MUSE observations. [O {\small III}] (left, from the wavelength range $4988-5022 \AA$ rest-frame), \ha (centre, $6556-6572 \AA$), and [S {\small II}] (right, $6704-6747 \AA$) images have been obtained by subtracting continuum emission using the adjacent regions at shorter and longer wavelengths with respect to the emission line systemics.}
\label{IRAS10190_2}
\end{figure*}

\begin{figure*}[t]
\vspace{0.5cm}
\centering

\includegraphics[width=13.cm,trim= 295 10 90 395,clip]{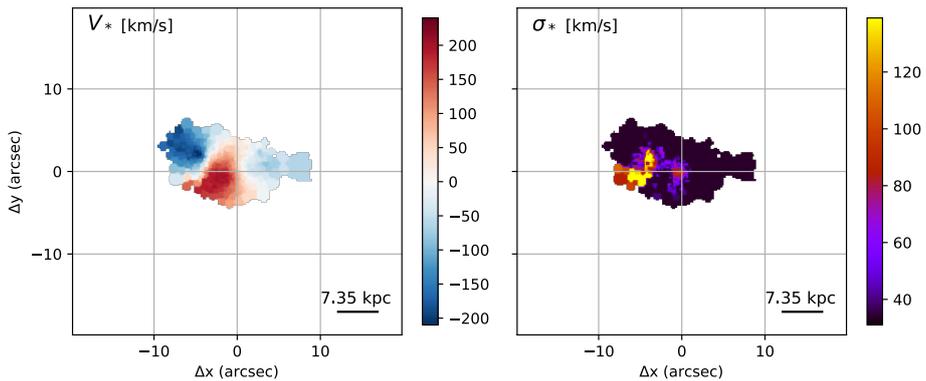}

\caption{\small IRAS 10190+1322 stellar kinematic maps from the pPXF analysis. The left panel shows the stellar velocity $V_*$, and the right panel represents the velocity dispersion $\sigma_*$.}
\label{IRAS10190_3}
\end{figure*}

\begin{figure*}[t]
\vspace{0.5cm}

\centering
\includegraphics[width=18.cm,trim= 0 0 0 0,clip]{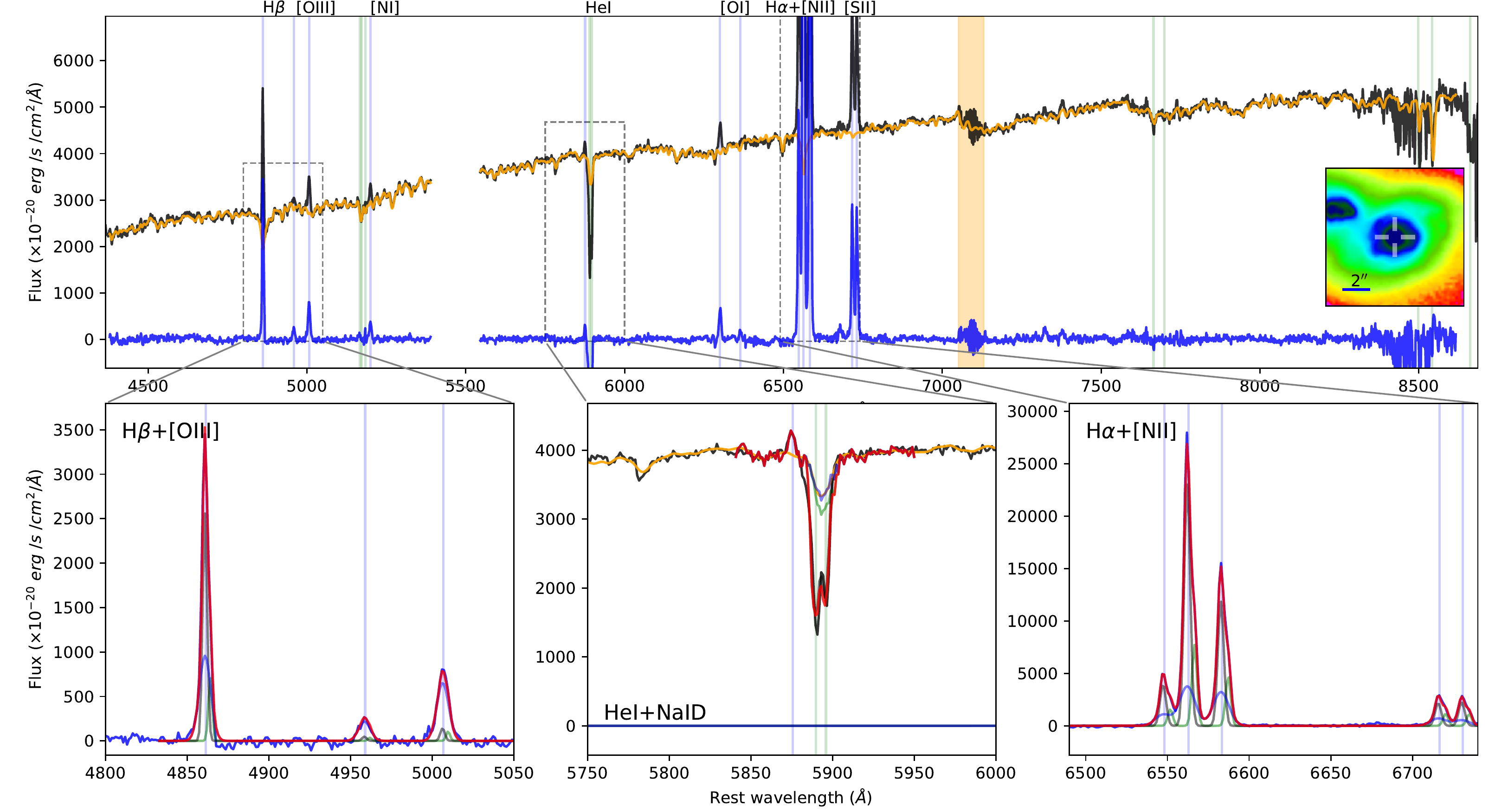}

\caption{\small IRAS 10190+1322 W nuclear spectrum extracted from a circular aperture with $r < 0.4''$, with the corresponding pPXF (top panel) and multi-component (bottom insets) best-fit models. See Fig. \ref{IRAS00188_n} for details.  }
\label{IRAS10190_n_w}
\end{figure*}

\begin{figure*}[t]
\centering
\includegraphics[width=18.cm,trim= 0 0 0 0,clip]{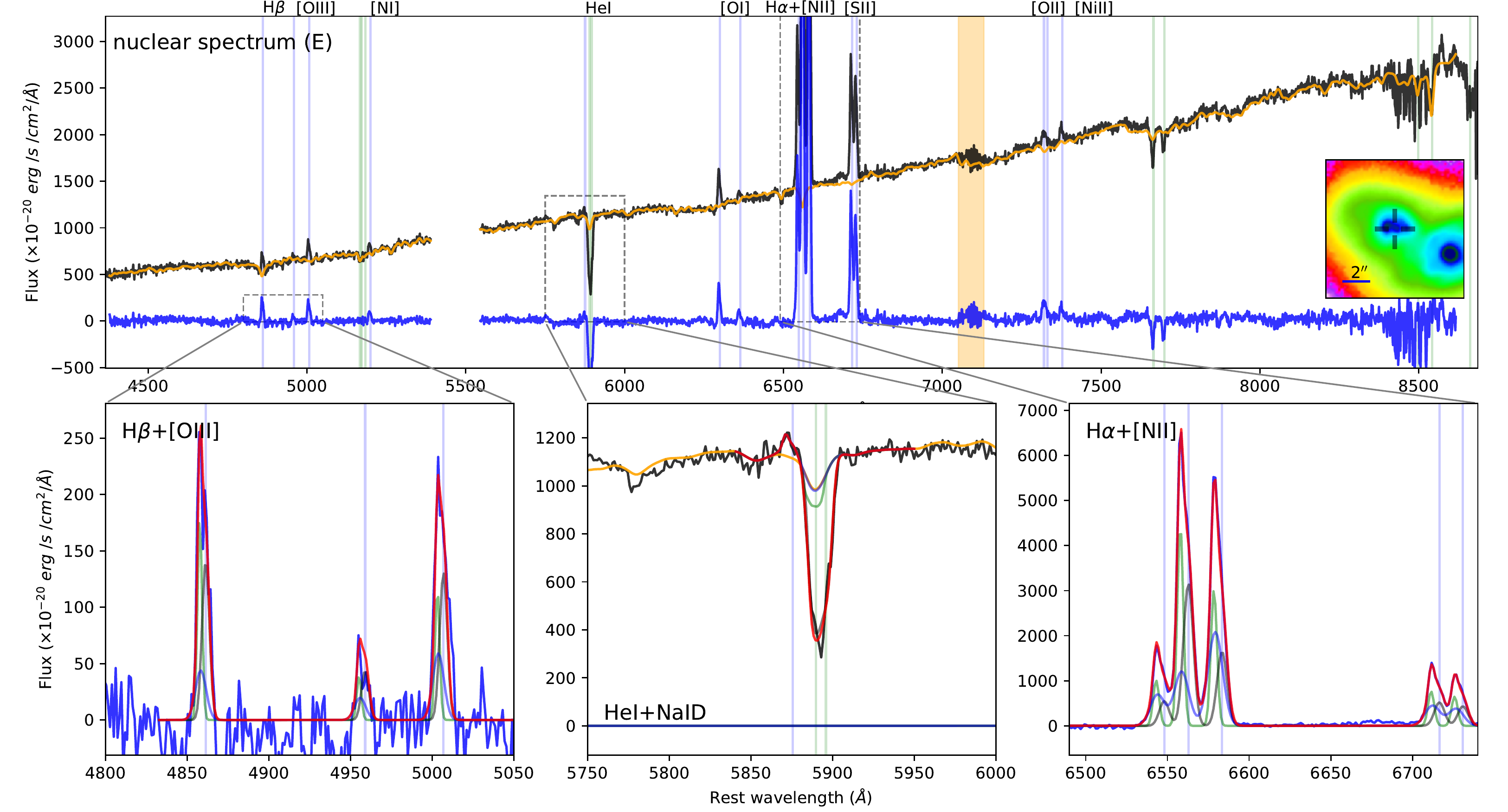}

\caption{\small IRAS 10190+1322 E nuclear spectrum extracted from a circular aperture with $r < 0.4''$, with the corresponding pPXF (top panel) and multi-component (bottom insets) best-fit models. See Fig. \ref{IRAS00188_n} for details. }
\label{IRAS10190_n_e}
\end{figure*}

\clearpage 

\begin{figure*}[t]
\centering
\includegraphics[width=21.cm,trim= 100 380 0 15,clip]{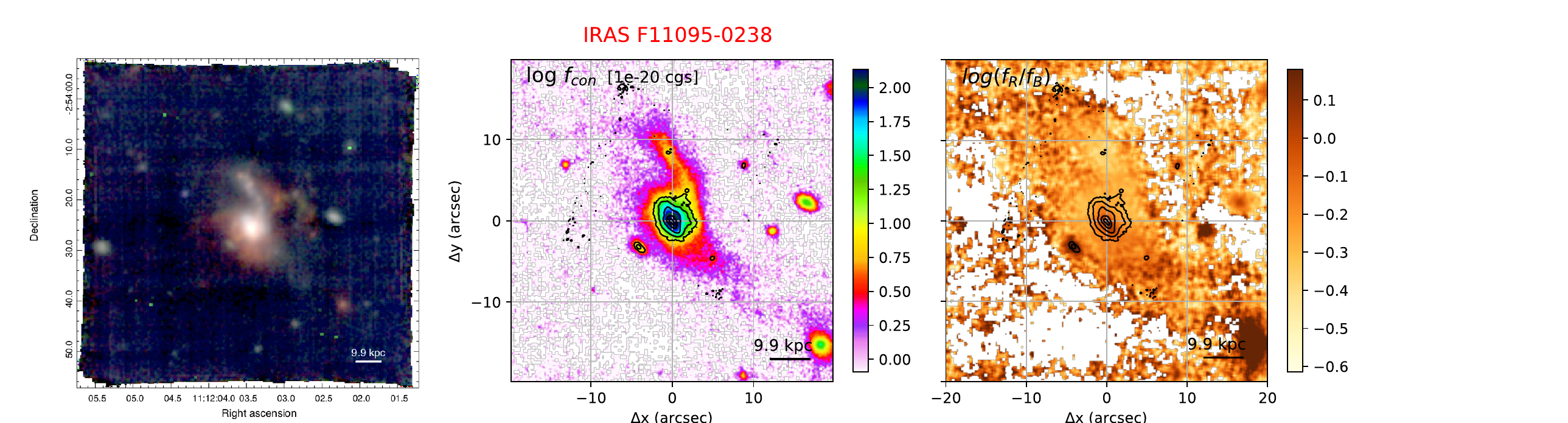}

\caption{\small IRAS F11095-0238 images from MUSE observations with TOT $= 2.72$ hr. {\it Left}: Colour-composite optical image showing [O {\small III}] (green, from the wavelength range $4975-5032 \AA$ rest-frame), \ha (red, $6549-6576 \AA$), and continuum (blue, $4410-4490 \AA$). {\it Centre}: Red ($7690-7760 \AA$) continuum image from MUSE, with contours from HST/F160W. {\it Right}: Continuum colour map obtained from MUSE by dividing the red continuum image (central panel) by a blue image obtained by collapsing the stellar emission in the range $4410-4490\AA$; contours from HST/F160W.}
\label{IRAS11095_1}
\end{figure*}

\begin{figure*}[t]
\vspace{0.5cm}
\centering

\includegraphics[width=21.cm,trim= 90 198 0 218,clip]{{IRAS11095_Appendix}.pdf}

\caption{\small IRAS F11095-0238 emission line images from MUSE observations. [O {\small III}] (left, from the wavelength range $4975-5032 \AA$ rest-frame), \ha (centre, $6549-6576 \AA$), and [S {\small II}] (right, $6695-6749 \AA$) images have been obtained by subtracting continuum emission using the adjacent regions at shorter and longer wavelengths with respect to the emission line systemics.}
\label{IRAS11095_2}
\end{figure*}

\begin{figure*}[t]
\vspace{0.5cm}
\centering

\includegraphics[width=13.cm,trim= 299 20 90 395,clip]{{IRAS11095_Appendix}.pdf}

\caption{\small IRAS F11095-0238 stellar kinematic maps from the pPXF analysis with contours from HST/F160W. The left panel shows the stellar velocity $V_*$, and the right panel represents the velocity dispersion $\sigma_*$.}
\label{IRAS11095_3}
\end{figure*}

\begin{figure*}[t]
\vspace{0.5cm}

\centering
\includegraphics[width=18.cm,trim= 0 0 0 0,clip]{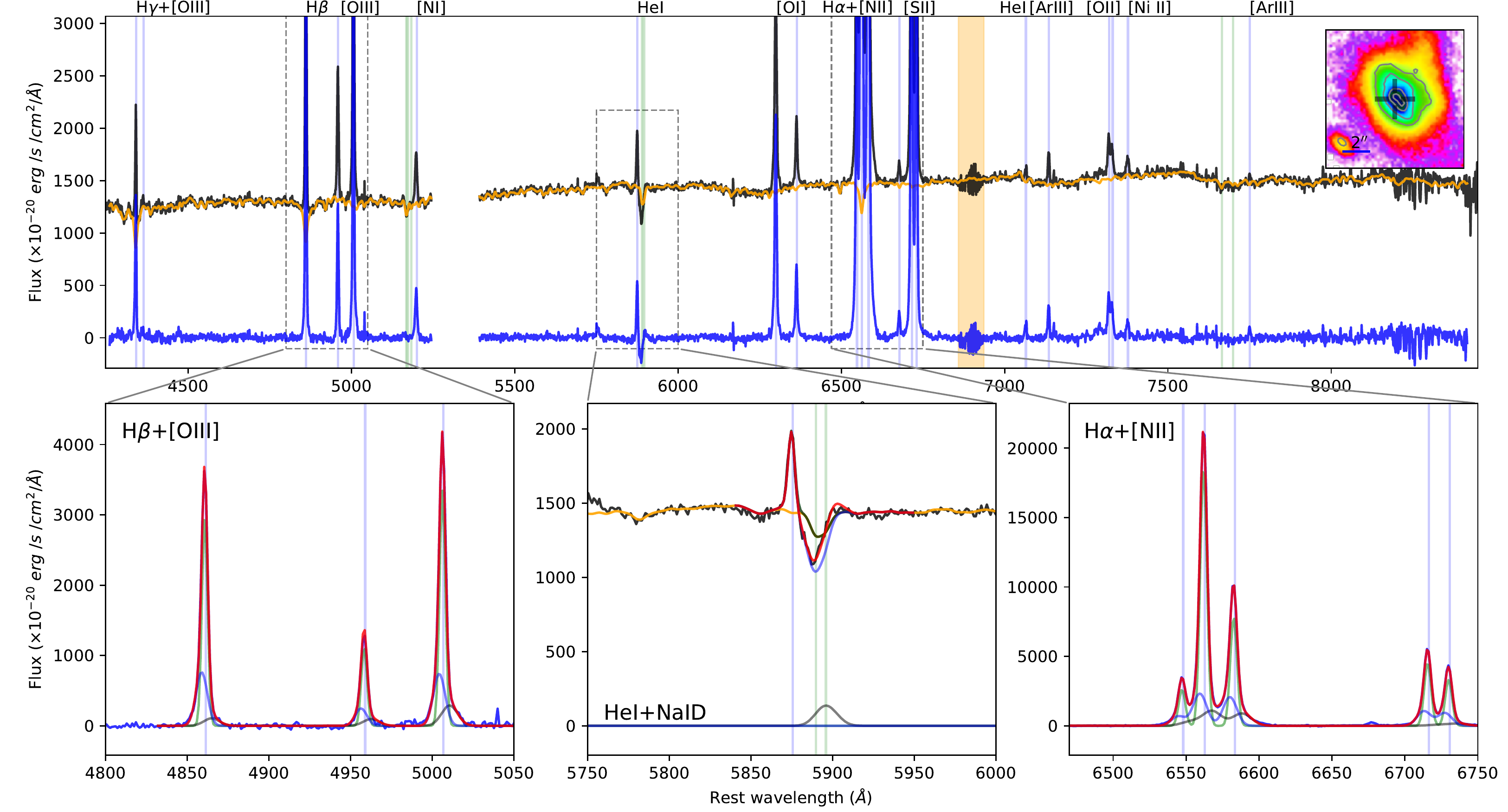}

\caption{\small IRAS F11095-0238 NE nuclear spectrum extracted from a circular aperture with $r < 0.4''$, with the corresponding pPXF (top panel) and multi-component (bottom insets) best-fit models. See Fig. \ref{IRAS00188_n} for details. }
\label{IRAS11095_n_n}
\end{figure*}

\begin{figure*}[t]
\centering
\includegraphics[width=18.cm,trim= 0 0 0 0,clip]{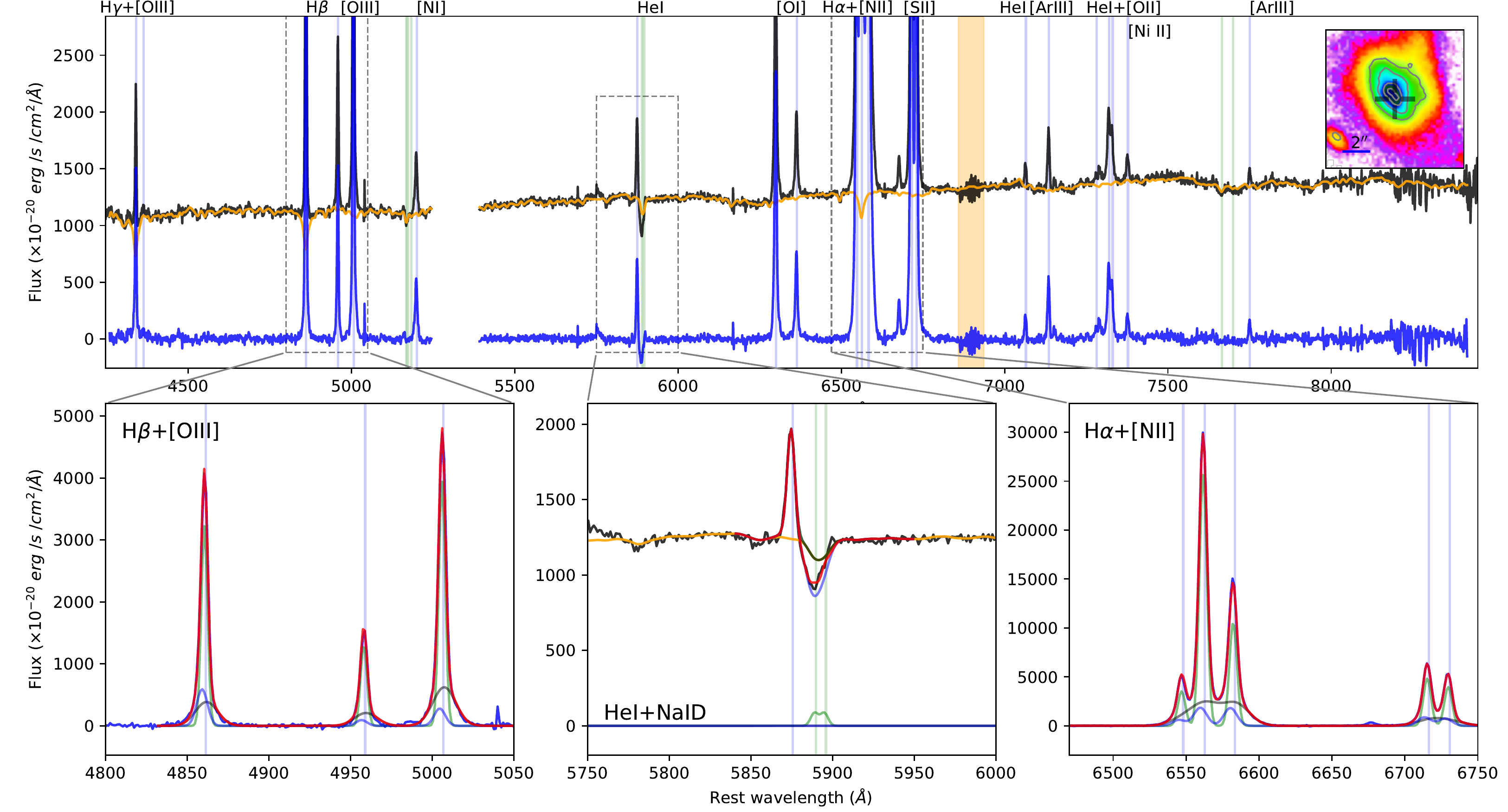}

\caption{\small IRAS F11095-0238 SW nuclear spectrum extracted from a circular aperture with $r < 0.4''$, with the corresponding pPXF (top panel) and multi-component (bottom insets) best-fit models. See Fig. \ref{IRAS00188_n} for details. }
\label{IRAS11095_n_s}
\end{figure*}

\clearpage 

\begin{figure*}[t]
\centering
\includegraphics[width=21.cm,trim= 100 380 0 19,clip]{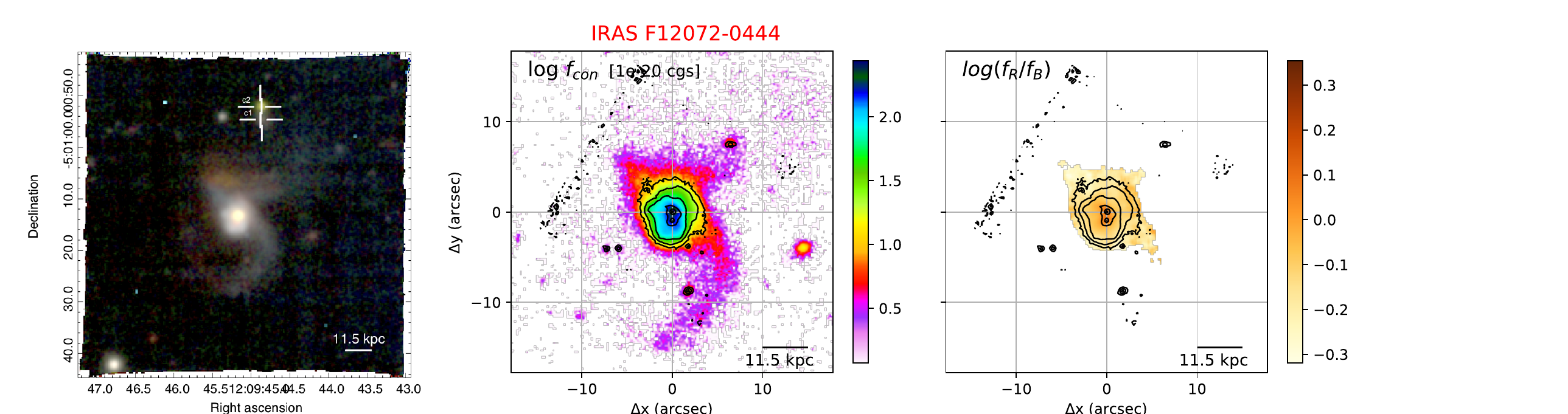}

\caption{\small IRAS F12072-0444 images from MUSE observations with TOT $= 0.68$ hr. {\it Left}: Colour-composite optical image showing [O {\small III}] (green, from the wavelength range $4980-5020\AA$ rest-frame), \ha (red, $6553-6576\AA$), and  continuum (blue, $4550-4650\AA$). {\it Centre}: Red ($7540-7610\AA$) continuum image from MUSE, with contours from HST/F160W. {\it Right}:  Continuum colour map obtained from MUSE by dividing the red continuum image (central panel) by a blue image obtained by collapsing the stellar emission in the range $4550-4650\AA$; contours from HST/F160W. In the first panel, we display the IRAS F12072-0444 companion galaxies with crosses.}
\label{IRAS12072_1}
\end{figure*}

\begin{figure*}[t]
\vspace{0.5cm}
\centering

\includegraphics[width=21.cm,trim= 90 197 0 218,clip]{{IRAS12072_Appendix}.pdf}

\caption{\small IRAS F12072-0444 emission line images from MUSE observations. [O {\small III}] (left, from the wavelength range $4980-5020\AA$ rest-frame), \ha (centre, $6553-6576\AA$), and [S {\small II}] (right, $6695-6747\AA$) images have been obtained by subtracting continuum emission using the adjacent regions at shorter and longer wavelengths with respect to the emission line systemics.}
\label{IRAS12072_2}
\end{figure*}

\begin{figure*}[t]
\vspace{0.5cm}
\centering

\includegraphics[width=13.cm,trim= 299 10 90 395,clip]{{IRAS12072_Appendix}.pdf}

\caption{\small IRAS F12072-0444 stellar kinematic maps from the pPXF analysis with contours from HST/F160W. The left panel shows the stellar velocity $V_*$, and the right panel represents the velocity dispersion $\sigma_*$.}
\label{IRAS12072_3}
\end{figure*}

\begin{figure*}[t]
\vspace{0.5cm}

\centering
\includegraphics[width=18.cm,trim= 0 0 0 0,clip]{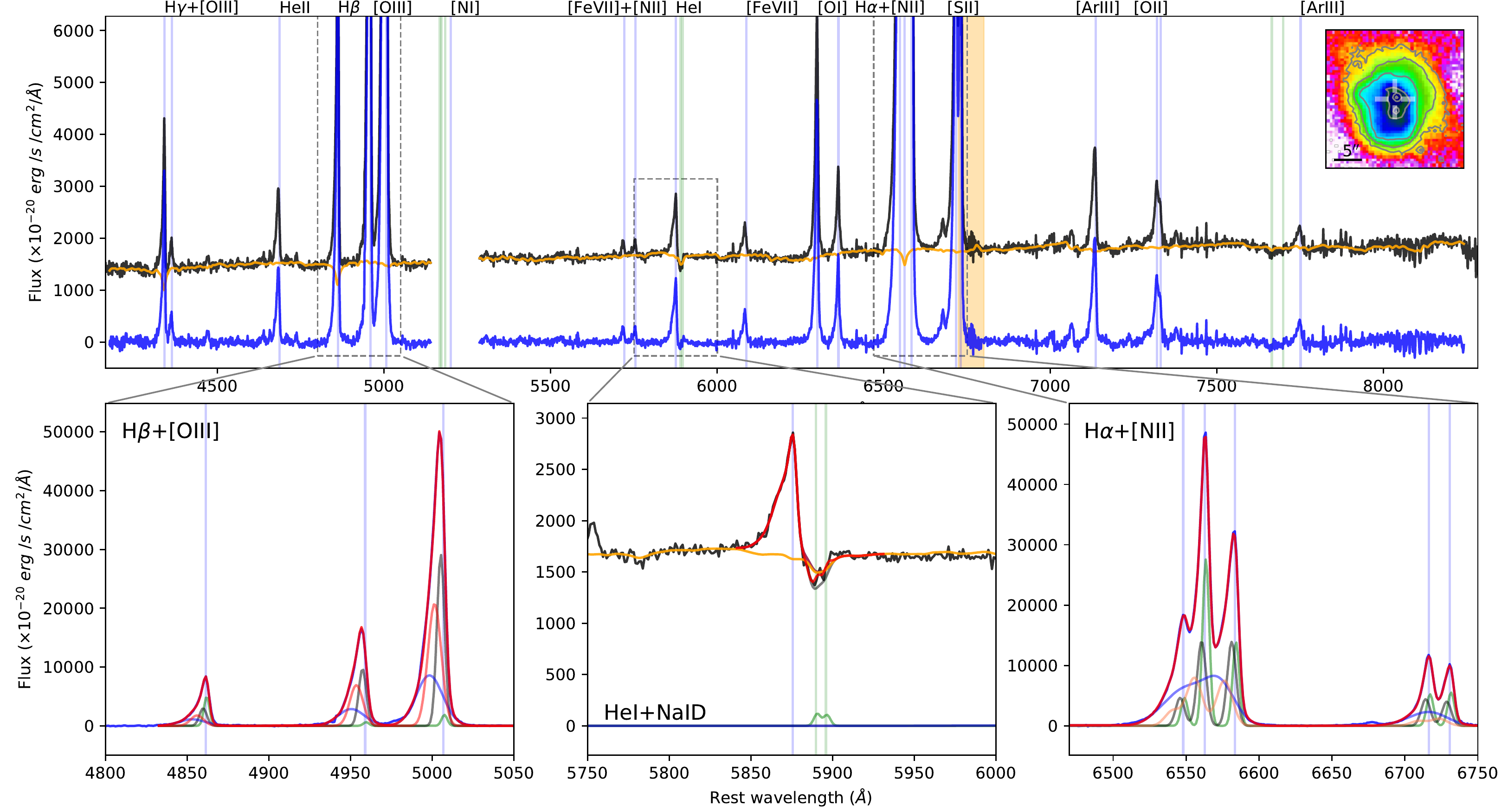}

\caption{\small  IRAS F12072-0444 NE nuclear spectrum extracted from a circular aperture with $r < 0.4''$, with the corresponding pPXF (top panel) and multi-component (bottom insets) best-fit models. See Fig. \ref{IRAS00188_n} for details.}
\label{IRAS12072_n_n}
\end{figure*}

\begin{figure*}[t]
\centering
\includegraphics[width=18.cm,trim= 0 0 0 0,clip]{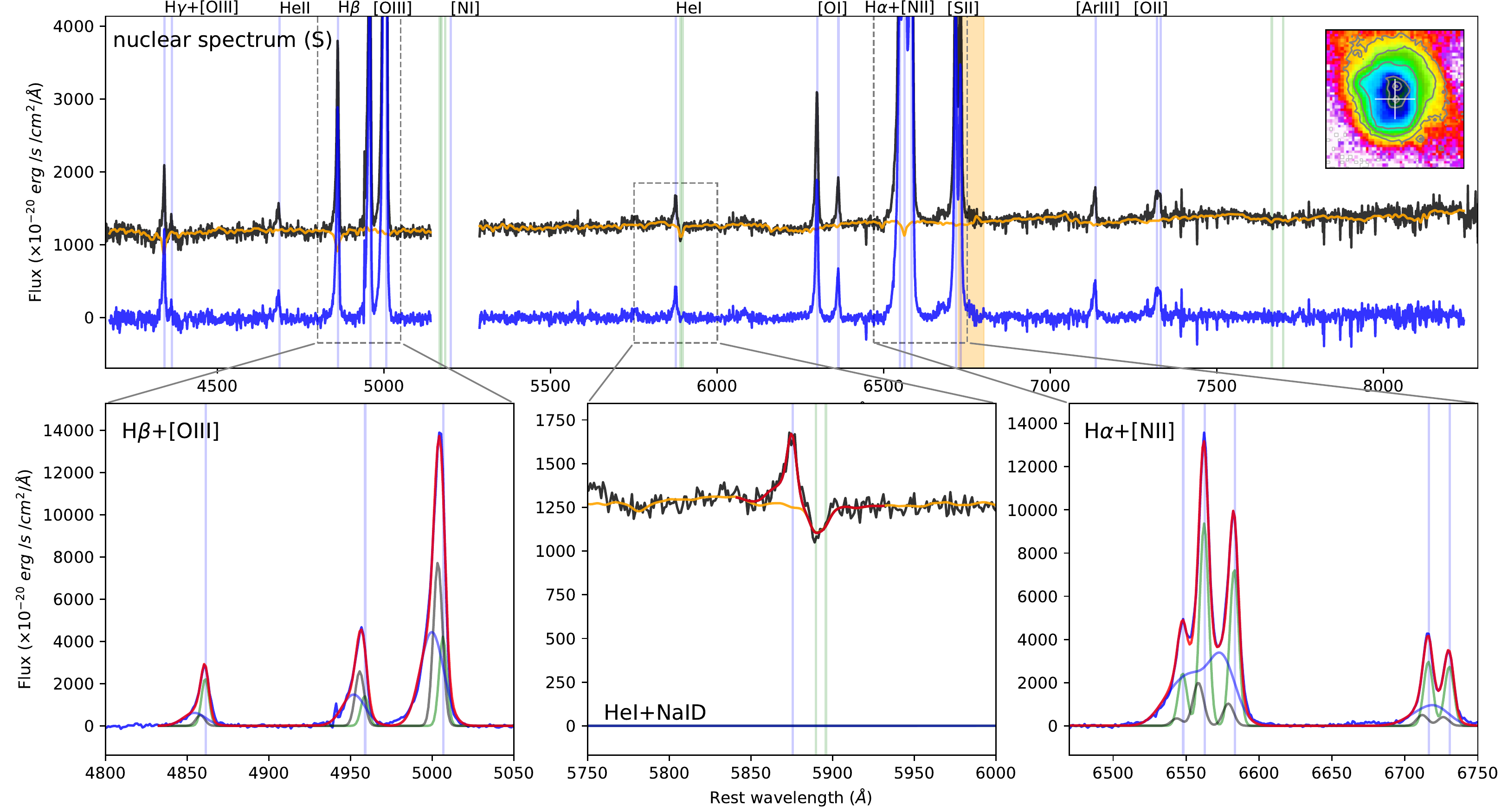}

\caption{\small IRAS F12072-0444 SW nuclear spectrum extracted from a circular aperture with $r < 0.4''$, with the corresponding pPXF (top panel) and multi-component (bottom insets) best-fit models. See Fig. \ref{IRAS00188_n} for details. }
\label{IRAS12072_n_s}
\end{figure*}

\clearpage 

\begin{figure*}[t]
\centering
\includegraphics[width=21.cm,trim= 100 380 0 15,clip]{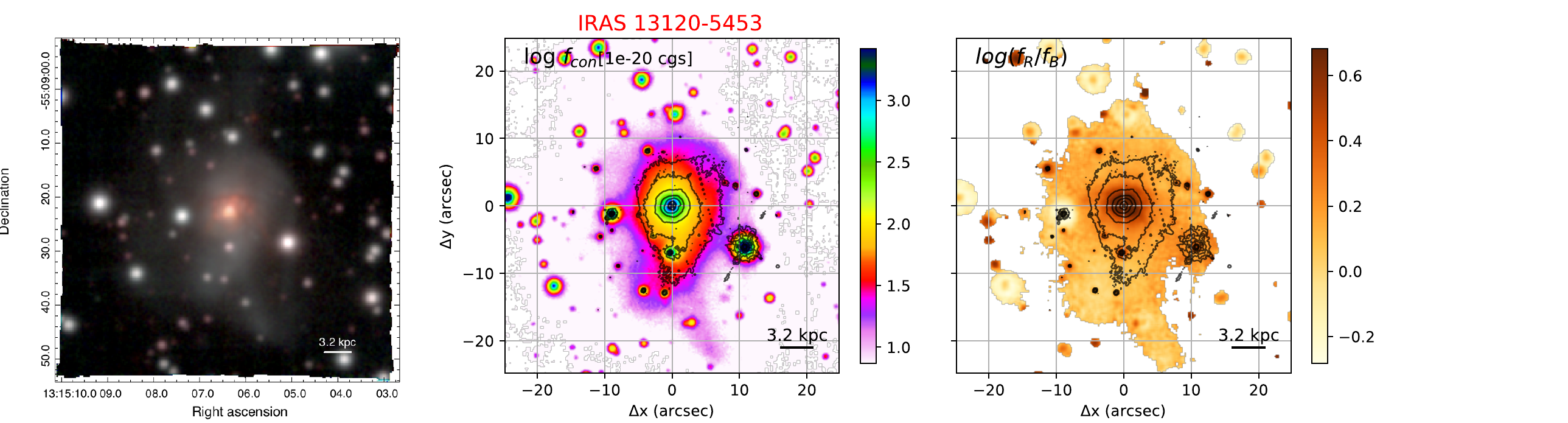}

\caption{\small IRAS 13120-5453 images from MUSE observations with TOT $= 0.52$ hr. {\it Left}: Colour-composite optical image showing [O {\small III}] (green, from the wavelength range $4980-5014\AA$ rest-frame), \ha (red, $6555-6572\AA$), and continuum (blue, $4600-4700\AA$). {\it Centre}: red ($7500-7600\AA$) continuum image from MUSE with contours from HST/F160W. {\it Right}:  Continuum colour map obtained from MUSE by dividing the red continuum image (central panel) by a blue image obtained by collapsing the stellar emission in the range $4600-4700\AA$; contours from HST/F160W. }
\label{IRAS13120_1}
\end{figure*}

\begin{figure*}[t]
\vspace{0.5cm}
\centering

\includegraphics[width=21.cm,trim= 90 197 0 216,clip]{{IRAS13120_Appendix}.pdf}

\caption{\small IRAS 13120-5453 emission line images from MUSE observations. [O {\small III}] (left, from the wavelength range $4980-5014\AA$ rest-frame), \ha (centre, $6555-6572\AA$), and [S {\small II}] (right, $6694-6746\AA$) images have been obtained by subtracting continuum emission using the adjacent regions at shorter and longer wavelengths with respect to the emission line systemics.  In all panels, we display the position of the  companion galaxy with a cross.}
\label{IRAS13120_2}
\end{figure*}

\begin{figure*}[t]
\vspace{0.5cm}
\centering

\includegraphics[width=13.cm,trim= 299 10 90 399,clip]{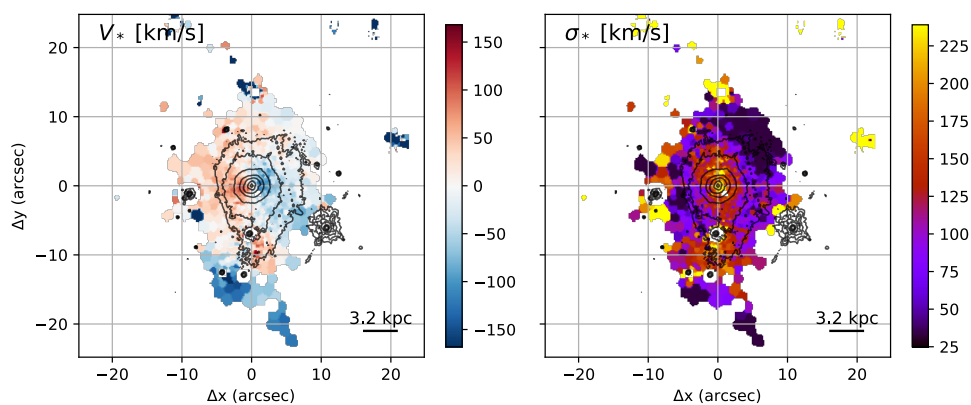}

\caption{\small IRAS 13120-5453 stellar kinematic maps from the pPXF analysis with contours from HST/F160W. The left panel shows the stellar velocity $V_*$, and the right panel represents the velocity dispersion $\sigma_*$. The inner 3.5 kpc may indicate a low-amplitude rotational pattern; he outside regions contain more irregular kinematics.}
\label{IRAS13120_3}
\end{figure*}

\begin{figure*}[t]
\centering
\includegraphics[width=18.cm,trim= 0 0 0 0,clip]{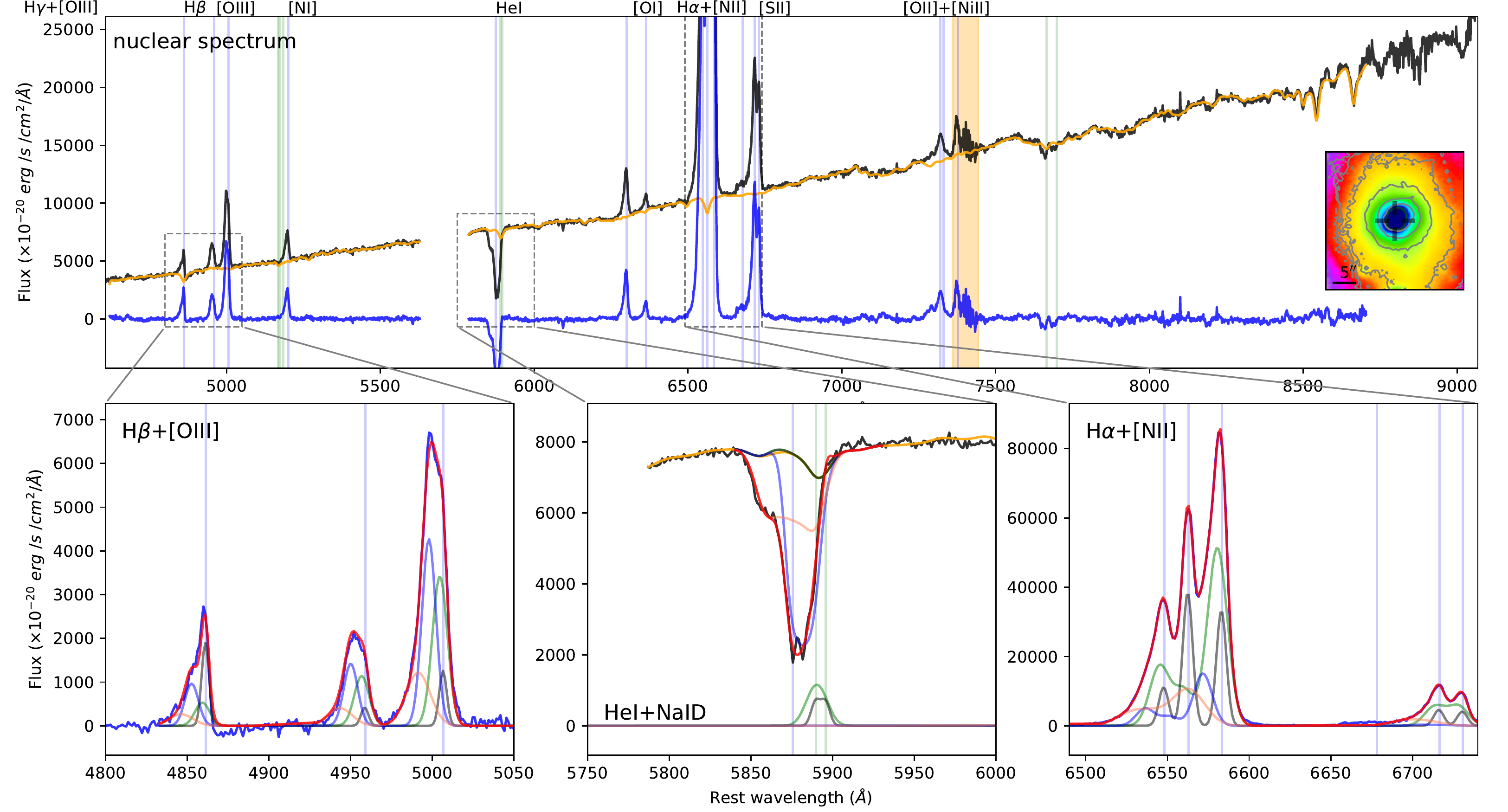}

\caption{\small IRAS 13120-5453 nuclear spectrum extracted from a circular aperture with $r < 0.4''$, with the corresponding pPXF (top panel) and multi-component (bottom insets) best-fit models. See Fig. \ref{IRAS00188_n} for details. }
\label{IRAS13120_n}
\end{figure*}

\clearpage 

\begin{figure*}[t]
\centering
\includegraphics[width=21.cm,trim= 100 375 0 15,clip]{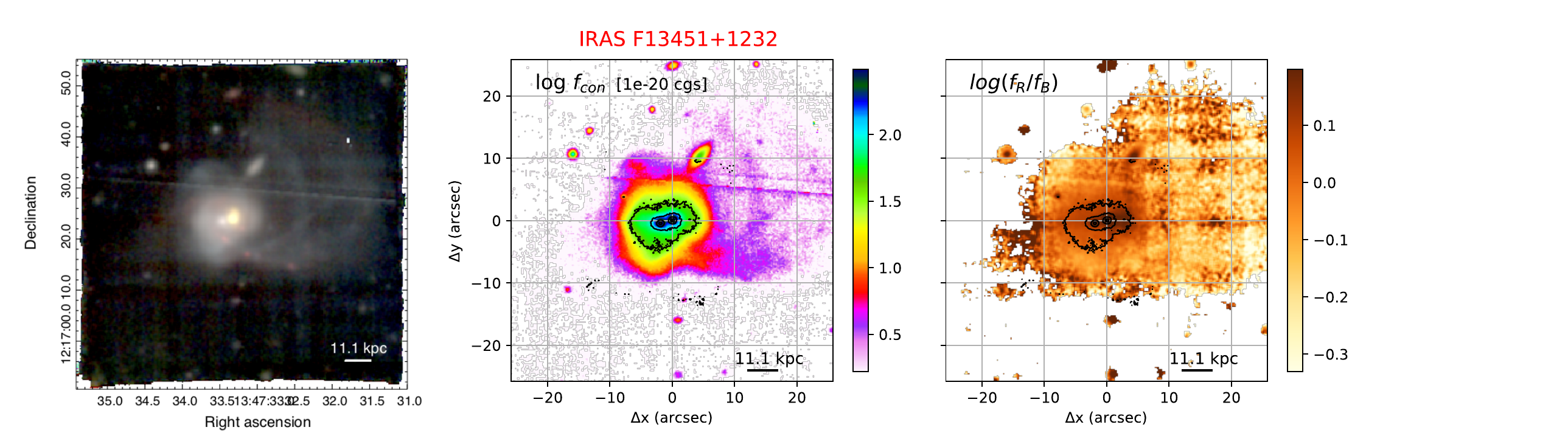}

\caption{\small IRAS F13451+1232 images from MUSE observations with TOT $= 2.04$ hr. {\it Left}: Colour-composite optical image, showing [O {\small III}] (green, from the wavelength range $4975-5047\AA$ rest-frame), \ha (red, $6554-6576\AA$), and stellar continuum (blue, $4410-4490\AA$). {\it Centre}: Red ($7590-7665\AA$) continuum image from MUSE with contours from HST/F160W. {\it Right}:  Continuum colour map obtained from MUSE by dividing the red continuum image (central panel) by a blue image obtained by collapsing the stellar emission in the range $4410-4490\AA$; contours from HST/F160W.
In all panels, the top part of the ULIRG shows an almost horizontal feature that is due to the removal of a satellite trail.}
\label{IRAS13451_1}
\end{figure*}

\begin{figure*}[t]
\vspace{0.5cm}
\centering

\includegraphics[width=21.cm,trim= 90 197 0 218,clip]{{IRAS13451_Appendix}.pdf}

\caption{\small IRAS F13451+1232 emission line images from MUSE observations. [O {\small III}] (left, from the wavelength range $4975-5047\AA$ rest-frame), \ha (centre, $6554-6576\AA$), and [S {\small II}] (right, $6662-6757\AA$) images have been obtained by subtracting continuum emission using the adjacent regions at shorter and longer wavelengths with respect to the emission line systemics. }
\label{IRAS13451_2}
\end{figure*}

\begin{figure*}[t]
\vspace{0.5cm}
\centering

\includegraphics[width=13.cm,trim= 299 10 90 395,clip]{{IRAS13451_Appendix}.pdf}

\caption{\small IRAS F13451+1232 stellar kinematic maps from the pPXF analysis with contours from HST/F160W. The left panel shows the stellar velocity $V_*$, and the right panel represents the velocity dispersion $\sigma_*$.}
\label{IRAS13451_3}
\end{figure*}

\begin{figure*}[t]
\vspace{0.5cm}

\centering
\includegraphics[width=18.cm,trim= 0 0 0 0,clip]{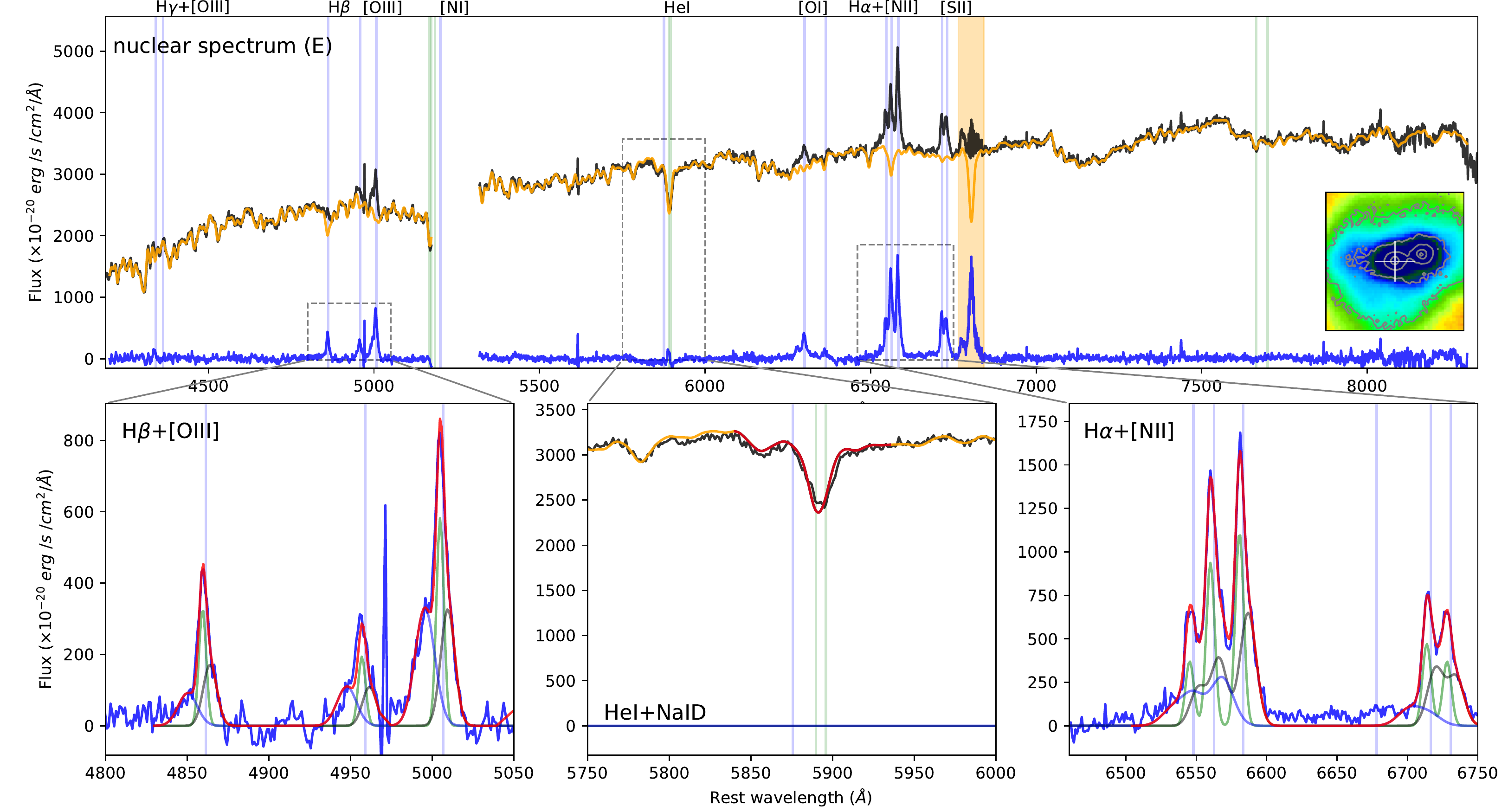}

\caption{\small IRAS F13451+1232 E nuclear spectrum extracted from a circular aperture with $r < 0.4''$, with the corresponding pPXF (top panel) and multi-component (bottom insets) best-fit models. See Fig. \ref{IRAS00188_n} for details. }
\label{IRAS13451_n_e}
\end{figure*}

\begin{figure*}[t]
\centering
\includegraphics[width=18.cm,trim= 0 0 0 0,clip]{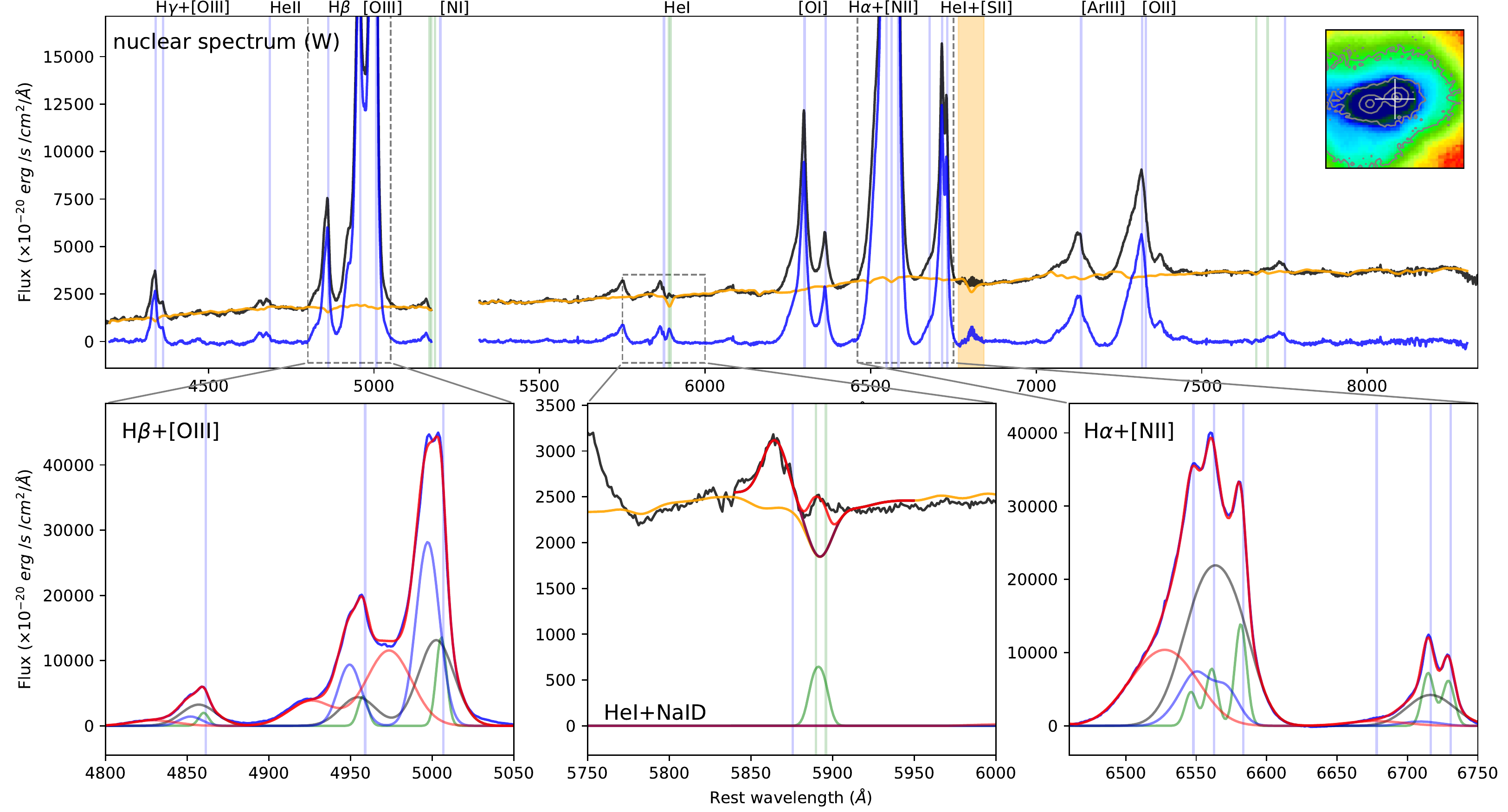}

\caption{\small IRAS F13451+1232 W nuclear spectrum extracted from a circular aperture with $r < 0.4''$, with the corresponding pPXF (top panel) and multi-component (bottom insets) best-fit models. See Fig. \ref{IRAS00188_n} for details.}
\label{IRAS13451_n_w}
\end{figure*}

\clearpage 

\begin{figure*}[t]
\centering
\includegraphics[width=21.cm,trim= 100 380 0 15,clip]{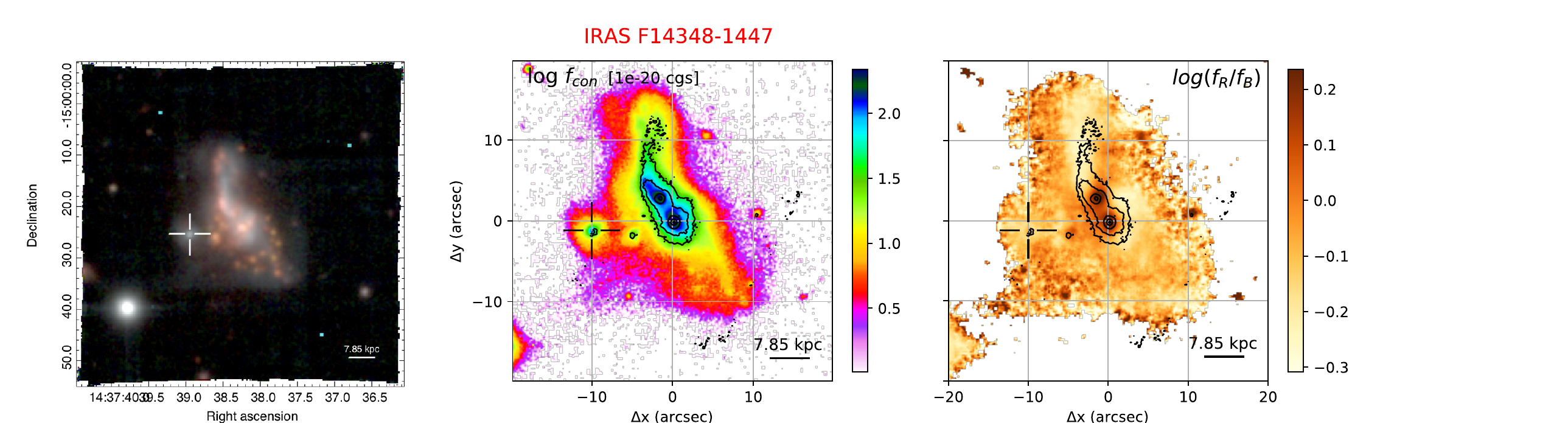}

\caption{\small IRAS F14348-1447 images from MUSE observations with TOT $= 0.68$ hr. {\it Left}: Colour-composite optical image showing [O {\small III}] (green, from the wavelength range $4980-5020\AA$ rest-frame), \ha (red, $6555-6575\AA$), and  continuum (blue, $4590-4690\AA$). {\it Centre}: Red ($7890-8000\AA$) continuum image from MUSE with contours from HST/F160W. {\it Right}:  Continuum colour map obtained from MUSE by dividing the red continuum image (central panel) by a blue image obtained by collapsing the stellar emission in the range $4590-4690\AA$; contours from HST/F160W. In all panels, we display the IRAS F14348-1447 companion galaxy with a cross.}
\label{IRAS14348_1}
\end{figure*}

\begin{figure*}[t]
\vspace{0.5cm}
\centering

\includegraphics[width=21.cm,trim= 90 197 0 216,clip]{{IRAS14348_Appendix}.pdf}

\caption{\small IRAS F14348-1447 emission line images from MUSE observations. [O {\small III}] (left, from the wavelength range $4980-5020\AA$ rest-frame), \ha (centre, $6555-6575\AA$), and [S {\small II}] (right, $6686-6750\AA$) images have been obtained by subtracting continuum emission using the adjacent regions at shorter and longer wavelengths with respect to the emission line systemics.  In all panels, we display the position of the  companion galaxy with a cross; the  line emission in the three maps is absent because they are faint.}
\label{IRAS14348_2}
\end{figure*}

\begin{figure*}[t]
\vspace{0.5cm}
\centering

\includegraphics[width=13.cm,trim= 299 10 90 396,clip]{{IRAS14348_Appendix}.pdf}

\caption{\small IRAS F14348-1447 stellar kinematic maps from the pPXF analysis with contours from HST/F160W. The left panel shows the stellar velocity $V_*$, and the right panel represents the velocity dispersion $\sigma_*$.}
\label{IRAS14348_3}
\end{figure*}

\begin{figure*}[t]
\vspace{0.5cm}

\centering
\includegraphics[width=18.cm,trim= 0 0 0 0,clip]{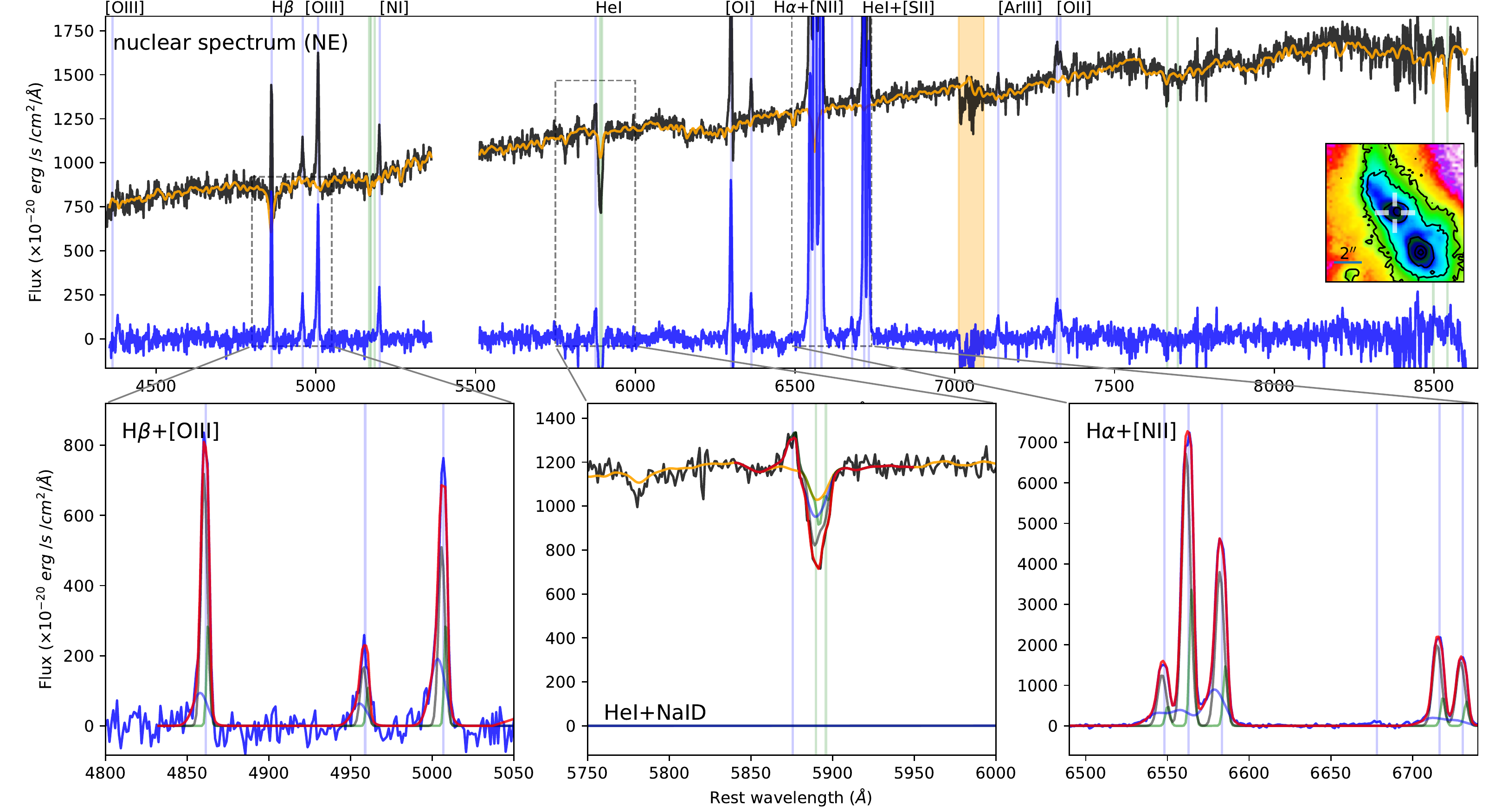}

\caption{\small IRAS F14348-1447 NE nuclear spectrum extracted from a circular aperture with $r < 0.4''$, with the corresponding pPXF (top panel) and multi-component (bottom insets) best-fit models. See Fig. \ref{IRAS00188_n} for details. }
\label{pPXF1}
\end{figure*}

\begin{figure*}[t]
\centering
\includegraphics[width=18.cm,trim= 0 0 0 0,clip]{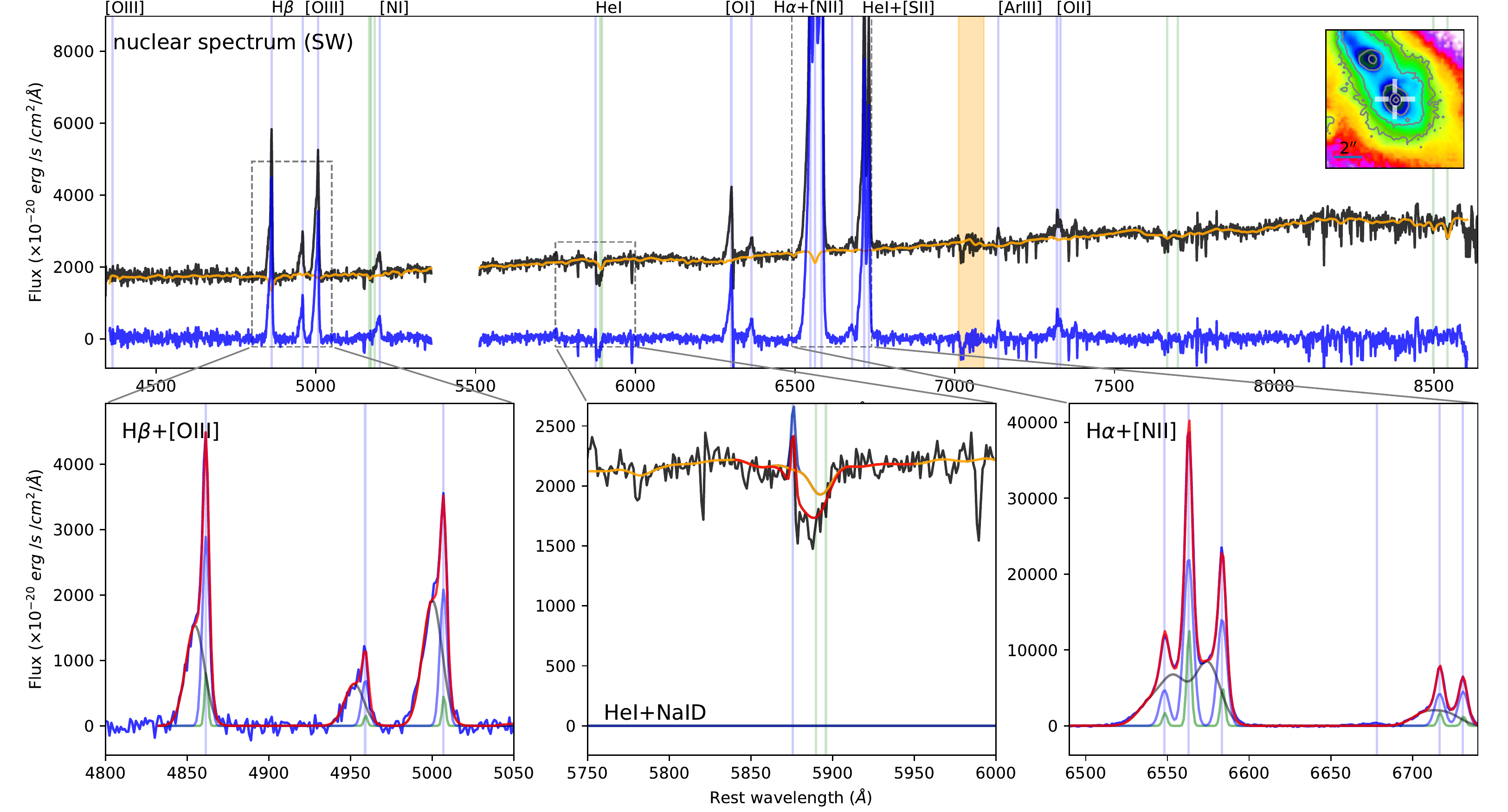}

\caption{\small IRAS F14348-1447 SW nuclear spectrum extracted from a circular aperture with $r < 0.4''$, with the corresponding pPXF (top panel) and multi-component (bottom insets) best-fit models. See Fig. \ref{IRAS00188_n} for details. }
\label{pPXF1}
\end{figure*}

\clearpage 

\begin{figure*}[t]
\centering
\includegraphics[width=21.cm,trim= 100 380 0 15,clip]{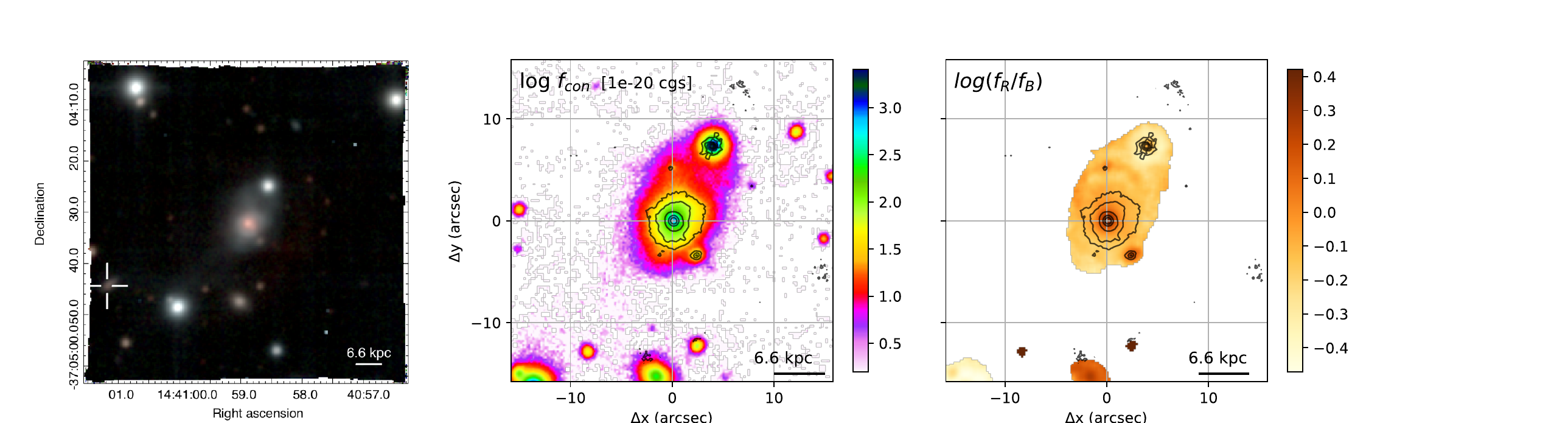}

\caption{\small IRAS F14378-3651 images from MUSE-noAO observations with TOT $= 0.47$ hr. {\it Left}: Colour-composite optical image showing [O {\small III}] (green, from the wavelength range $4980-5015\AA$ rest-frame), \ha (red, $6556-6572\AA$), and stellar continuum (blue, $4590-4790\AA$). {\it Centre}: Red ($7500-7600\AA$) continuum image from MUSE with contours from HST/F160W. {\it Right}:  Continuum colour map obtained from MUSE by dividing the red continuum image (central panel) by a blue image obtained by collapsing the stellar emission in the range $4590-4790\AA$; contours from HST/F160W. In the left panel, we display the IRAS F14378-3651 companion galaxy with a cross.}
\label{IRAS14378_1}
\end{figure*}

\begin{figure*}[t]
\vspace{0.5cm}
\centering

\includegraphics[width=21.cm,trim= 90 197 0 216,clip]{{IRAS14378_Appendix}.pdf}

\caption{\small IRAS F14378-3651 emission line images from MUSE observations. [O {\small III}] (left, from the wavelength range $4980-5015\AA$ rest-frame), \ha (centre, $6556-6572\AA$), and [S {\small II}] (right, $6695-6747\AA$) images have been obtained by subtracting continuum emission using the adjacent regions at shorter and longer wavelengths with respect to the emission line systemics.}
\label{IRAS14378_2}
\end{figure*}

\begin{figure*}[t]
\vspace{0.5cm}
\centering

\includegraphics[width=13.cm,trim= 299 10 90 396,clip]{{IRAS14378_Appendix}.pdf}

\caption{\small IRAS F14378-3651 stellar kinematic maps from the pPXF analysis with contours from HST/F160W. The left panel shows the stellar velocity $V_*$, and the right panel represents the velocity dispersion $\sigma_*$.}
\label{IRAS14378_3}
\end{figure*}

\begin{figure*}[t]
\vspace{0.5cm}

\centering
\includegraphics[width=18.cm,trim= 0 0 0 0,clip]{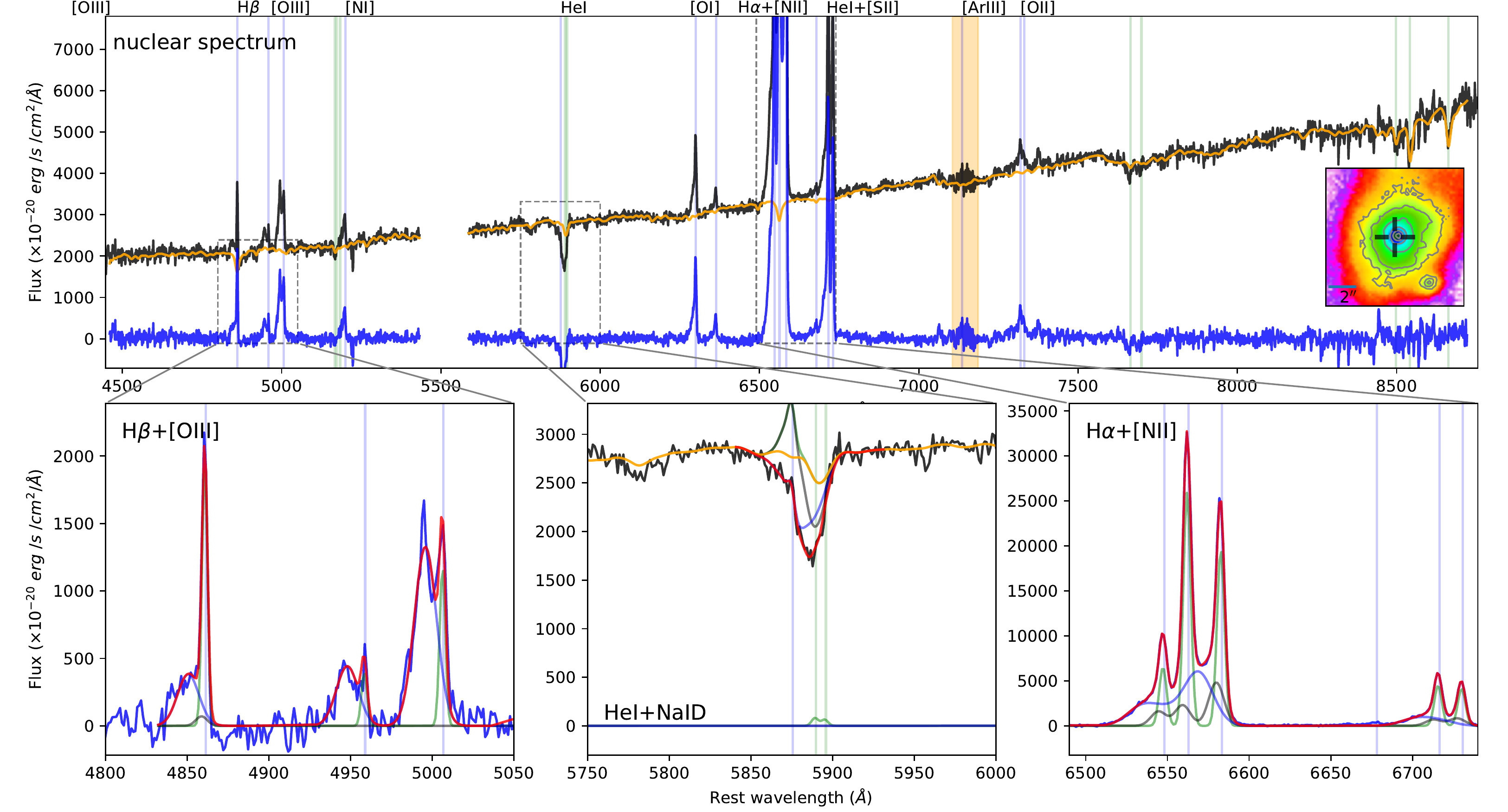}

\caption{\small IRAS F14378-3651 nuclear spectrum extracted from a circular aperture with $r < 0.4''$, with the corresponding pPXF (top panel) and multi-component (bottom insets) best-fit models. See Fig. \ref{IRAS00188_n} for details. }
\label{pPXF1}
\end{figure*}

\clearpage 

\begin{figure*}[t]
\centering
\includegraphics[width=21.cm,trim= 100 380 0 15,clip]{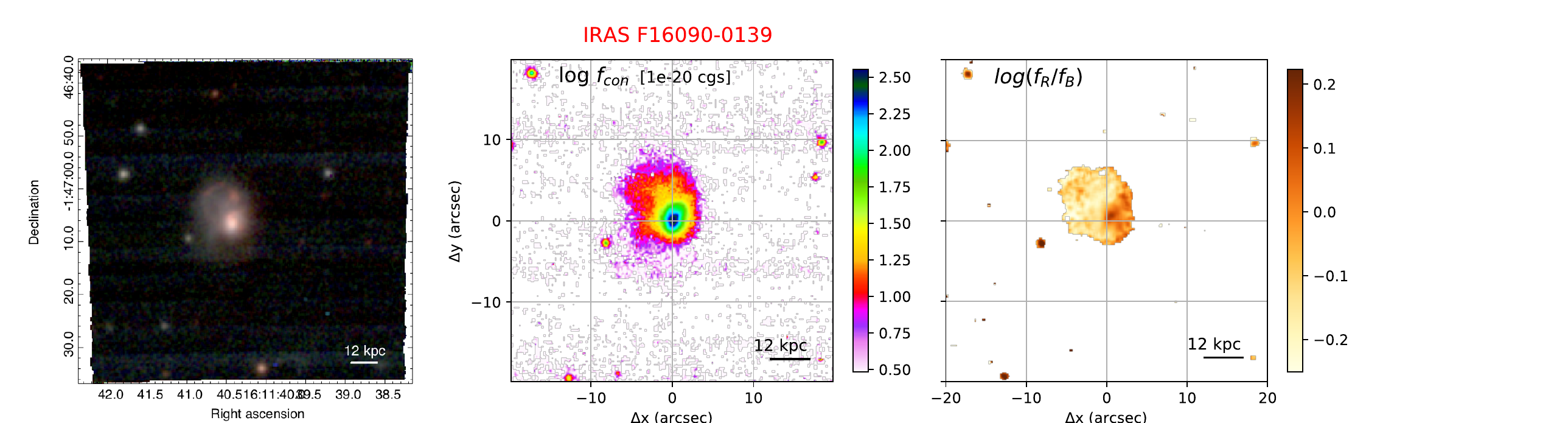}

\caption{\small IRAS F16090-0139 images from MUSE observations with TOT $= 0.17$ hr. {\it Left}: Colour-composite optical image showing [O {\small III}] (green, from the wavelength range $4983-5016 \AA$ rest-frame), \ha (red, $6556-6573 \AA$), and stellar continuum (blue, $4420-4480 \AA$). {\it Centre}: Red ($7670-6730 \AA$) continuum image from MUSE with contours from HST/F160W. {\it Right}: Stellar continuum colour map obtained from MUSE by dividing the red continuum image (central panel) by a blue image obtained by collapsing the stellar emission in the range $4420-4480\AA$. }
\label{IRAS16090_1}
\end{figure*}

\begin{figure*}[t]
\vspace{0.5cm}
\centering

\includegraphics[width=21.cm,trim= 90 197 0 216,clip]{{IRAS16090_Appendix}.pdf}

\caption{\small IRAS F16090-0139 emission line images from MUSE observations. [O {\small III}] (left, from the wavelength range $4983-5016 \AA$ rest-frame), \ha (centre, $6556-6573 \AA$), and [S {\small II}] (right, $6690-6747 \AA$) images have been obtained by subtracting continuum emission using the adjacent regions at shorter and longer wavelengths with respect to the emission line systemics. }
\label{IRAS16090_2}
\end{figure*}

\begin{figure*}[t]
\vspace{0.5cm}
\centering

\includegraphics[width=13.cm,trim= 299 10 90 395,clip]{{IRAS16090_Appendix}.pdf}

\caption{\small IRAS F16090-0139 stellar kinematic maps from the pPXF analysis with contours from HST/F160W. The left panel shows the stellar velocity $V_*$, and the right panel represents the velocity dispersion $\sigma_*$.}
\label{IRAS16090_3}
\end{figure*}

\begin{figure*}[t]
\vspace{0.5cm}

\centering
\includegraphics[width=18.cm,trim= 0 0 0 0,clip]{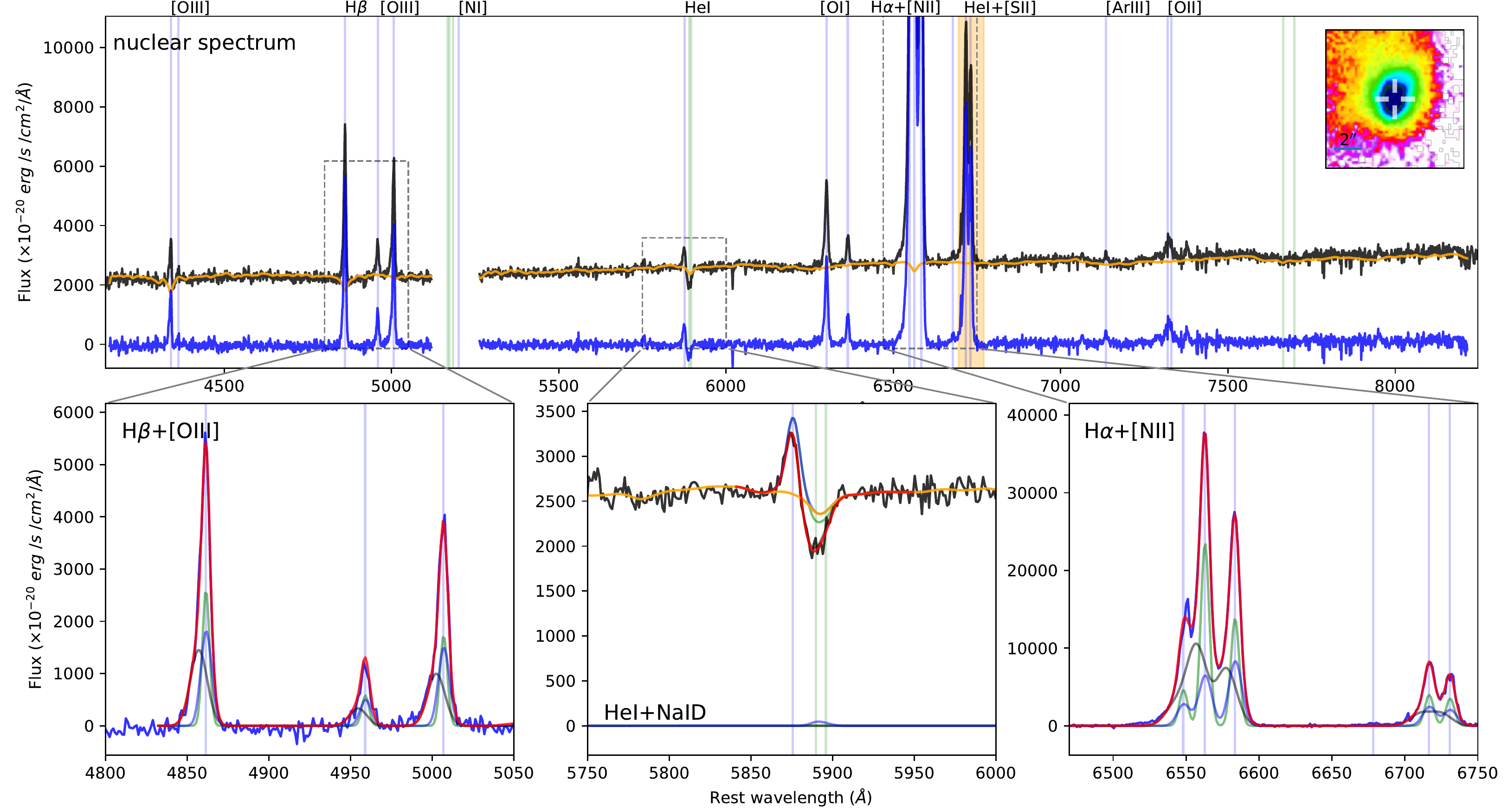}

\caption{\small IRAS F16090-0139 nuclear spectrum extracted from a circular aperture with $r < 0.4''$, with the corresponding pPXF (top panel) and multi-component (bottom insets) best-fit models. See Fig. \ref{IRAS00188_n} for details. }
\label{IRAS16090_n}
\end{figure*}

\clearpage 

\begin{figure*}[t]
\centering
\includegraphics[width=21.cm,trim= 100 375 0 15,clip]{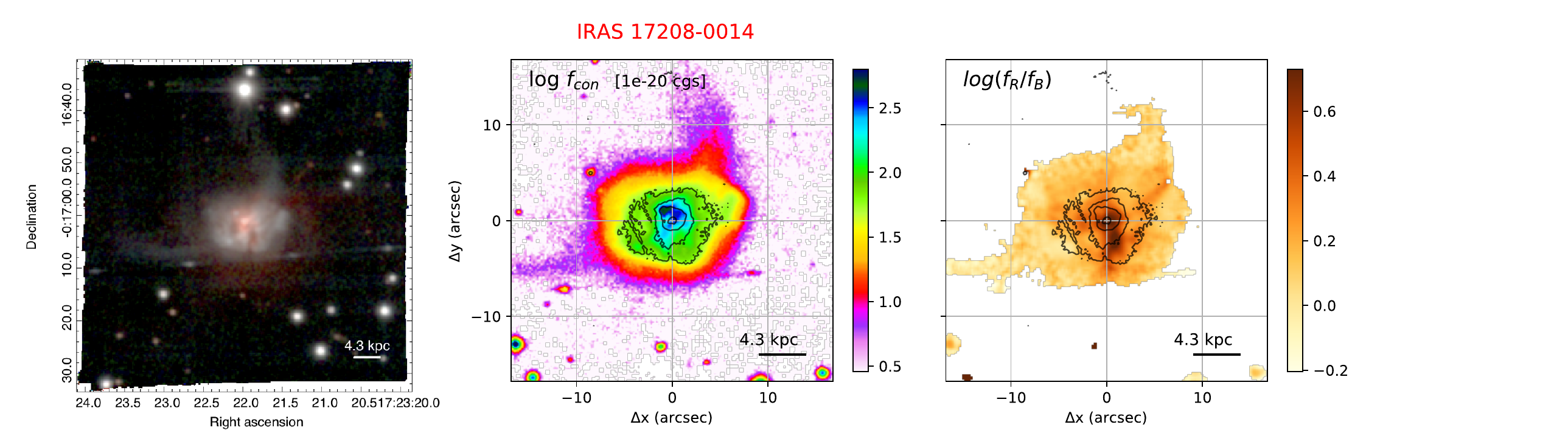}

\caption{\small IRAS 17208-0014 images from MUSE observations with TOT $= 0.39$ hr. {\it Left}: Colour-composite optical image showing [O {\small III}] (green, from the wavelength range $4975-5032 \AA$ rest-frame), \ha (red, $6549-6576 \AA$), and stellar continuum (blue, $4520-4490 \AA$). {\it Centre}: Red ($7690-7760 \AA$) continuum image from MUSE with contours from HST/F160W. {\it Right}: Stellar continuum colour map obtained from MUSE by dividing the red continuum image (central panel) by a blue image obtained by collapsing the stellar emission in the range $4520-4490\AA$. In all panels, the bottom part of the ULIRG shows a satellite trail resulting in three artificial blobs in the continuum emission.}
\label{IRAS17208_1}
\end{figure*}

\begin{figure*}[t]
\vspace{0.5cm}
\centering

\includegraphics[width=21.cm,trim= 90 199 0 219,clip]{{IRAS17208_Appendix}.pdf}

\caption{\small IRAS 17208-0014 emission line images from MUSE observations. [O {\small III}] (left, from the wavelength range $4975-5032\AA$ rest-frame), \ha (centre, $6549-6576\AA$), and [S {\small II}] (right, $6695-6749\AA$) images have been obtained by subtracting continuum emission using the adjacent regions at shorter and longer wavelengths with respect to the emission line systemics. }
\label{IRAS17208_2}
\end{figure*}

\begin{figure*}[t]
\vspace{0.5cm}
\centering

\includegraphics[width=13.cm,trim= 299 10 90 395,clip]{{IRAS17208_Appendix}.pdf}

\caption{\small IRAS 17208-0014 stellar kinematic maps from the pPXF analysis with contours from HST/F160W. The left panel shows the stellar velocity $V_*$, and the right panel represents the velocity dispersion $\sigma_*$.}
\label{IRAS17208_3}
\end{figure*}

\begin{figure*}[t]
\vspace{0.5cm}

\centering
\includegraphics[width=18.cm,trim= 0 0 0 0,clip]{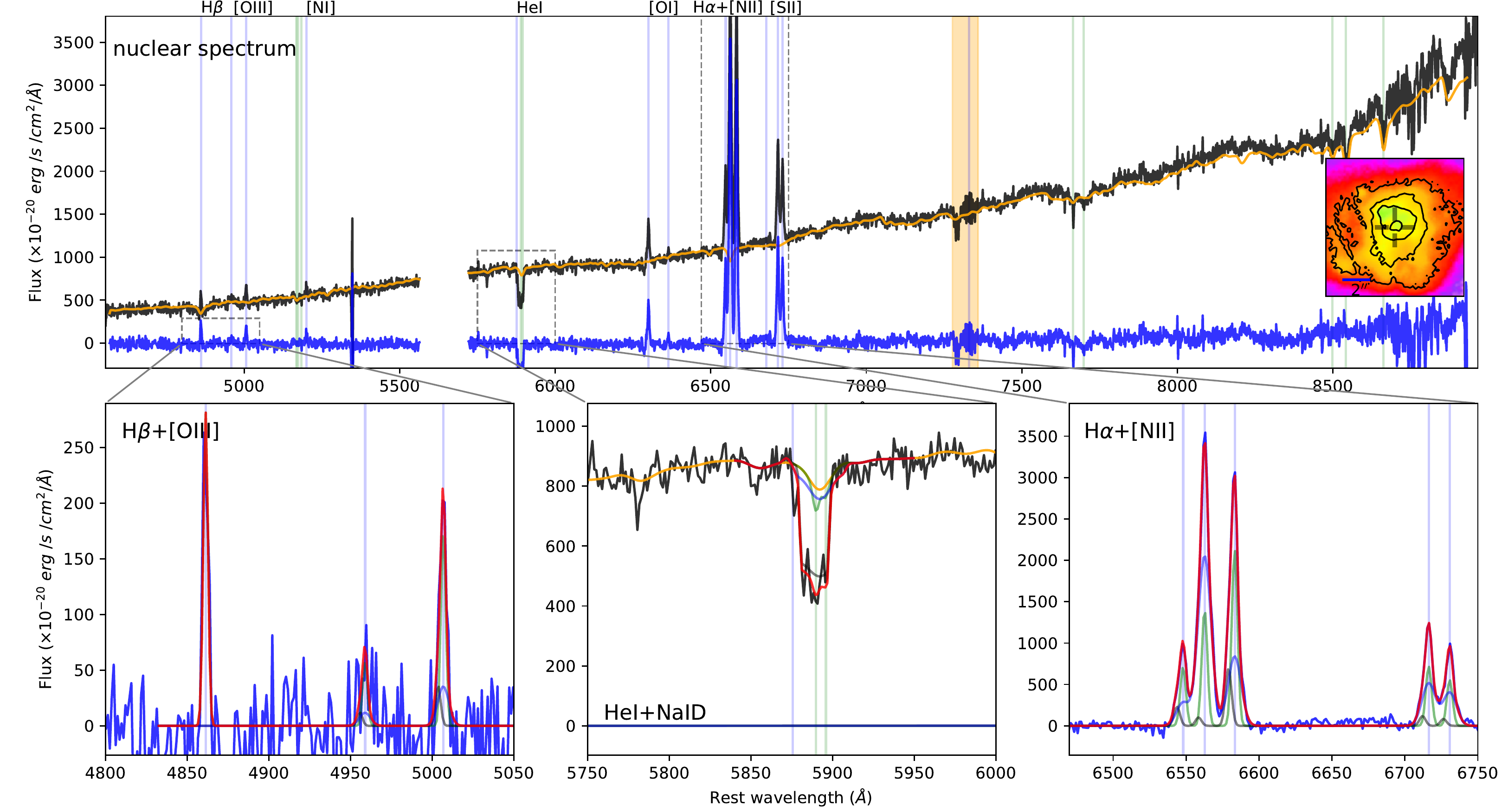}

\caption{\small IRAS 17208-0014  nuclear spectrum extracted from a circular aperture with $r < 0.4''$, with the corresponding pPXF (top panel) and multi-component (bottom insets) best-fit models. See Fig. \ref{IRAS00188_n} for details.}
\label{IRAS17208_n}
\end{figure*}

\clearpage 

\begin{figure*}[t]
\centering
\includegraphics[width=21.cm,trim= 100 380 0 19,clip]{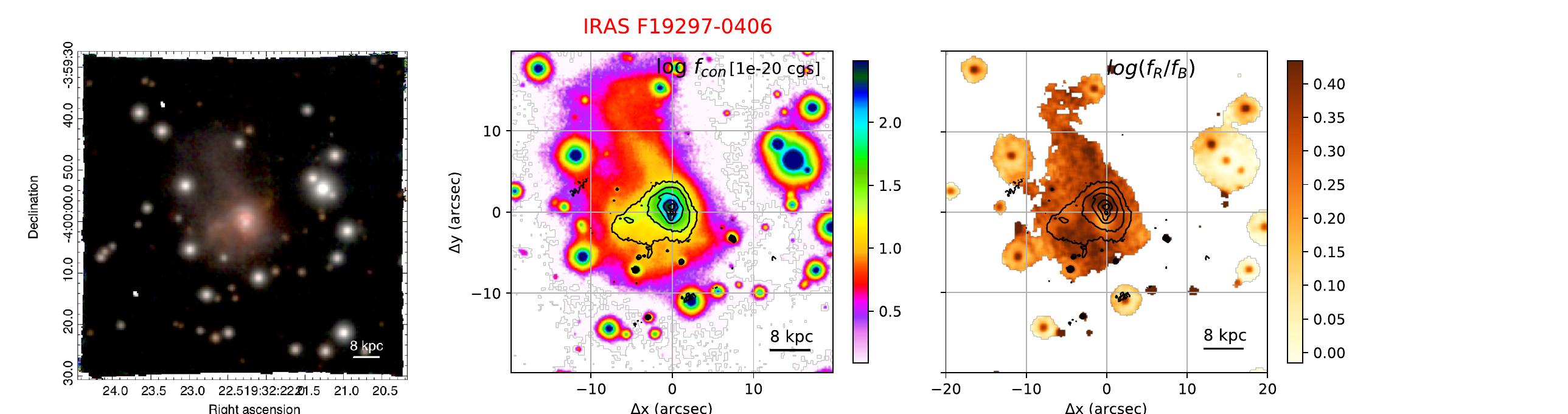}

\caption{\small IRAS F19297-0406 images from MUSE observations with TOT $= 2.04$ hr. {\it Left}: Colour-composite optical image showing [O {\small III}] (green, from the wavelength range $4991-5021\AA$ rest-frame), \ha (red, $6557-6579\AA$), and stellar continuum (blue, $4550-4620\AA$). {\it Centre}: Red ($7440-7590\AA$) continuum image from MUSE  with contours from HST/F160W. {\it Right}: Stellar continuum colour map obtained from MUSE by dividing the red continuum image (central panel) by a blue image obtained by collapsing the stellar emission in the range $4550-4620\AA$. }
\label{IRAS19297_1}
\end{figure*}

\begin{figure*}[t]
\vspace{0.5cm}
\centering

\includegraphics[width=21.cm,trim= 90 197 0 216,clip]{{IRAS19297_Appendix}.pdf}

\caption{\small IRAS F19297-0406 emission line images from MUSE observations. [O {\small III}] (left, from the wavelength range $4991-5021\AA$ rest-frame), \ha (centre, $6557-6579\AA$), and [S {\small II}] (right, $6700-6752\AA$) images have been obtained by subtracting continuum emission using the adjacent regions at shorter and longer wavelengths with respect to the emission line systemics. }
\label{IRAS19297_2}
\end{figure*}

\begin{figure*}[t]
\vspace{0.5cm}
\centering

\includegraphics[width=13.cm,trim= 299 10 90  395,clip]{{IRAS19297_Appendix}.pdf}

\caption{\small IRAS F19297-0406 stellar kinematic maps from the pPXF analysis with contours from HST/F160W. The left panel shows the stellar velocity $V_*$, and the right panel represents the velocity dispersion $\sigma_*$.}
\label{IRAS19297_3}
\end{figure*}

\begin{figure*}[t]
\vspace{0.5cm}

\centering
\includegraphics[width=18.cm,trim= 0 0 0 0,clip]{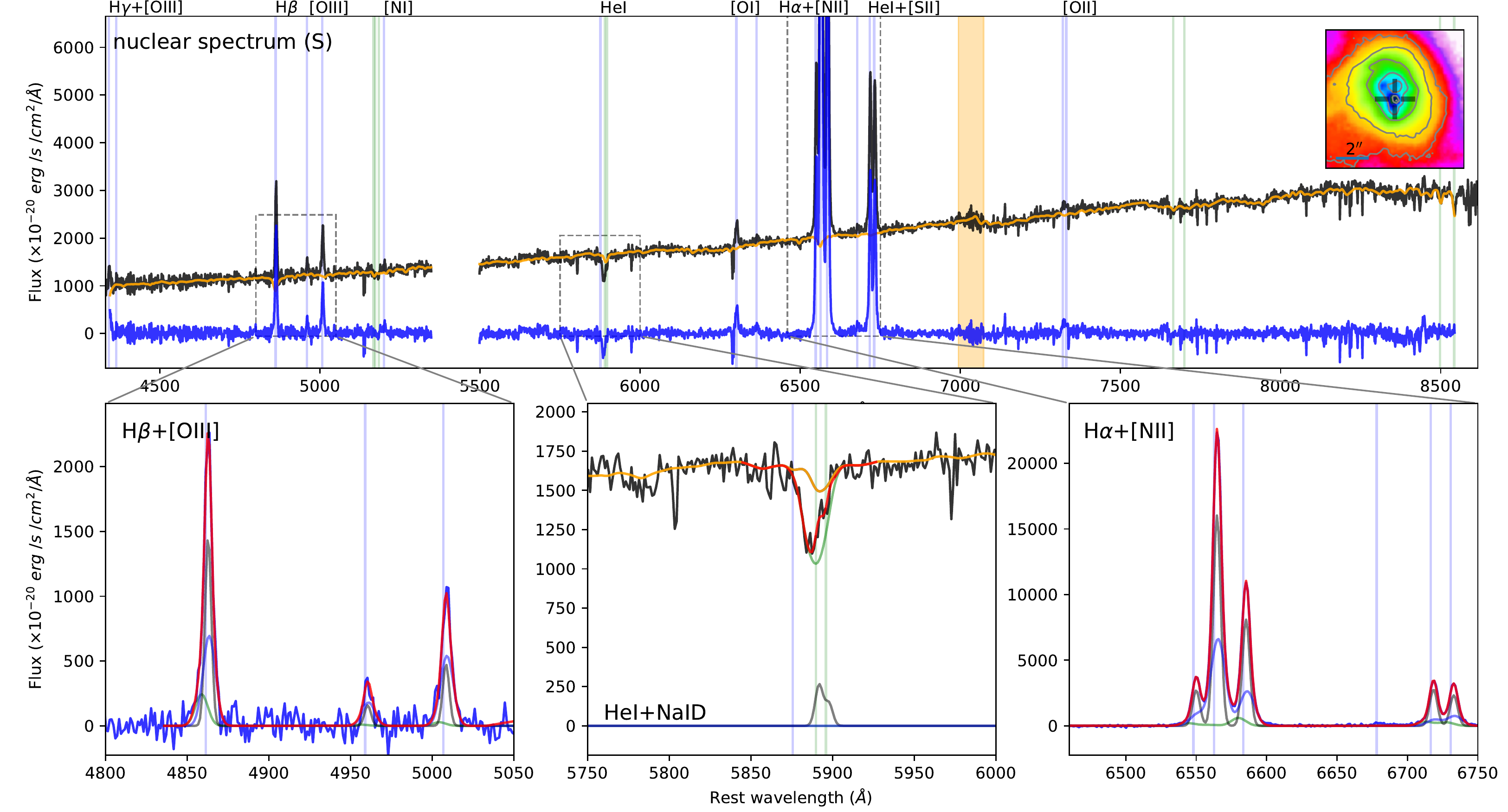}

\caption{\small IRAS F19297-0406 (S)  nuclear spectrum extracted from a circular aperture with $r < 0.4''$, with the corresponding pPXF (top panel) and multi-component (bottom insets) best-fit models. See Fig. \ref{IRAS00188_n} for details.}
\label{IRAS19297_n}
\end{figure*}

\begin{figure*}[t]
\vspace{0.5cm}

\centering
\includegraphics[width=18.cm,trim= 0 0 0 0,clip]{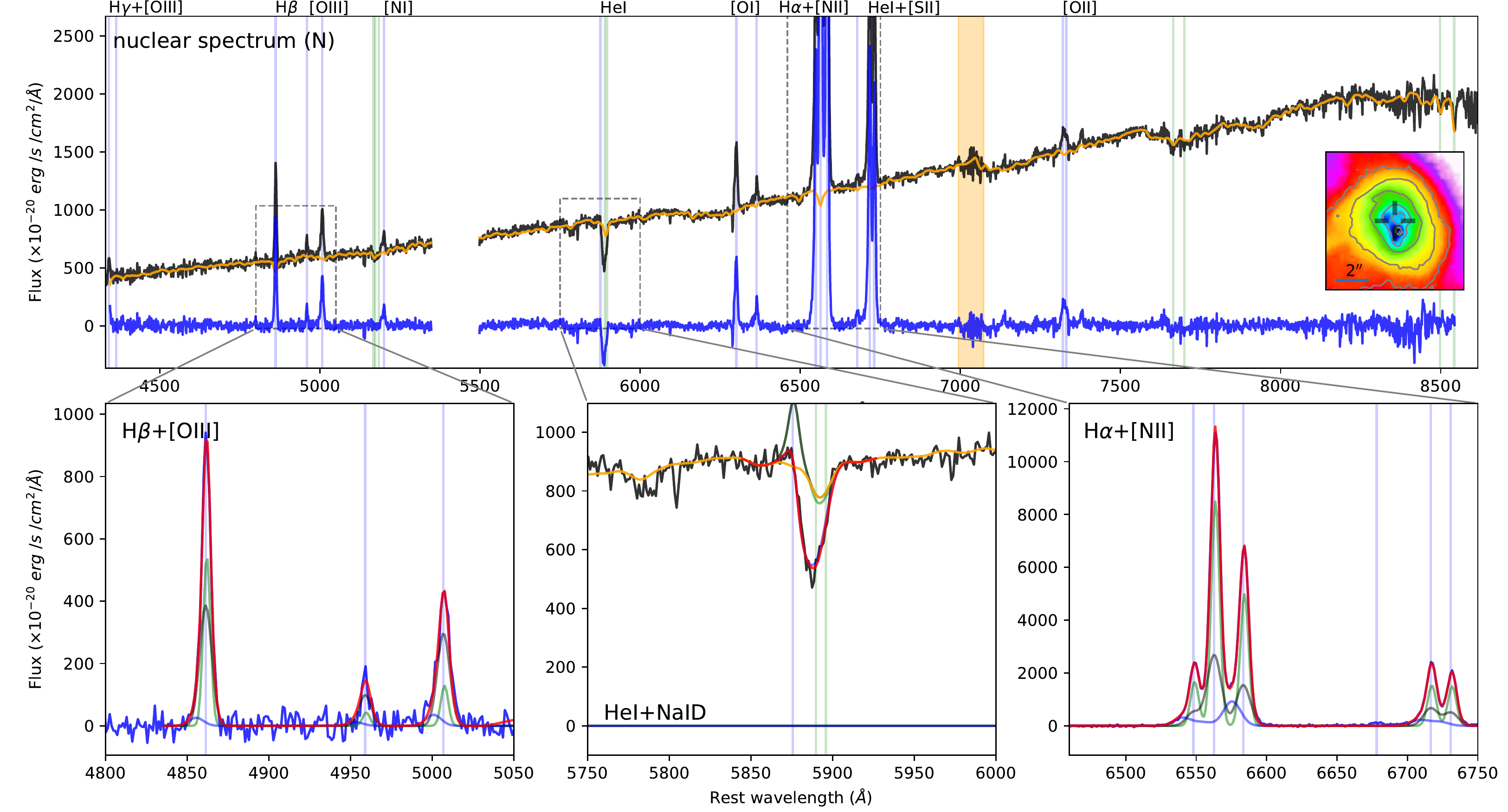}

\caption{\small IRAS F19297-0406 (N)  nuclear spectrum extracted from a circular aperture with $r < 0.4''$, with the corresponding pPXF (top panel) and multi-component (bottom insets) best-fit models. See Fig. \ref{IRAS00188_n} for details.}
\label{IRAS19297_n}
\end{figure*}

\clearpage 

\begin{figure*}[t]
\centering
\includegraphics[width=21.cm,trim= 100 377 0 24,clip]{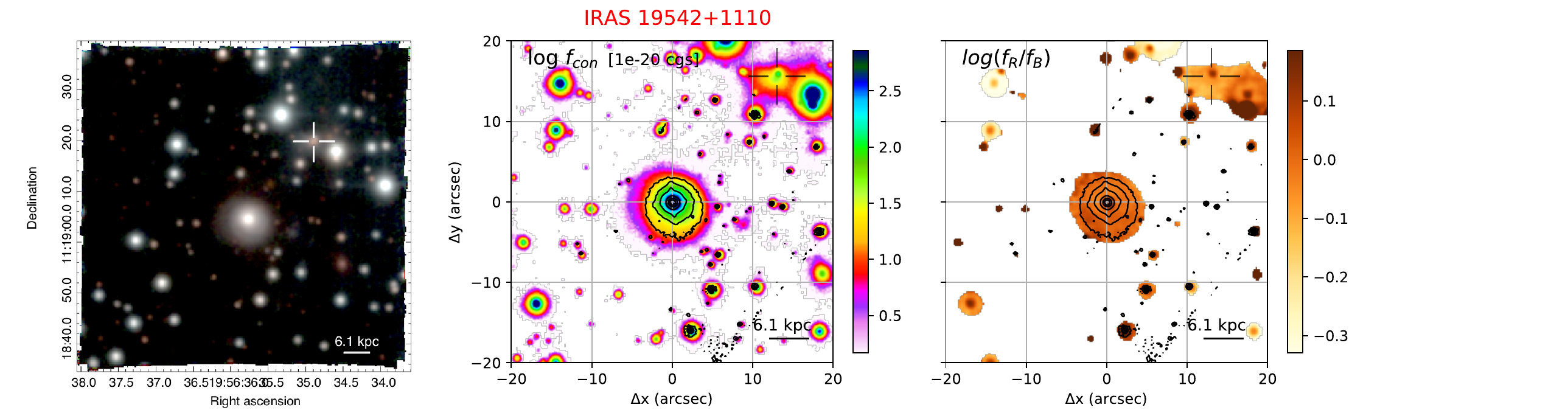}

\caption{\small IRAS 19542+1110 images from MUSE observations with TOT $= 2.04$ hr. {\it Left}: Colour-composite optical image showing [O {\small III}] (green, from the wavelength range $4985-5016 \AA$ rest-frame), \ha (red, $6552-6573 \AA$), and stellar continuum (blue, $4630-4740 \AA$). {\it Centre}: Red ($7650-7750 \AA$) continuum image from MUSE with contours from HST/F160W. {\it Right}: Stellar continuum colour map obtained from MUSE by dividing the red continuum image (central panel) by a blue image obtained by collapsing the continuum emission in the range $4630-4740\AA$; contours from HST/F160W. In all panels, we display the companion galaxy of IRAS 19542+1110 with a cross.}
\label{IRAS19542_1}
\end{figure*}

\begin{figure*}[t]
\vspace{0.5cm}
\centering

\includegraphics[width=21.cm,trim= 90 198 0 217,clip]{{IRAS19542_Appendix}.pdf}

\caption{\small IRAS 19542+1110 emission line images from MUSE observations. [O {\small III}] (left, from the wavelength range $4985-5016\AA$ rest-frame), \ha (centre, $6552-6573\AA$), and [S {\small II}] (right, $6700-6744\AA$) images have been obtained by subtracting continuum emission using the adjacent regions at shorter and longer wavelengths with respect to the emission line systemics. }
\label{IRAS19542_2}
\end{figure*}

\begin{figure*}[t]
\vspace{0.5cm}
\centering

\includegraphics[width=13.cm,trim= 299 10 90 395,clip]{{IRAS19542_Appendix}.pdf}

\caption{\small IRAS 19542+1110 stellar kinematic maps from the pPXF analysis with contours from HST/F160W. The left panel shows the stellar velocity $V_*$, and the right panel represents the velocity dispersion $\sigma_*$.}
\label{IRAS19542_1}
\end{figure*}

\begin{figure*}[t]
\vspace{0.5cm}

\centering
\includegraphics[width=18.cm,trim= 0 0 0 0,clip]{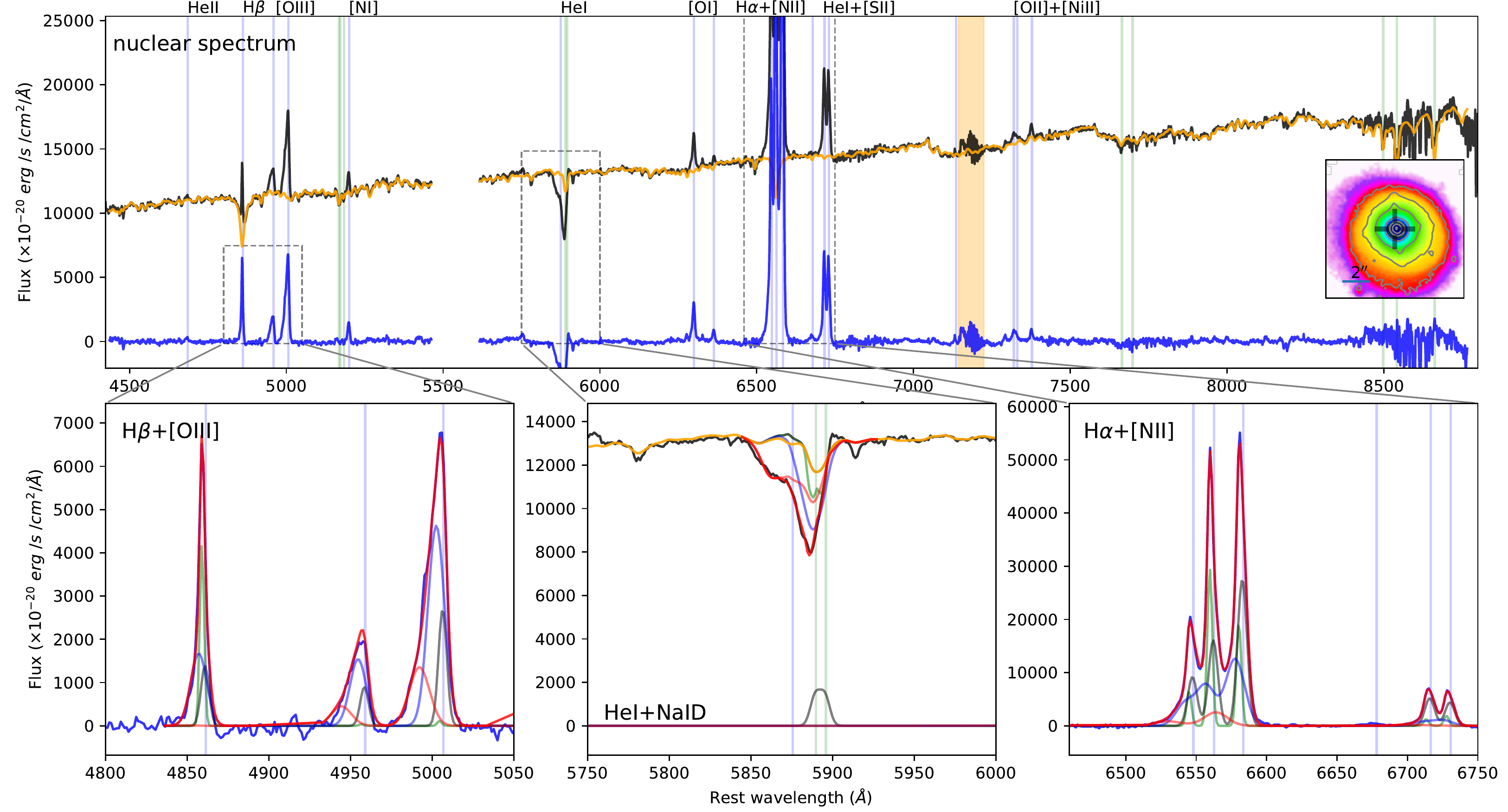}

\caption{\small IRAS 19542+1110 nuclear spectrum extracted from a circular aperture with $r < 0.4''$, with the corresponding pPXF (top panel) and multi-component (bottom insets) best-fit models. See Fig. \ref{IRAS00188_n} for details.}
\label{IRAS19542_n}
\end{figure*}

\clearpage 

\begin{figure*}[t]
\centering
\includegraphics[width=21.cm,trim= 100 376 0 15,clip]{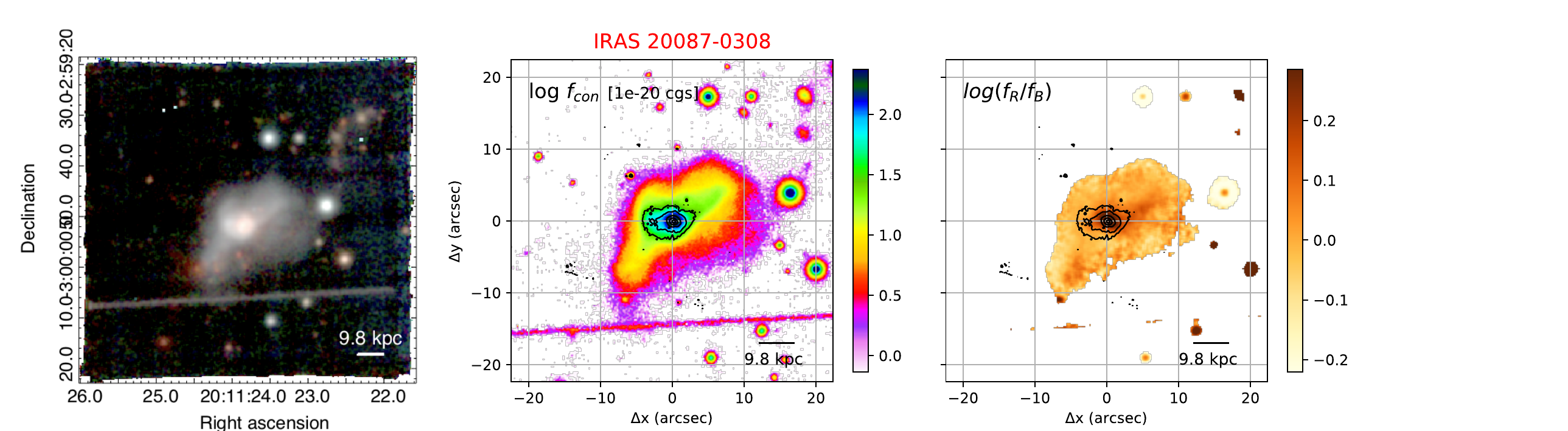}

\caption{\small IRAS 20087-0308 images from MUSE observations with TOT $= 2.04$ hr. {\it Left}: Colour-composite optical image showing [O {\small III}] (green, from the wavelength range $4986-5020 \AA$ rest-frame), \ha (red, $6555-6574\AA$), and  continuum (blue, $4600-4700 \AA$). {\it Centre}: Red ($7780-7880 \AA$) continuum image from MUSE with contours from HST/F160W. {\it Right}: Stellar continuum colour map obtained from MUSE by dividing the red continuum image (central panel) by a blue image obtained by collapsing the stellar emission in the range $4600-4700 \AA$; contours from HST/F160W. In all panels, the bottom part of the FOV shows a satellite trail.}
\label{IRAS20087_1}
\end{figure*}

\begin{figure*}[t]
\vspace{0.5cm}
\centering

\includegraphics[width=21.cm,trim= 90 197 0 219,clip]{{IRAS20087_Appendix}.pdf}

\caption{\small IRAS 20087-0308 emission line images from MUSE observations. [O {\small III}] (left, from the wavelength range $4986-5020\AA$ rest-frame), \ha (centre, $6555-6574\AA$), and [S {\small II}] (right, $6700-6747\AA$) images have been obtained by subtracting continuum emission using the adjacent regions at shorter and longer wavelengths with respect to the emission line systemics.}
\label{IRAS20087_2}
\end{figure*}

\begin{figure*}[t]
\vspace{0.5cm}
\centering

\includegraphics[width=13.cm,trim= 299 10 90 395,clip]{{IRAS20087_Appendix}.pdf}

\caption{\small IRAS 20087-0308 stellar kinematic maps from the pPXF analysis with contours from HST/F160W. The left panel shows the stellar velocity $V_*$, and the right panel represents the velocity dispersion $\sigma_*$.}
\label{IRAS20087_3}
\end{figure*}

\begin{figure*}[t]
\vspace{0.5cm}

\centering
\includegraphics[width=18.cm,trim= 0 0 0 0,clip]{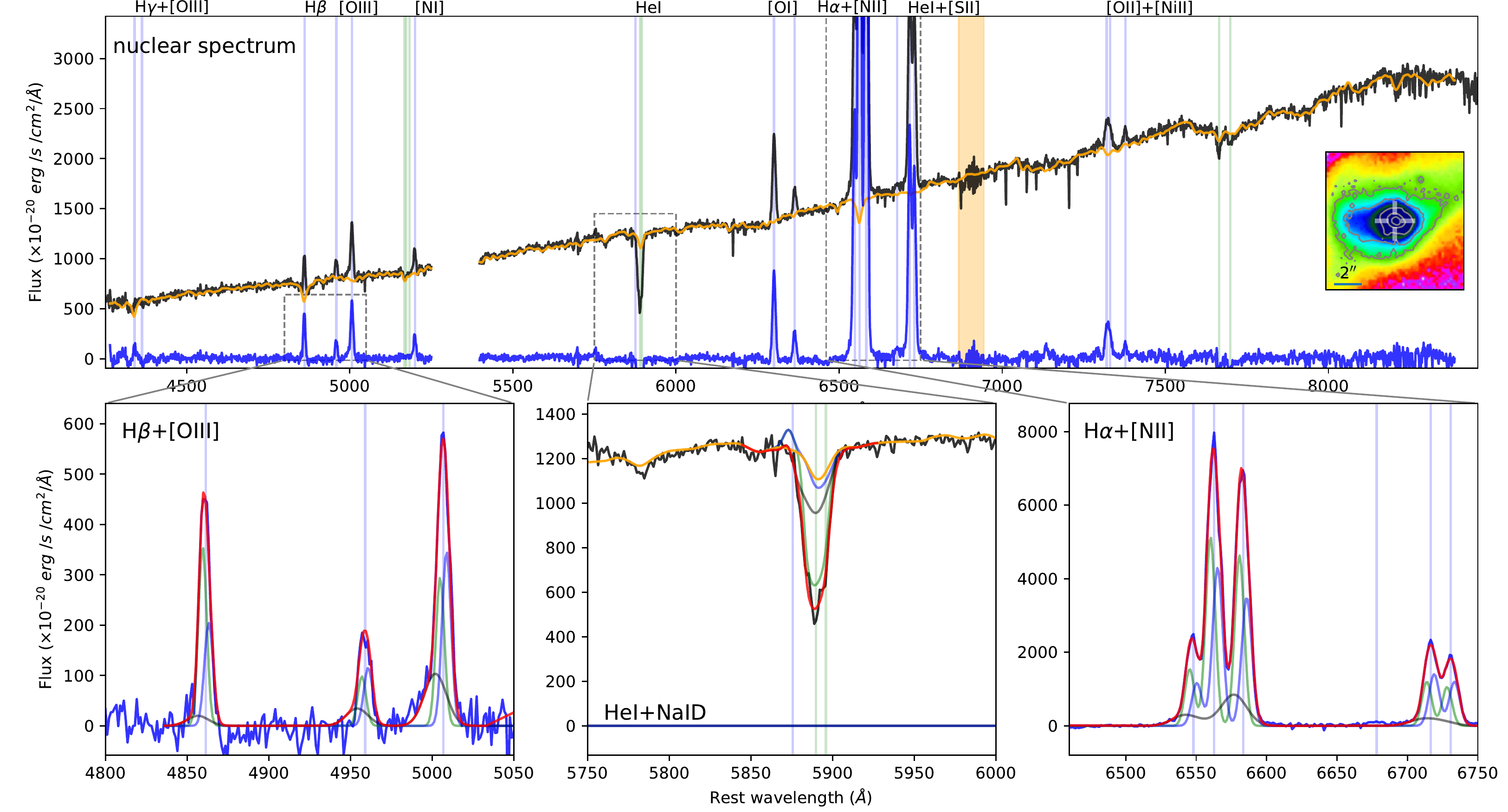}

\caption{\small IRAS 20087-0308  nuclear spectrum extracted from a circular aperture with $r < 0.4''$, with the corresponding pPXF (top panel) and multi-component (bottom insets) best-fit models. See Fig. \ref{IRAS00188_n} for details.}
\label{IRAS20087_n}
\end{figure*}

\clearpage 

\begin{figure*}[t]
\centering
\includegraphics[width=21.cm,trim= 100 378 0 19,clip]{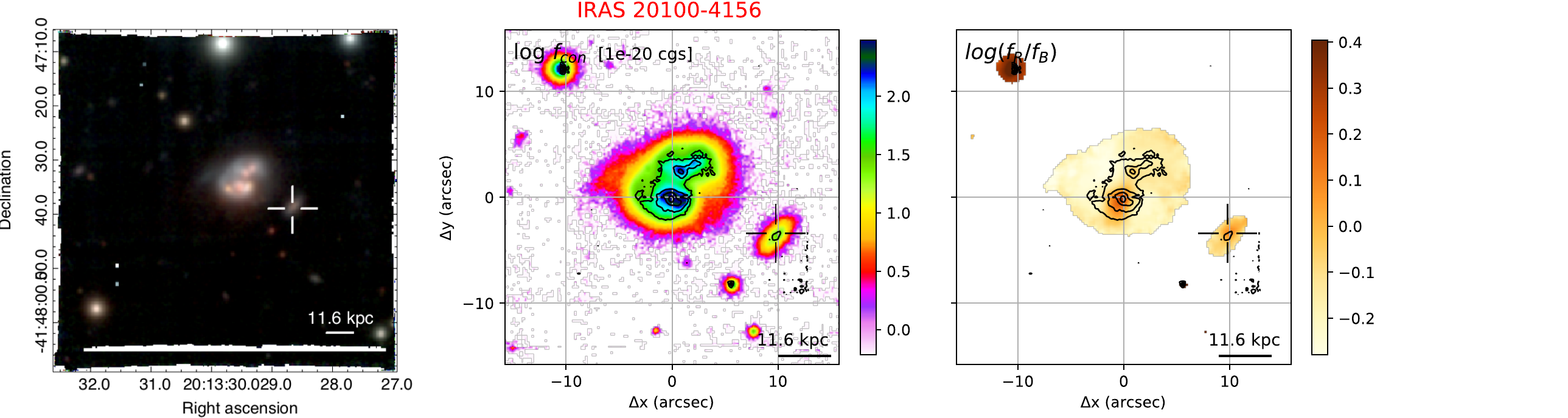}

\caption{\small IRAS 20100-4156 images from MUSE observations with TOT $= 2.04$ hr. {\it Left}: Colour-composite optical image showing [O {\small III}] (green, from the wavelength range $4985-5022\AA$ rest-frame), \ha (red, $6554-6579\AA$), and  continuum (blue, $4450-4600\AA$). {\it Centre}: Red ($7500-7700\AA$) continuum image from MUSE with contours from HST/F160W. {\it Right}: Stellar continuum colour map obtained from MUSE by dividing the red continuum image (central panel) by a blue image obtained by collapsing the stellar emission in the range $4450-4600\AA$; contours from HST/F160W. In all panels, we display the IRAS 20100-4156 companion with a cross.}
\label{IRAS20100_1}
\end{figure*}

\begin{figure*}[t]
\vspace{0.5cm}
\centering

\includegraphics[width=21.cm,trim= 90 197 0 215,clip]{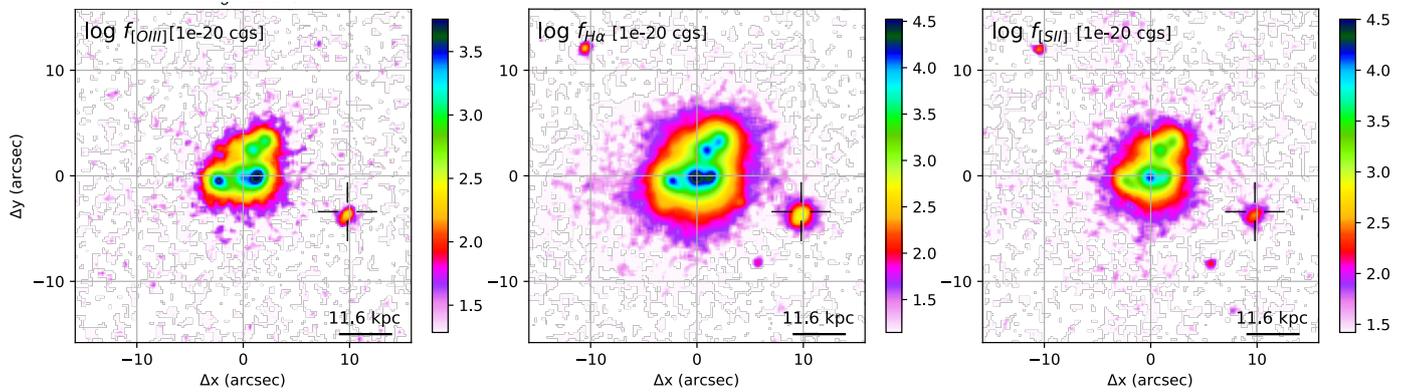}

\caption{\small IRAS 20100-4156 emission line images from MUSE observations. [O {\small III}] (left, from the wavelength range $4985-5022\AA$ rest-frame), \ha (centre, $6554-6579\AA$), and [S {\small II}] (right, $6691-6743\AA$) images have been obtained by subtracting continuum emission using the adjacent regions at shorter and longer wavelengths with respect to the emission line systemics. In all panels, we display the 20100-4156 companion with a cross.}
\label{IRAS20100_2}
\end{figure*}

\begin{figure*}[t]
\vspace{0.5cm}
\centering

\includegraphics[width=13.cm,trim= 299 10 90 400,clip]{{IRAS20100_Appendix}.pdf}

\caption{\small IRAS 20100-4156 stellar kinematic maps from the pPXF analysis with contours from HST/F160W. The left panel shows the stellar velocity $V_*$, and the right panel represents the velocity dispersion $\sigma_*$.}
\label{IRAS20100_3}
\end{figure*}

\begin{figure*}[t]
\vspace{0.5cm}

\centering
\includegraphics[width=18.cm,trim= 0 0 0 0,clip]{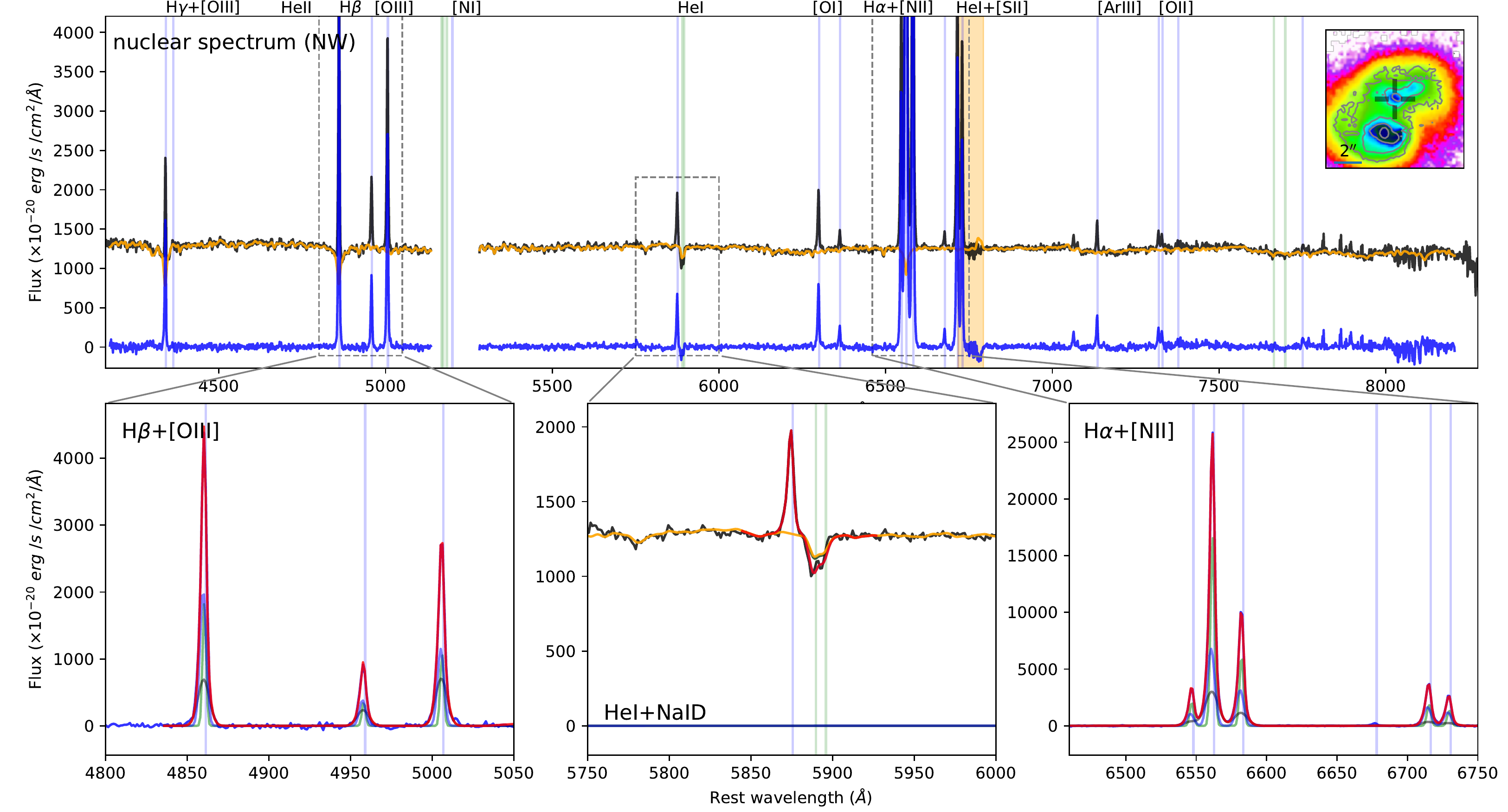}

\caption{\small IRAS 20100-4156 NW nuclear spectrum extracted from a circular aperture with $r < 0.4''$, with the corresponding pPXF (top panel) and multi-component (bottom insets) best-fit models. See Fig. \ref{IRAS00188_n} for details. }
\label{IRAS20100_n_n}
\end{figure*}

\begin{figure*}[t]
\vspace{0.5cm}

\centering
\includegraphics[width=18.cm,trim= 0 0 0 0,clip]{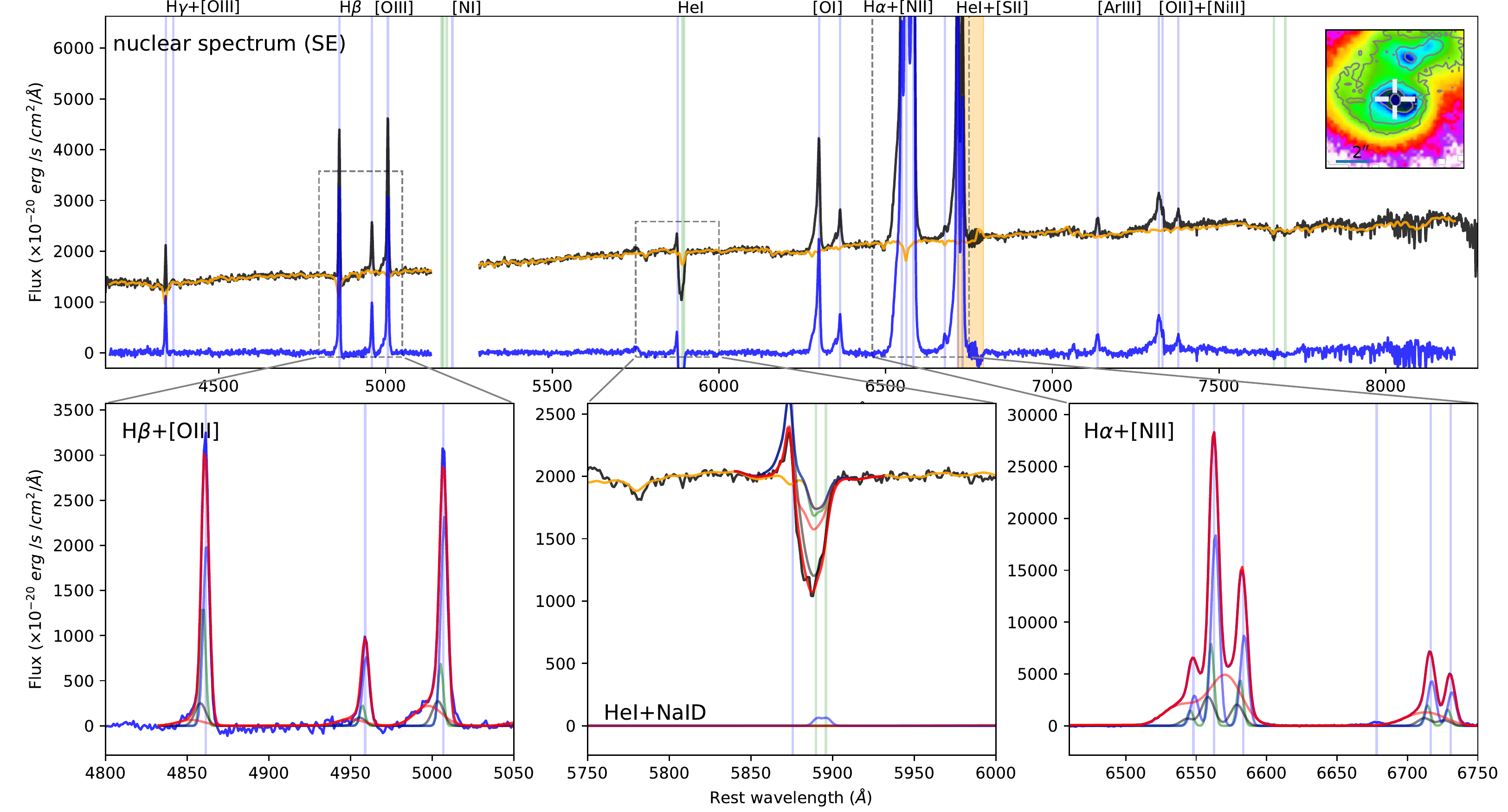}

\caption{\small IRAS 20100-4156 SE nuclear spectrum extracted from a circular aperture with $r < 0.4''$, with the corresponding pPXF (top panel) and multi-component (bottom insets) best-fit models. See Fig. \ref{IRAS00188_n} for details.}
\label{IRAS20100_n_s}
\end{figure*}

\clearpage 

\begin{figure*}[t]
\centering
\includegraphics[width=21.cm,trim= 100 380 0 15,clip]{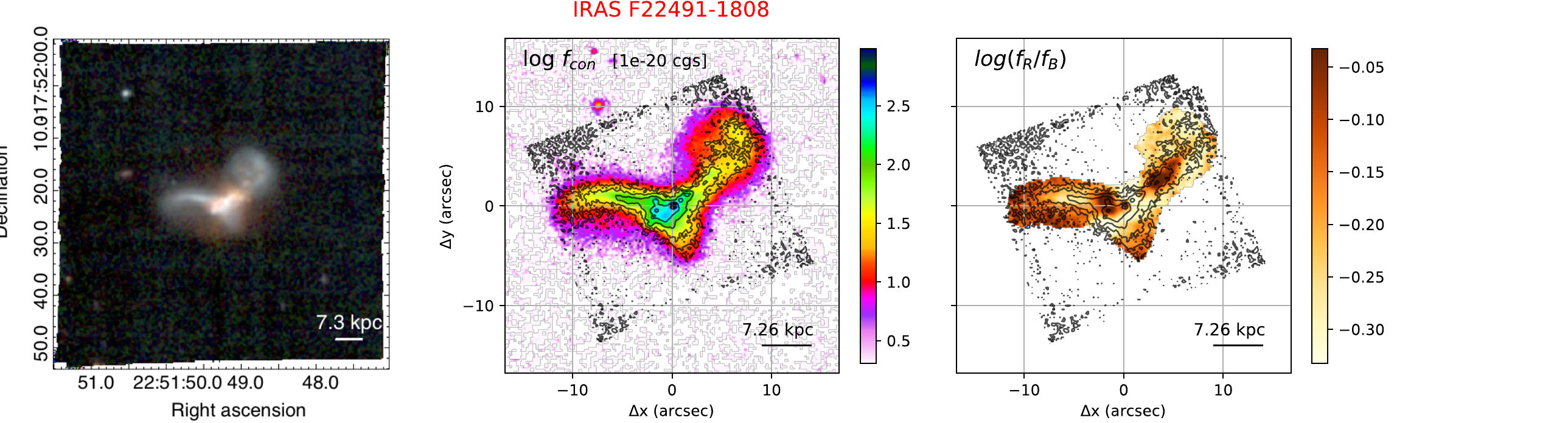}

\caption{\small IRAS F22491-1808 images from MUSE observations with TOT $= 0.41$ hr. {\it Left}: Colour-composite optical image showing [O {\small III}] (green, from the wavelength range $4998-5017\AA$ rest-frame), \ha (red, $6556-6575\AA$), and stellar continuum (blue, $4460-4740\AA$). {\it Centre}: Red ($8020-8120\AA$) stellar continuum image from MUSE with contours from HST/F160W. {\it Right}: Stellar continuum colour map obtained from MUSE by dividing the red continuum image (central panle) by a blue image obtained collapsing the stellar emission in the range $4460-4740\AA$; contours from HST/F160W.}
\label{IRAS22491_1}
\end{figure*}

\begin{figure*}[t]
\vspace{0.5cm}
\centering

\includegraphics[width=21.cm,trim= 90 195 0 214,clip]{{IRAS22491_Appendix}.pdf}

\caption{\small IRAS F22491-1808 emission line images from MUSE observations. [O {\small III}] (left, from the wavelength range $4998-5017\AA$ rest-frame), \ha (centre, $6556-6575\AA$), and [S {\small II}] (right, $6707-6739\AA$) images have been obtained by subtracting continuum emission using the adjacent regions at shorter and longer wavelengths with respect to the emission line systemics.}
\label{IRAS22491_2}
\end{figure*}

\begin{figure*}[t]
\vspace{0.5cm}
\centering

\includegraphics[width=13.cm,trim= 299 10 90 400,clip]{{IRAS22491_Appendix}.pdf}

\caption{\small IRAS F22491-1808 stellar kinematic maps from the pPXF analysis with contours from HST/F160W. The left panel shows the stellar velocity $V_*$, and the right panel represents the velocity dispersion $\sigma_*$.}
\label{IRAS22491_3}
\end{figure*}

\begin{figure*}[t]
\vspace{0.5cm}

\centering
\includegraphics[width=18.cm,trim= 0 0 0 0,clip]{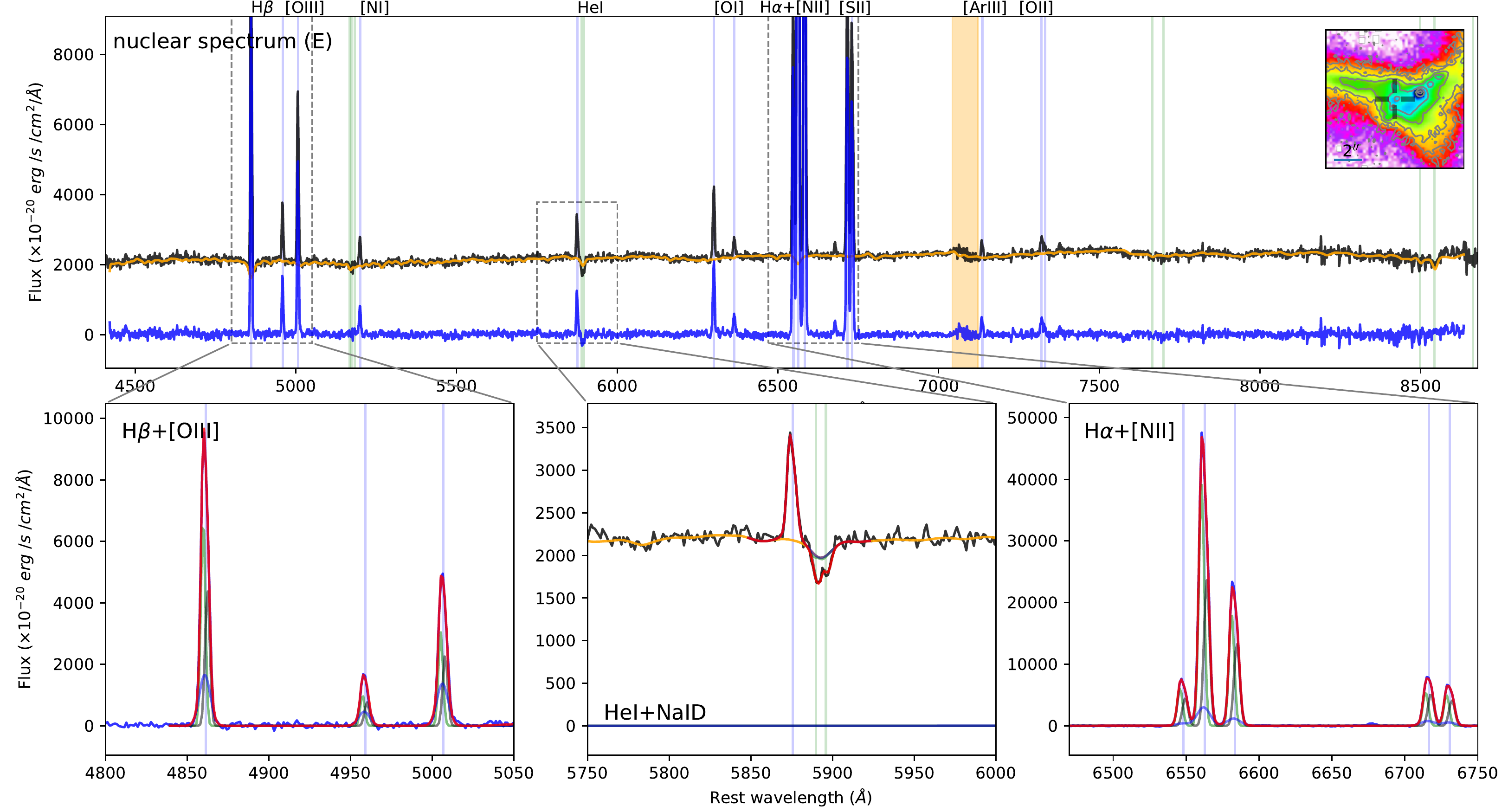}

\caption{\small IRAS F22491-1808 E nuclear spectrum extracted from a circular aperture with $r < 0.4''$, with the corresponding pPXF (top panel) and multi-component (bottom insets) best-fit models. See Fig. \ref{IRAS00188_n} for details.}
\label{IRAS22491_n_e}
\end{figure*}

\begin{figure*}[t]
\vspace{0.5cm}

\centering
\includegraphics[width=18.cm,trim= 0 0 0 0,clip]{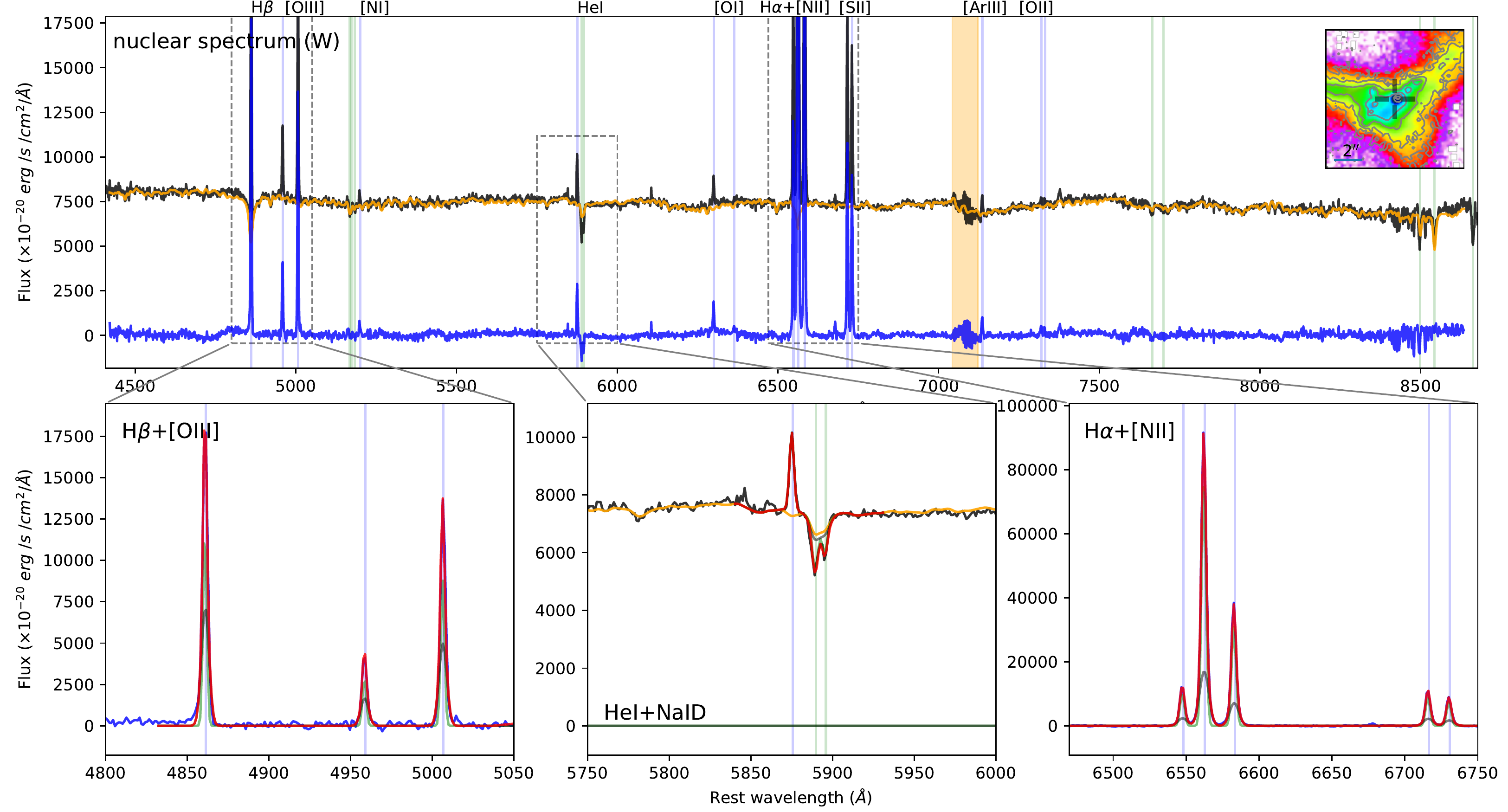}

\caption{\small IRAS F22491-1808 W nuclear spectrum extracted from a circular aperture with $r < 0.4''$, with the corresponding pPXF (top panel) and multi-component (bottom insets) best-fit models. See Fig. \ref{IRAS00188_n} for details.}
\label{IRAS22491_n_w}
\end{figure*}

\end{appendix}

\end{document}